\documentclass[a4paper]{article}
\pdfoutput=1 

\usepackage{setspace}

\usepackage{jheppub}

\usepackage[T1]{fontenc}

\usepackage[latin9]{inputenc}

\newcommand{\noun}[1]{\textsc{#1}}

\usepackage[mathscr]{euscript}

\preprint{\\\\MCnet-15-35\\TUM-HEP-1031/15}

\title{{Extending the M\Large{INLO} \LARGE method}}

\author[a]{Rikkert Frederix,}
\author[b]{Keith Hamilton}

\affiliation[a]{Physik Department T31, Technische Universität München,
\\James-Franck-Str.~1, 85748 Garching, Germany}

\affiliation[b]{Department of Physics and Astronomy, University College London,
\\London, WC1E 6BT, UK}

\emailAdd{rikkert.frederix@tum.de}
\emailAdd{keith.hamilton@ucl.ac.uk}

\abstract{We consider improving \noun{Powheg+Minlo} simulations,
so as to also render them NLO accurate in the description of observables receiving
contributions from events with lower parton multiplicity than present in their
underlying NLO calculation. On a conceptual level we follow the strategy of the
so-called \noun{Minlo}$^{\prime}$  programs.
Whereas the existing \noun{Minlo}$^{\prime}$ framework requires explicit
analytic input from higher order resummation, here we derive
an effective numerical approximation to these ingredients, by imposing unitarity.
This offers a way of extending the  \noun{Minlo}$^{\prime}$ method to more complex
processes, complementary to the known route which uses explicit computations
of high-accuracy resummation inputs. Specifically, we have focused on Higgs-plus-two-jet production (\noun{Hjj}) and related processes. We also consider how one can
cover three units of multiplicity at NLO accuracy, i.e.~we consider how the
\noun{Hjj-Minlo} simulation may yield NLO accuracy for inclusive \noun{H},
\noun{Hj} and \noun{Hjj} quantities. We perform a feasibility study assessing
the potential of these ideas.}


\begin{document}
\maketitle
\flushbottom

\section{Introduction}

In recent years, next-to-leading order parton shower (\noun{Nlops})
matching techniques have been developed and realized as practical
simulation tools, routinely used in LHC data analysis \cite{Frixione:2002ik,Nason:2004rx,Frixione:2007vw,Alioli:2010xd,Platzer:2011bc,Hoeche:2011fd,Alwall:2014hca}.
By now \noun{Nlops} methods have been applied to many processes involving
the production of a primary colourless system, e.g.~a massive boson,
$\mathcal{B}$, in association with jets ($\mathcal{B}n$\noun{j})
\cite{Alioli:2010qp,Frederix:2011ig,Hoeche:2012ft,Campbell:2012am,Re:2012zi}.
A $\mathcal{B}n$\noun{j Nlops} simulation yields NLO accuracy for
$\mathcal{B}+n$-jet inclusive observables, and LO accuracy for $\mathcal{B}+m$-jet
ones ($m=n+1$), while its predictions for more inclusive observables
are divergent. Motivated by the success of leading order matrix element-parton
shower multi-jet merging approaches in the earlier part of the last
decade \cite{Mangano:2001xp,Catani:2001cc,Lonnblad:2001iq,Mrenna:2003if},
it has been considered highly desirable to combine \noun{Nlops} generators
for $\mathcal{B}n$\noun{j} processes corresponding to different jet
multiplicity, $n$, to obtain a unified simulation output, consistently
describing inclusive $\mathcal{B}$, $\mathcal{B}+1$-jet ($\mathcal{B}$\noun{j})
and $\mathcal{B}+2$-jet ($\mathcal{B}$\noun{jj}) observables, simultaneously,
with NLO accuracy.

This merging problem has been addressed by a number of groups in the
last few years~\cite{Lavesson:2008ah,Alioli:2011nr,Hoeche:2012yf,Gehrmann:2012yg,Frederix:2012ps,Alioli:2012fc,Platzer:2012bs}.
All of these approaches separate the output of each component simulation
($\mathcal{B}$, $\mathcal{B}$\noun{j} or $\mathcal{B}$\noun{jj})
according to the jet multiplicity of the events it produces, discarding
those having a multiplicity for which the generator does not possess
the relevant NLO corrections. Having processed the output of each
simulation in this way, the event samples are joined to give an inclusive
sample. Loosely speaking, each generator can be regarded as contributing
a single exclusive jet bin to the final inclusive sample, the magnitude
of each bin being predominantly determined by the jet resolution scale
used in performing the merging, the so-called merging scale. Different
approaches use different means to mitigate the dependence on this
unphysical scale.

If the merging scale is too high one loses the benefits of the higher
multiplicity generators, describing relatively hard jets with tree-level
accuracy, or the parton-shower approximation. If the merging scale
is too low, the inclusive sample is dominated by the higher-multiplicity
generators, which in general leads to \emph{unitarity} \emph{violation},
whereby more inclusive quantities like the total inclusive cross section,
exhibit spurious differences with respect to their corresponding conventional
NLO predictions. The \noun{Geneva }approach \cite{Alioli:2012fc}
can completely avoid unitarity violation, and even the introduction
of a merging scale, by employing very high accuracy resummations.
The method has been demonstrated for effectively merging two units
of multiplicity (without a merging scale) in the context of an\noun{
NNLL$^{\prime}$+NNLO} parton\noun{ }shower matched simulation of
the Drell-Yan process \cite{Alioli:2015toa}. In the sense that it
proposes to resolve the merging problem through implementing sufficiently
high order resummation, \noun{Geneva} represents the best solution
of merging problem. However, details pertaining to exactly how this
is done are subject to debate in the community. In the \noun{Unlops}
approach \cite{Lonnblad:2012ix} unitarity is exactly maintained for
sufficiently inclusive quantities, through what the authors refer
to as `subtract-what-you-add' approach. Nevertheless, the \noun{Unlops}
method is affected by other complications connected to the presence
of a merging scale.

\noun{I}n the\noun{ Minlo }framework \cite{Hamilton:2012np}, fully
differential NLO cross sections for processes of type $\mathcal{B}+n-\mathrm{jets}$
are matched onto a leading log resummation of the exclusive $n-$jet
cross section, as defined by the $k_{t}$-jet algorithm, with $\mathcal{B}$
here referring typically to a given collection of colourless final-state
particles. This is the generalization, to the NLO level, of the resummation
applied in the \noun{Ckkw} formalism \cite{Catani:2001cc,Krauss:2002up}
to the highest multiplicity tree level matrix element \cite{Mrenna:2003if}.
Essentially the $n$-hardest pseudopartons found by the $k_{t}$-jet
algorithm have a distribution which is equivalent to that of a parton
shower simulation of $\mathcal{B}$-production with, in addition,
matching to the exact NLO $\mathcal{B}n$\noun{j} matrix elements.

Whereas previously the leading order parts of these cross sections
themselves would exhibit unphysical divergences when the Born partons
became soft and/or collinear to one another, with the \noun{Minlo
}prescription applied their behavior is instead regular, physical
and sensible, i.e.~$\mathcal{B}n$\noun{j} computations with the \noun{Minlo
}prescription also yield physical results for $\mathcal{B}m$\noun{j
(}$m<n$) and even fully inclusive $\mathcal{B}$-production observables.
In the case of $\mathcal{B}$\noun{j}, with $\mathcal{B}$ a W/Z/Higgs
boson, it was found that the standard \noun{Minlo }procedure yielded
results for inclusive $\mathcal{B}$ production observables equivalent
to conventional NLO ones up to terms $\mathcal{O}(\alpha_{{\scriptscriptstyle \mathrm{S}}}^{3/2})$
relative to the LO component \cite{Hamilton:2012rf}. In the same
article it was proven how, by delicate adjustment of the \noun{Minlo
}Sudakov form factor and clustering procedure, the spurious $\mathcal{O}(\alpha_{{\scriptscriptstyle \mathrm{S}}}^{3/2})$
terms could be eliminated, with the subsequent $\mathcal{B}$\noun{j-Minlo$^{\prime}$}
calculations achieving NLO accuracy for both $\mathcal{B}$ and $\mathcal{B}$\noun{j}
inclusive observables; in the following we call the improved \noun{Minlo}
procedure of ref.~\cite{Hamilton:2012rf} \noun{Minlo$^{\prime}$}
to distinguish it from the original \noun{Minlo }prescription \cite{Hamilton:2012np}.
Thus one obtains, from the single NLO calculation of $\mathcal{B}$\noun{j}
production, also the fully differential NLO calculation of $\mathcal{B}$
production. The \noun{Minlo}$^{\prime}$ calculation can then be matched
to a parton shower using the standard techniques \cite{Frixione:2002ik,Nason:2004rx,Frixione:2007vw}.
When viewed in the context of the recent work in merging multiple
NLO calculations together this amounts to merging without any unphysical
merging scale. It was also demonstrated in refs.~\cite{Hamilton:2013fea,Karlberg:2014qua,Astill:2016hpa}
how to promote the \noun{Minlo$^{\prime}$} simulations to \noun{Nnlops}
simulations.

While the\noun{ }modifications made in going from \noun{Minlo }to
\noun{Minlo$^{\prime}$} involve including higher order terms in the
Sudakov form factor, and lead to the recovery of NLO accuracy also
for inclusive $\mathcal{B}$ observables, the related resummation
is not improved in accuracy. The resummation of the $\mathcal{B}$
system's transverse momentum is $\mathrm{NNLL}_{\sigma}$ accurate\footnote{ $\mathrm{LL}_{\sigma}$ refers to the resummation of the leading
log tower in the spectrum, containing terms $\bar{\alpha}_{{\scriptscriptstyle \mathrm{S}}}^{n}L^{2n-1}$,
$\mathrm{NLL}_{\sigma}$ refers to the next-to-leading log tower,
$\bar{\alpha}_{{\scriptscriptstyle \mathrm{S}}}^{n}L^{2n-2}$, and
$\mathrm{NNLL}_{\sigma}$ the next-to-next-to-leading log series,
$\bar{\alpha}_{{\scriptscriptstyle \mathrm{S}}}^{n}L^{2n-3}$. %
} 
\cite{Ellis:1997ii,Frixione:1998dw} before and after the inclusion
of the latter terms in the Sudakov form factor \cite{Hamilton:2012rf}
(and before \noun{Nlops} matching adds ambiguities). Thus, 
\noun{Minlo$^{\prime}$} amounts to the \noun{Minlo }method with
additional, subtle, unitarization. This is in difference to the \noun{Geneva
}approach, wherein higher order resummation is taken as the main defining
specification, with unitarization coming `for free' along with the
latter \cite{Alioli:2012fc,Alioli:2013hqa,Alioli:2015toa}. In this
sense \noun{Minlo$^{\prime}$} is, minimally, the same as the \noun{Powheg }method. Indeed, in the \noun{Powheg} method
a very specific Sudakov form factor is required to achieve
an exact unitarization, needing terms in the exponent that are sub-leading
with respect to the resummation accuracy to do so, including even
power suppressed terms that are nothing to do with resummation.

To realize the \noun{Minlo$^{\prime}$} method one needs to know the
$v\rightarrow0$ singular part of the $\mathcal{B}$\noun{j} cross
section differential in the underlying Born variables, $\Phi_{{\scriptscriptstyle \mathcal{B}}}$,
describing the kinematics of the produced $\mathcal{B}$-final state,
where $v$ is a variable measuring radiation hardness, at NLO; this
information may be obtained from suitably integrated NLO predictions
for the spectrum, or from fixed order expansion of $\mathrm{N}^{3}\mathrm{LL}_{\sigma}$
resummation.
In the case of ref.~\cite{Hamilton:2012rf} $v$ was given by the
transverse momentum of the produced $W/Z/$Higgs boson, for which
the latter NLO distributions have long been known in the literature
\cite{Arnold:1990yk,deFlorian:2001zd,Glosser:2002gm}. Recently these
distributions have also become available to the same level of accuracy
for the transverse momentum of the hardest produced jet \cite{Banfi:2012yh,Becher:2012qa,Tackmann:2012bt,Banfi:2012jm,Becher:2013xia,Banfi:2013eda,Banfi:2015pju},
with which an equally accurate \noun{Minlo$^{\prime}$ }calculation could have been developed.
For more complicated observables, such as those which might be used
for implementing the \noun{Minlo$^{\prime}$} method in the context
of higher-multiplicity processes, with the exception of the $N$-jettiness
variable \cite{Stewart:2009yx,Stewart:2010qs,Stewart:2010tn,Berger:2010xi,Jouttenus:2011wh,Boughezal:2015dva,Boughezal:2015aha,Gaunt:2015pea},
these distributions are (so far) not available in the literature.
We note, however, that important progress is being made in the direction
of automated approaches to final-state resummation at NNLL
\cite{Banfi:2014sua}, valid for broad classes of observables, including
those we consider in the present work.
Whenever these theoretical ingredients become
available, the \noun{Minlo$^{\prime}$} method is in principle straightforward
to apply, to make a NLO $\mathcal{B}n$\noun{j} calculation simultaneously
NLO accurate for $\mathcal{B}m$\noun{j} ($m=n-1$) observables etc;
many of the details for that are clarified by the present work.

Nevertheless, even when all the necessary theoretical ingredients
are at hand, experience with the $\mathcal{B}$\noun{j-Minlo$^{\prime}$}
calculations tells that the results of implementations are sometimes
not as ideal as one might have liked. The \noun{Minlo$^{\prime}$
}codes are proven to return conventional NLO results for inclusive
$\mathcal{B}$ and $\mathcal{B}$\noun{j} observables up to NNLO sized
ambiguities and power corrections. In practice the \noun{Hj-Minlo$^{\prime}$}
calculation was found to give very satisfactory agreement with the
regular NLO predictions for inclusive Higgs boson production.
On the other hand, comparing \noun{Wj-Minlo$^{\prime}$ }and \noun{Zj-Minlo$^{\prime}$}
predictions for inclusive W and Z production to those of regular
NLO calculations, one can see numerical differences between the two
sets of formally equivalent results, which don't really sit easily
with the fact that the two formally agree up to NNLO-sized ambiguities.
In the \noun{Wj-Minlo$^{\prime}$ }case, the inclusion of the relevant
NNLL terms in the Sudakov form factor do not lead to noticeably better
agreement with the conventional NLO cross sections than those obtained
with the original \noun{Minlo} prescription. We also point out that
the true NNLO corrections to Higgs boson production are large, $\sim20-30\%$,
thus the almost perfect agreement of \noun{Hj-Minlo$^{\prime}$} with
conventional NLO calculations for inclusive Higgs boson production
--- which looks to be a striking vindication of the theoretical framework
--- should be considered fortuitous.

Some people (like us) may be dismissive of numerical disagreement
between $\mathcal{B}$\noun{j-Minlo$^{\prime}$ }and standard NLO
predictions for fully inclusive observables, since the \noun{Minlo$^{\prime}$}
method has been rigorously proven. Others may be less comfortable
accepting the fact that these differences arise from contributions
beyond the formal accuracy of either type of calculation, given their
size in some cases. If $5-6\%$ differences can be found in total
inclusive cross sections for inclusive W and Z production, it does
not seem unreasonable to expect that larger differences may be found
in more complex processes, with more powers of the strong coupling
associated to the LO contribution and a richer kinematic content.
Assuming one is content to dismiss differences due to higher order
ambiguities, for complex processes, with even more complex calculations
underlying them, it will be difficult to satisfy oneself that the
level of numerical disagreement is of this kind.

A final motivation for considering to extend the reach of the \noun{Minlo$^{\prime}$}
method is that of merging NLO calculations differing by more than
one unit of jet multiplicity. Specifically, one would ultimately like
a \noun{Minlo$^{\prime}$} procedure applied to $\mathcal{B}$\noun{jj-Minlo
}such that it retrieves NLO accuracy for inclusive $\mathcal{B}$,
$\mathcal{B}$\noun{j }and $\mathcal{B}$\noun{jj }observables. Naive
extension of the \noun{Minlo$^{\prime}$} method then implies having
a $\mathrm{N}^{3}\mathrm{LL}_{\sigma}$-accurate nested resummation
with which to base it on. While the resummation community is making
impressive progress in recent years \cite{Pietrulewicz:2016nwo}, the
prospects for obtaining the high-accuracy ingredients needed to tackle
this issue in the near future are unclear to us.

Noting these desirable and undesirable features of the existing \noun{Minlo$^{\prime}$}
method, we investigated extending it in a number of ways: 
\begin{enumerate}
\item The \noun{Minlo$^{\prime}$} specification can be reached with only
limited knowledge of the required Sudakov form factors (at least 
$\mathrm{NLL}_{\sigma}$).

Thus, one can begin to make \noun{Minlo$^{\prime}$} simulations for
more complex processes.

\item \noun{$\mathcal{B}$jj/$\mathcal{B}$j-Minlo$^{\prime}$} predictions
for \noun{$\mathcal{B}$j/}inclusive\noun{ $\mathcal{B}$ }observables
agree precisely with those of the corresponding conventional NLO calculations.

Numerical ambiguities between conventional inclusive NLO predictions
and the associated \noun{Minlo$^{\prime}$ }ones can be largely eliminated.

\item NLO \noun{$\mathcal{B}$jj} calculations can yield simultaneously
NLO predictions for $\mathcal{B}$\noun{j} and inclusive $\mathcal{B}$
production observables.

The method to produce\noun{ Nnlops }simulations of $\mathcal{B}$
production can follow exactly as in ref.~\cite{Hamilton:2013fea}.

\end{enumerate}
In the present work we suggest an alternative approach to \noun{Minlo$^{\prime}$},
meeting the objectives listed overhead, and we present a feasibility
study confirming its potential. The basic concept is very close to that of the
original \noun{Minlo$^{\prime}$} method and, more broadly speaking,
the \noun{Powheg} approach itself. As with the original \noun{Minlo$^{\prime}$}
formulation, we attribute discrepancies of \noun{$\mathcal{B}m$j-Minlo}
predictions for \noun{$\mathcal{B}n$j}  ($m=n+1$) inclusive quantities, with
respect to conventional fixed order results, as owing to deficiencies
in the Sudakov form factor employed in the former. In the
existing \noun{Minlo$^{\prime}$} framework, the correction to the
Sudakov form factor which leads to the elimination of these discrepancies,
is derived from highly intricate, third-party, analytic computations,
of the NLO \noun{$\mathcal{B}m$j} radiation spectrum. Here, instead,
we determine the relation between \noun{$\mathcal{B}m$j-Minlo}
predictions for \noun{$\mathcal{B}n$j} Born kinematics and those of
conventional (N)NLO, in terms of the a priori unknown correction factor
to the \noun{$\mathcal{B}m$j-Minlo} Sudakov form factor. Since both
the \noun{$\mathcal{B}m$j-Minlo} and conventional fixed order predictions
for the Born kinematics are to-hand, we can then solve this relation
for the unknown Sudakov correction, numerically, to sufficent accuracy,
bringing the \noun{$\mathcal{B}n$j} Born kinematics of
\noun{$\mathcal{B}m$j-Minlo} into complete agreement with regular (N)NLO
predictions. This then renders \noun{$\mathcal{B}m$j-Minlo} (N)NLO accurate
for \noun{$\mathcal{B}n$j} inclusive observables, while maintaining NLO
accuracy for \noun{$\mathcal{B}m$j} ones. We arrive at the aforesaid
defining equation for our method by manipulating the original
\noun{Minlo$^{\prime}$} computation, neglecting terms which lead to
irrelevant higher order ambiguities
(sects.~\ref{sub:MiNLO-jet-resolution-spectrum}-\ref{sub:Removal-of-spurious-terms}). With such an approach the \noun{Minlo$^{\prime}$} and \noun{Nnlops}
methods become much more easily/widely applicable than before,
being no longer contingent on the existence of high accuracy, observable-
and process-dependent analytic ingredients. To implement this approach
it is sufficient to have only $\mathrm{NLL}_{\sigma}$ accuracy in the
initial, uncorrected, \noun{Minlo} simulation and an NLO (or NNLO)
prediction for the Born kinematics of the associated lower multiplicity
process. At the same time, residual ambiguities between
\noun{Minlo$^{\prime}$} predictions and conventional NLO/\noun{Nlops}
are brought under much tighter control, and can be completely removed,
if so desired.

We do not
propose to replace the existing \noun{Minlo$^{\prime}$} method, but
rather to supplement it and, as such, we don't question the importance
of efforts to provide further resummation input which that fundamentally
needs --- on the contrary, it's clear that work is, at the very least,
complementary to the improvements discussed here. The study of the
problem of combining multiple NLO simulations together is not long
started though, so we consider more options, understanding and investigations
in this direction, to be still welcome.

In section \ref{sec:Merging-without-a-merging-scale-two-units} we
discuss the \noun{Minlo} procedure and its extension(s), in the context
of recovering NLO accuracy for $\mathcal{B}m\noun{j}$ processes
from NLO $\mathcal{B}n\noun{j}$ calculations, $m=n-1$, without any
merging scale, focusing on $\mathcal{B}\noun{j}$ and $\mathcal{B}\noun{jj}$. We derive \noun{$\mathrm{NNLL}_{\sigma}$ }resummation
formulas from the \noun{Caesar} formalism \cite{Banfi:2004yd} and
compare these to the equivalent \noun{Minlo}/C\noun{kkw} results.
This reveals conventional ways in which to improve the $\mathcal{B}$\noun{jj-Minlo}
procedure, in particular it gives details for improving the accuracy
of the Sudakov form factor to $\mathrm{NNLL}_{\sigma}$; we leave
the implementation of such improvements for future work. With the
true resummation clarified at $\mathrm{NNLL}_{\sigma}$, by the latter
comparison to \noun{Caesar}, we proceed with clarity to propose how
one could infer an effective approximation to higher order Sudakov
terms, needed by \noun{Minlo$^{\prime}$, }by imposing unitarity.
In section \ref{sec:Merging-without-a-merging-scale-three-units}
we propose that the latter method can also be used for the purposes
of rendering $\mathcal{B}\noun{jj}$ \noun{Minlo }simulations simultaneously
NLO accurate in the description of inclusive $\mathcal{B}$ and $\mathcal{B}\noun{j}$
production. Section \ref{sec:Feasibility-study} presents a feasibility
study assessing the potential of these ideas. We summarize our findings
and conclude in sect.~\ref{sec:Conclusion}.

\section{Merging two units of multiplicity\label{sec:Merging-without-a-merging-scale-two-units}}

In the following we ultimately present a method for merging \noun{Nlops
}simulations of $\mathcal{B}$- and $\mathcal{B}$\noun{j-}production
and, separately, $\mathcal{B}$\noun{j}- and $\mathcal{B}$\noun{jj-}production
without any actual merging. More precisely, the improved \noun{Minlo}
procedure will render the $\mathcal{B}$\noun{j }simulation also NLO
accurate for $\mathcal{B}$-production and, in the case of $\mathcal{B}$\noun{jj
}it will build in NLO precision for $\mathcal{B}$\noun{j}.

We remind that in this work we refer to the leading tower of logarithms
in the spectrum, terms $\mathcal{O}(\bar{\alpha}_{{\scriptscriptstyle \mathrm{S}}}^{n}L^{2n-1})$,
as $\mathrm{LL}_{\sigma}$, with $\mathrm{NLL}_{\sigma}$ denoting
the next-to-leading log tower, $\mathcal{O}(\bar{\alpha}_{{\scriptscriptstyle \mathrm{S}}}^{n}L^{2n-2})$,
$\mathrm{NNLL}_{\sigma}$ for the next-to-next-to-leading log tower
$\mathcal{O}(\bar{\alpha}_{{\scriptscriptstyle \mathrm{S}}}^{n}L^{2n-3})$,
and so forth.

In section \ref{sub:Definitions-jet-resolutions} we introduce preliminary
notation and definitions, in particular regarding the clustering variables
which the \noun{Minlo} procedure is to resum, and the so-called underlying
Born kinematics that the resummation is performed about. In section
\ref{sub:NNLLsigma-jet-resolution-spectrum} we present a formula
for $\mathrm{NNLL}_{\sigma}$ resummation of these clustering variables
($k_{t}$-jet resolutions) based on the \noun{Caesar} resummation
framework \cite{Banfi:2004yd}. The Sudakov form factors of the latter
are compared to the corresponding \noun{Minlo} formulae in section \ref{subsec:comparison-to-ckkw-minlo}. In section \ref{sub:MiNLO-jet-resolution-spectrum}
we derive the fixed order expansion of the $\mathrm{NNLL}_{\sigma}$
\noun{Caesar} formula and from this we show how our \noun{Minlo }procedure
applied to the $\mathcal{B}$\noun{j(j) }NLO computations returns
a matched, resummed, NLO accurate jet resolution spectrum. In doing
so, we also assume that spurious, unknown, $\mathrm{NNLL}_{\sigma}$
or $\mathrm{N^{3}LL}_{\sigma}$ terms may arise in the \noun{Minlo}
resummation, owing to a lack of understanding of the true $\mathrm{N}^{3}\mathrm{LL}_{\sigma}$
spectrum truncated at NLO, and we closely monitor how these propagate
through the \noun{Minlo} procedure, to better understand and eradicate
them, as needed. In section \ref{sub:MiNLO-integrated-jet-resolution-spectrum}
we integrate over the \noun{Minlo} jet resolution spectrum, determining
that the distribution of the Born kinematics differs from that of
conventional NLO owing to the latter spurious terms, which we have
tracked and quarantined. In section \ref{sub:Removal-of-spurious-terms}
we first demonstrate how the original \noun{Minlo$^{\prime}$} approach
removes such terms, by explicitly correcting the \noun{Minlo} Sudakov
form factor, leading to NLO accurate Born kinematics. In the second
part of section \ref{sub:Removal-of-spurious-terms}, we present our
new proposal. Over-simplifying somewhat, this amounts to using the
constraint that the corrected \noun{Minlo} predictions must recover
NLO (or NNLO) results for the inclusive Born kinematics, as a defining equation
for the elusive Sudakov correction factors, needed to promote \noun{Minlo}
to \noun{Minlo}$^{\prime}$. This equation can no doubt be solved for the
latter in several ways, and we present one basic, simple, way to do so.
In order to maintain NLO accuracy in the higher multiplicity phase space,
it  is necessary that the initial \noun{Minlo} resummation be at least
$\mathrm{NLL}_{\sigma}$ accurate, however, this is an easily obtainable
threshold by today's standards.

We underline now that it is a working assumption of the \noun{Caesar} 
framework, that the underlying Born kinematics, about which
the resummation of soft radiation is carried out, are not themselves
associated with large logarithmic corrections, i.e.~it is assumed
that the radiating particles in the hard underlying Born kinematics
are well separated. For the case of $\mathcal{B}$ production, with
only two radiating particles in the initial-state, the latter criterion
is fulfilled automatically. On the other hand, for $\mathcal{B}$\noun{j}
production it implies that we have to restrict ourselves to a regime in
which the final-state (pseudo-)parton in the underlying Born has transverse
momentum of order the mass of $\mathcal{B}$, or greater. In other
words, in this section \ref{sec:Merging-without-a-merging-scale-two-units}
it should be understood that the $y_{12}$ resummation in $\mathcal{B}$\noun{j}-production
assumes $y_{01}\gtrsim\mathcal{O}(m_{{\scriptscriptstyle \mathcal{B}}}^{2})$.
Only in section \ref{sec:Merging-without-a-merging-scale-three-units}
will we consider extending down into the region where the transverse
momentum of the final-state Born (pseudo-)parton is small.

\subsection{Definitions: jet resolutions and underlying Born kinematics\label{sub:Definitions-jet-resolutions}}

Since it is underlies the whole discussion we first quickly present
a reminder of the exclusive $k_{t}$-jet clustering procedure (for
brevity we henceforth refer to pseudopartons obtained in the clustering
sequence as just partons): 
\begin{enumerate}
\item In an $n$-parton final-state we determine the smallest distance 
\[
d^{\left(n\right)}=\mathrm{min}\left(\left\{ d_{ij}\right\} ,\left\{ d_{iB}\right\} \right)\,,
\]
where $\left\{ d_{ij}\right\} $ is the set of measures 
\[
d_{ij}=\mathrm{min}(k_{t,i}^{2},k_{t,j}^{2})\,(\left(\mathrm{y}_{i}-\mathrm{y}_{j}\right)^{2}+\left(\phi_{i}-\phi_{j}\right)^{2})/R^{2}\,,
\]
obtained by considering all pairwise combinations of final-state partons
$i$ and $j$, with $k_{t,i}$, $\mathrm{y}_{i}$ and $\phi_{i}$
being, respectively, the transverse momentum, rapidity and azimuth
of parton $i$, and $\{d_{iB}\}$ the set of all final-state transverse
momenta:
\[
d_{iB}=k_{t,i} \, .
\]
$R$ is the so-called jet radius parameter
which we take equal to be $1$. 
\item If $d^{\left(n\right)}=d_{ij}$ partons $i$ and $j$ are replaced
in the event by a single mother parton with four momentum $p_{i}+p_{j}$,
otherwise, if $d^{\left(n\right)}=d_{i}$, particle $i$ is considered
to have been similarly absorbed in one of the beam jets and is deleted
from the event. 
\item If further partons remain return to step 1, otherwise the clustering
sequence terminates. 
\end{enumerate}
In order to have the $\mathcal{B}$\noun{j-Minlo} calculation return
NLO accuracy for inclusive $\mathcal{B}$ production and $\mathcal{B}$\noun{jj-Minlo}
likewise reproduce NLO accuracy for $\mathcal{B}$\noun{j }quantities,
we are interested to resum the $k_{t}$-jet resolution variables $y_{01}$
and $y_{12}$, which we here define as

\begin{eqnarray*}
y_{01}=\max_{n\ge1}\{d^{\left(n\right)}\}\,, & \quad\quad & v_{01}=y_{01}/Q_{{\scriptscriptstyle \mathcal{B}\mathrm{\phantom{J}}}}^{2}\,,\\
y_{12}=\max_{n\ge2}\{d^{\left(n\right)}\}\,, & \quad\quad & v_{12}=y_{12}/Q_{{\scriptscriptstyle \mathcal{B}\mathrm{J}}}^{2}\,,
\end{eqnarray*}
where $Q_{{\scriptscriptstyle \mathcal{B}}}$ and $Q_{{\scriptscriptstyle \mathcal{B}\mathrm{J}}}$
are, in the context of the \noun{Caesar }resummation framework \cite{Banfi:2004yd},
the hard scales of the problem, largely determined by the respective
Born kinematics. We take $Q_{{\scriptscriptstyle \mathcal{B}}}^{2}=m_{{\scriptscriptstyle \mathcal{B}}}^{2}$,
where $m_{{\scriptscriptstyle \mathcal{B}}}$ is the invariant mass
of the system $\mathcal{B}$, and $Q_{{\scriptscriptstyle \mathcal{B}\mathrm{J}}}^{2}=Q_{{\scriptscriptstyle \mathcal{B}}}^{2}v_{01}=y_{01}$.
With this definition the resummation of $y_{12}$ amounts to --- up
to corrections owing to the lack of monotonicity of the clustering
sequence in $d^{\left(n\right)}$, which we neglect --- a resummation
of large logarithms of the transverse momentum of the second hardest
relative to the hardest branching in the event. For what follows we
notate these large logarithms 
\[
L_{01}=\log\frac{1}{v_{01}},\quad\quad L_{12}=\log\frac{1}{v_{12}}\,.
\]
In discussions and formulae that apply equally well to the $\mathcal{B}$\noun{j-Minlo
}and $\mathcal{B}$\noun{jj-Minlo }computations we simply use $L$
to refer to either $L_{01}$ or $L_{12}$. Equally, we will use $y$
to ambiguously mean $y_{01}$ and $y_{12}$, and $v$ to mean $v_{01}$
and $v_{12}$, when safe to do so.

We now introduce the kinematic variables specifying the hard configurations
about which we intend \noun{Minlo} to resum the $y_{01}$ and $y_{12}$
variables. First consider applying the $k_{t}$-jet algorithm to events
such that they are clustered to the point of containing just a single
jet (pseudoparton) and the system/particle $\mathcal{B}$. We define
directly from such ensembles $\mathcal{B}$\noun{j} underlying Born
variables, $\Phi_{{\scriptscriptstyle \mathcal{B}\mathrm{J}}}=\{\hat{\Phi}_{{\scriptscriptstyle \mathcal{B}}},\mathrm{y}_{{\scriptscriptstyle \mathcal{B}}},\mathrm{y}_{{\scriptscriptstyle \mathrm{J}}},p_{{\scriptscriptstyle \mathrm{T}}}^{{\scriptscriptstyle \mathrm{J}}},\phi_{{\scriptscriptstyle \mathrm{J}}}\}$,
where the set $\hat{\Phi}_{{\scriptscriptstyle \mathcal{B}}}$ specifies
the configuration of $\mathcal{B}$ in its rest frame, including its
invariant mass,%
\footnote{If $\mathcal{B}$ is a single particle, as in the case of Higgs boson
production, $\hat{\Phi}_{{\scriptscriptstyle \mathcal{B}}}$, is just
the invariant mass of the particle.%
} $\mathrm{y}_{{\scriptscriptstyle \mathrm{J}}}$ is the rapidity of
the jet, $\mathrm{y}_{{\scriptscriptstyle \mathcal{B}}}$ is the rapidity
of $\mathcal{B}$, $p_{{\scriptscriptstyle \mathrm{T}}}^{{\scriptscriptstyle \mathrm{J}}}$
the transverse momentum of the jet, and $\phi_{{\scriptscriptstyle \mathrm{J}}}$ its azimuthal angle. After a subsequent clustering
with the $k_{t}$-jet algorithm the jet/pseudoparton is also removed
leaving just the system $\mathcal{B}$, for which we further define
$\mathcal{B}$ underlying Born variables $\Phi_{{\scriptscriptstyle \mathcal{B}}}=\{\hat{\Phi}_{{\scriptscriptstyle \mathcal{B}}},\mathrm{y}_{{\scriptscriptstyle \mathcal{B}}}\}$.
Thus we can also write $\Phi_{{\scriptscriptstyle \mathcal{B}\mathrm{J}}}=\{\Phi_{{\scriptscriptstyle \mathcal{B}}},\mathrm{y}_{{\scriptscriptstyle \mathrm{J}}},p_{{\scriptscriptstyle \mathrm{T}}}^{{\scriptscriptstyle \mathrm{J}}},\phi_{{\scriptscriptstyle \mathrm{J}}}\}$.

The definitions of $\Phi_{{\scriptscriptstyle \mathcal{B}\mathrm{J}}}$
and $\Phi_{{\scriptscriptstyle \mathcal{B}}}$ can also be considered
as projections from real (or multiple emission) kinematics onto Born
kinematics for $\mathcal{B}$\noun{j }and $\mathcal{B}$ final-states
respectively; note that, strictly speaking, in that context the jet
in the projected $\mathcal{B}$\noun{j} kinematics should be understood
as being massless. The choices of $\mathrm{y}_{{\scriptscriptstyle \mathcal{B}}}$
and $\mathrm{y}_{{\scriptscriptstyle \mathrm{J}}}$ are motivated
by our expectation that even basic formulations of the \noun{$\mathcal{B}$j-}
and \noun{$\mathcal{B}$jj-Minlo}$^{\prime}$ calculations will reproduce
well the shapes of these quantities, as they are predicted in the
respective (conventional) NLO $\mathcal{B}$ and $\mathcal{B}$\noun{j
}computations. Our choosing of $p_{{\scriptscriptstyle \mathrm{T}}}^{{\scriptscriptstyle \mathrm{J}}}$
in $\Phi_{{\scriptscriptstyle \mathcal{B}\mathrm{J}}}$ is made in
light of the fact that this variable, as defined here, is equal to
$\sqrt{y_{01}}$, which we expect to greatly increase the consistency
between the $\mathcal{B}$\noun{j-Minlo}$^{\prime}$ (which resums
precisely $v_{01}$) and the $\mathcal{B}$\noun{jj-Minlo}$^{\prime}$
calculations, which we intend to have identically reproduce NLO $\mathcal{B}$\noun{j}
predictions such as $p_{{\scriptscriptstyle \mathrm{T}}}^{{\scriptscriptstyle \mathrm{J}}}$,
as defined here, according to the exclusive $k_{t}$-jet clustering
algorithm. The latter consideration is important in the context of
nesting the $v_{01}$ and $v_{12}$ resummations, with a view to having
a $\mathcal{B}$\noun{jj-Minlo}$^{\prime}$ simulation NLO accurate
for $\mathcal{B}$\noun{jj}, $\mathcal{B}$\noun{j}, and inclusive
$\mathcal{B}$\noun{ }production processes.\noun{ }

\subsection{\texorpdfstring{$\mathrm{NNLL}_{\sigma}$}{NNLLo} resummation \label{sub:NNLLsigma-jet-resolution-spectrum}}

By differentiating and expanding the master resummation formula in
ref.~\cite{Banfi:2004yd}, we are able to derive simultaneously $\mathrm{LL}$
and $\mathrm{NNLL}_{\sigma}$ accurate expressions for the $v_{01}$
and $v_{12}$ spectra in $\mathcal{B}$ and $\mathcal{B}$\noun{j
}production processes respectively. In a nutshell, one takes the $\mathrm{NLL}$
resummation of ref.~\cite{Banfi:2004yd}, matched to NLO, and proceeds
to omit NLL terms $\mathcal{O}(\bar{\alpha}_{{\scriptscriptstyle \mathrm{S}}}^{3})$
and beyond in the resummed exponent, specifically those due to observable-dependent
multiple emission effects. Details on these manipulations can be found
in appendix \ref{sub:Derivation-of-NNLLsigma-jet-resolution-spectra}.
The general expression we derive can be written simply in
the form \cite{Dokshitzer:1978hw}:%
\footnote{The subscript ${\cal \mathcal{R}}$ is used here to distinguish the
cross section as the resummed cross section. %
} 
\begin{eqnarray}
\frac{d\sigma_{{\scriptscriptstyle \mathcal{R}}}}{d\Phi dL} & = & \frac{d\sigma_{0}}{d\Phi}\,\left[1+\bar{\alpha}_{{\scriptscriptstyle \mathrm{S}}}\left(\mu_{{\scriptscriptstyle R}}^{2}\right)\,\mathcal{H}_{1}\left(\mu_{{\scriptscriptstyle R}}^{2}\right)\right]\,\frac{d}{dL}\,\left[\exp\left[\,-R\left(v\right)\,\right]\,\mathcal{L}\left(\left\{ x_{\ell}\right\} ,\mu_{{\scriptscriptstyle F}},v\right)\right]\,,\label{eq:sect22-NNLL_sigma-differential-xsecn-i}
\end{eqnarray}
where $\mathcal{L}$ is our luminosity factor 
\begin{eqnarray}
\mathcal{L}\left(\left\{ x_{\ell}\right\} ,\mu_{{\scriptscriptstyle F}},v\right) & = & \prod_{\ell=1}^{n_{i}}\frac{q^{\left(\ell\right)}\left(x_{\ell},\mu_{{\scriptscriptstyle F}}^{2}v\right)}{q^{\left(\ell\right)}\left(x_{\ell},\mu_{{\scriptscriptstyle F}}^{2}\right)}\left[1+\sum_{\ell=1}^{n_{i}}\,\bar{\alpha}_{{\scriptscriptstyle \mathrm{S}}}\left(\mu_{{\scriptscriptstyle R}}^{2}v\right)\,\frac{\left[\boldsymbol{\mathcal{C}}_{1}\otimes\boldsymbol{\mathrm{q}}^{\left(\ell\right)}\right]_{i}\left(x_{\ell},\mu_{{\scriptscriptstyle F}}^{2}v\right)}{q^{\left(\ell\right)}\left(x_{\ell},\mu_{{\scriptscriptstyle F}}^{2}v\right)}\right]\,.\label{eq:sect22-NNLL_sigma-differential-xsecn-ii}
\end{eqnarray}
Since this form applies to both jet resolutions in $\mathcal{B}$
and $\mathcal{B}$\noun{j }production, the components inside it should
be understood as referring to one of these two processes, e.g.~$\Phi$
and $L$ refer to $\Phi_{{\scriptscriptstyle \mathcal{B}\mathrm{J}}}$
and $L_{\text{12}}$ in the $\mathcal{B}$\noun{j }case, and $\Phi_{{\scriptscriptstyle \mathcal{B}}}$
and $L_{01}$ in $\mathcal{B}$\noun{-}production. Equally, $v$ refers
to $v_{01}$ in the latter case and $v_{12}$ in the former.

First let's overview the resummation formula, eq.~\ref{eq:sect22-NNLL_sigma-differential-xsecn-i},
before disappearing into the details. The first factor in eq.~\ref{eq:sect22-NNLL_sigma-differential-xsecn-i},
$d\sigma_{0}/d\Phi$, denotes the leading order cross section for
$\mathcal{B}$ or $\mathcal{B}$\noun{j }processes as appropriate.
Within $d\sigma_{0}/d\Phi$ the scale used for the evaluation of the
parton distribution functions is $\mu_{{\scriptscriptstyle F}}$ and
the scale in any implicit strong coupling constant factors is $\mu_{{\scriptscriptstyle R}}$.
The function $\mathcal{H}_{1}$ includes hard virtual corrections
to $d\sigma_{0}/d\Phi$. The Sudakov form factor is present in eq.~\ref{eq:sect22-NNLL_sigma-differential-xsecn-i}
as $\exp\left[-R\left(v\right)\right]$; here we have made a simplification
with respect to the notation of ref.~\cite{Banfi:2004yd}, including
in its definition contributions from soft-wide angle radiation and
observable-dependent multiple emission corrections. The $q^{\left(\ell\right)}\left(x_{\ell},\mu^{2}\right)$
terms in the luminosity factor, eq.~\ref{eq:sect22-NNLL_sigma-differential-xsecn-ii},
are parton distribution functions (PDFs), for a given incoming leg,
$\ell$, with momentum fraction $x_{\ell}$, evaluated at scale $\mu$.
The product of PDF ratios runs over $n_{i}=2$ incoming legs. The
functions $\mathcal{C}_{1}$ involved in convolutions with PDFs in
the luminosity factor, $\mathcal{L}$, are due to universal hard-collinear
corrections. The renormalization and factorization scales $\mu_{{\scriptscriptstyle R}}$
and $\mu_{{\scriptscriptstyle F}}$ are understood as being $\sim Q$
\cite{Banfi:2004yd}.

To try to lighten the formulae we use the following abbreviations,
\begin{equation}
\bar{\alpha}_{{\scriptscriptstyle \mathrm{S}}}=\frac{\alpha_{{\scriptscriptstyle \mathrm{S}}}}{2\pi}\,,\quad\beta_{0}=\frac{11C_{A}-2n_{f}}{12\pi}\,,\quad\bar{\beta}_{0}=2\pi\beta_{0}\,.\label{eq:sect22-aSbar-beta0bar-beta0-defns}
\end{equation}

The Sudakov form factor exponent in the $\mathrm{NNLL}_{\sigma}$
differential cross section formula (eq.~\ref{eq:sect22-NNLL_sigma-differential-xsecn-i})
is given by 
\begin{equation}
-R\left(v\right)=\int_{0}^{L}dL^{\prime}\,\left[\bar{\alpha}_{{\scriptscriptstyle \mathrm{S}}}\left(y^{\prime}\right)\,\left[2G_{12}\, L^{\prime}+G_{11}+2S_{1}\right]+\bar{\alpha}_{{\scriptscriptstyle \mathrm{S}}}^{2}\left(y^{\prime}\right)\,2G_{12}\,\left[K+4\mathcal{F}_{2}G_{12}\right]\, L^{\prime}\right]\,.\label{eq:sect22-Big-Sudakov-minus-Ry}
\end{equation}
The $G_{ij}$ contributions are due to independent soft-collinear
/ collinear emission contributions, they are given by 
\begin{equation}
\begin{array}{lll}
G_{12}=-\frac{1}{2}\,\sum_{\ell}\, C_{\ell}\,, & \quad\quad & G_{11}=-2\,\sum_{\ell}\, B_{\ell}C_{\ell}\,,\end{array}\label{eq:sect22-G12-G11-defns}
\end{equation}
where 
\begin{equation}
\left.\begin{array}{l}
C_{\ell}=C_{F}\\
B_{\ell}=-\frac{3}{4}
\end{array}\right\} \,\mbox{ for a quark leg}\,,\quad\quad\left.\begin{array}{l}
C_{\ell}=C_{A}\\
B_{\ell}=-\frac{\bar{\beta}_{0}}{2C_{A}}
\end{array}\right\} \,\mbox{ for a gluon leg}.\label{eq:sect22-Cl-Bl-defns}
\end{equation}
The sum, $\sum_{\ell}$, runs over all $n$ hard colour-charged legs
$\ell$ --- $n=2$ for $\mathcal{B}$-production, $n=3$ for $\mathcal{B}$\noun{j}.
Single logarithmic soft-wide angle emission contributions are included
via the $S_{1}$ term. Soft-wide angle radiation is obviously sensitive
to the structure of the underlying hard event on large angular scales,
so in contrast to the collinear contributions above, this piece is
sensitive to the orientation of the hard external legs and not just
their charges. For $\mathcal{B}$ ($n=2$) and $\mathcal{B}$\noun{j($n=3$)}
processes we have 
\begin{eqnarray}
n=2:\quad S_{1} & = & -\left(C_{q}+C_{q^{\prime}}\right)\ln\frac{Q_{qq^{\prime}}}{Q_{{\scriptscriptstyle \mathcal{B}}}}\,,\quad\label{eq:sect22-n-eq-2-S1-defn}\\
n=3:\quad S_{1} & = & -\frac{1}{2}\left(C_{q}+C_{q^{\prime}}\right)\ln\frac{Q_{{\scriptscriptstyle \mathcal{B}}}^{2}}{y_{01}}+\frac{1}{2}\left(C_{q}+C_{q^{\prime}}\right)\ln\frac{m_{{\scriptscriptstyle \mathcal{B}}}^{2}}{Q_{qq^{\prime}}^{2}}-\frac{1}{2}C_{g}\ln\frac{Q_{qg}^{2}Q_{q^{\prime}g}^{2}}{Q_{qq^{\prime}}^{2}y_{01}}\label{eq:sect22-n-eq-3-S1-defn}\\
 &  & -\frac{1}{2}\left(C_{q}+C_{q^{\prime}}\right)\ln\frac{m_{{\scriptscriptstyle \mathcal{B}}}^{2}}{Q_{{\scriptscriptstyle \mathcal{B}}}^{2}}-\frac{1}{2}\sum_{\ell}\, C_{\ell}\ln\frac{y_{01}}{Q_{{\scriptscriptstyle \mathcal{B}\mathrm{J}}}^{2}}\,.\nonumber 
\end{eqnarray}
where $Q_{ij}=\sqrt{\left|2p_{i}.p_{j}\right|}$. For the case of
two/three hard gluon legs, we simply replace $C_{F}$ by $C_{A}$
in $S_{1}$ and, in addition, $q$, $q^{\prime}$, $g$ with $g_{1}$,
$g_{2}$, $g_{3}$ (see bottom of pg.~38 in ref.~\cite{Banfi:2004yd}).
By writing $S_{1}$ in this form for the $n=3$ case one can already
glimpse, in the first term, its interpretation in terms of coherent
emission from the $n=2$ kinematic underlying the $n=3$ one; we discuss
this in more depth later on.

The $K$ in the $\mathcal{O}\left(\bar{\alpha}_{{\scriptscriptstyle \mathrm{S}}}^{2}\right)$
part of $R$ is the two-loop cusp anomalous dimension 
\begin{equation}
K=C_{A}\left(\frac{67}{18}-\frac{\pi^{2}}{6}\right)-\frac{5}{9}n_{f}.\label{eq:sect22-the-magic-K-factor-defn}
\end{equation}

Concerning the Sudakov form factor, the only remaining part needing
introduction is $\mathcal{F}_{2}$. The $\mathcal{F}\left(R^{\prime}\right)$
function of ref.~\cite{Banfi:2004yd} accounts for NLL corrections
arising from a resummed observable's sensitivity to multiple emission
effects; for observables whose behavior is largely dictated by the
leading single emission $\mathcal{F}\left(R^{\prime}\right)\rightarrow1$.
The \noun{Caesar} $\mathcal{F}\left(R^{\prime}\right)$ factor is
understood to depend only on the flavours of the particles entering
and exiting the hard scattering and the multiple emission properties
of the observable, it does not depend on the kinematics of the underlying
hard scattering $\left(\Phi\right)$ \cite{Banfi:2004nk}. To $\mathrm{NLL}$/$\mathrm{NNLL}_{\sigma}$
accuracy $\mathcal{F}\left(R^{\prime}\right)=1$ for the $v_{01}$
resummation ($\mathcal{B}$-production). The combination of factors
$\mathcal{F}_{2}\left(2G_{12}\right)^{2}\,\bar{\alpha}_{{\scriptscriptstyle \mathrm{S}}}^{2}L^{2}$
is the next-to-leading term%
\footnote{The leading term in the expansion is just $1$. By only including the leading and next-to-leading terms for $\mathcal{F}\left(R^{\prime}\right)$ we break the $\mathrm{NLL}$/$\mathrm{NNLL}_{\sigma}$ accuracy down to $\mathrm{LL}$/$\mathrm{NNLL}_{\sigma}$.%
} in the fixed order expansion of the $\mathcal{F}\left(R^{\prime}\right)$
function and as such defines $\mathcal{F}_{2}$. From refs.~\cite{Banfi:2004nk,Banfi:2001bz,Banfi:2012yh}
we derive the following process-independent expression for $\mathcal{F}_{2}$,
for jet rates in the exclusive $k_{t}$ algorithm 
\begin{equation}
\mathcal{F}_{2}=-\frac{\pi^{2}}{16}\,\frac{\sum_{\ell=1}^{n}C_{\ell}^{2}-\sum_{\ell=1}^{n_{i}}C_{\ell}^{2}}{\left(\sum_{\ell=1}^{n}C_{\ell}\right)^{2}}\,,\label{eq:sect22-F2-general-kt-algo-formula}
\end{equation}
We have tested this expression using the numerical implementation
of the \noun{Caesar }formalism for resummation of $y_{23}$ in hadronic
jet production and $y_{12}$ in hadronic $Z$ boson production. With
the exception of the $qg$ and $gq$ channels in $Z$ production,
for which only 3\% differences were found, our $\mathcal{F}_{2}$
expression yielded agreement with the \noun{Caesar }program at the
per mille level in all tested processes and channels.

In the resummation formula, eqs.~\ref{eq:sect22-NNLL_sigma-differential-xsecn-ii}-\ref{eq:sect22-NNLL_sigma-differential-xsecn-i},
for the PDF dependent pieces we have adopted the notation 
\begin{equation}
\boldsymbol{\mathrm{q}}\left(x,\mu_{{\scriptscriptstyle F}}^{2}\right)=\left(\begin{array}{c}
q_{u}\left(x,\mu_{{\scriptscriptstyle F}}^{2}\right)\\
q_{\bar{u}}\left(x,\mu_{{\scriptscriptstyle F}}^{2}\right)\\
\vdots\\
g\left(x,\mu_{{\scriptscriptstyle F}}^{2}\right)
\end{array}\right)\,,\quad\boldsymbol{\mathrm{P}}\left(x\right)=\left(\begin{array}{cccc}
P_{qq}^{\left(0\right)}\left(x\right) & 0 & \cdots & P_{qg}^{\left(0\right)}\left(x\right)\\
0 & P_{qq}^{\left(0\right)}\left(x\right)\\
\vdots &  & \ddots\\
P_{gq}^{\left(0\right)}\left(x\right) &  &  & P_{gg}^{\left(0\right)}\left(x\right)
\end{array}\right)\,,\label{eq:sect22-vector-PDF-and-matrix-AP-fn-notation}
\end{equation}
where $P_{ij}^{\left(0\right)}\left(x\right)$ are the regularized
leading order Altarelli-Parisi splitting functions (see e.g.~appendix
A.3 of ref.~\cite{Banfi:2004yd}). We also identify $q^{\left(\ell\right)}\left(x_{\ell},\mu^{2}\right)=\boldsymbol{\mathrm{q}}_{i}^{\left(\ell\right)}\left(x_{\ell},\mu^{2}\right)$,
with $i$ the flavour of the hard parton with momentum $p_{\ell}$,
and we employ the following notation to denote matrix multiplication
and convolution in $x$-space 
\begin{equation}
\left[\,\mathrm{\boldsymbol{\mathrm{P}}}\otimes\boldsymbol{\mathrm{q}}\,\right]_{i}\left(x,\mu^{2}\right)=\int_{x}^{1}\frac{dz}{z}\,\boldsymbol{\mathrm{P}}_{ij}\left(\frac{x}{z}\right)\boldsymbol{\mathrm{q}}_{j}\left(z,\mu^{2}\right)=\int_{x}^{1}\frac{dz}{z}\,\boldsymbol{\mathrm{P}}_{ij}\left(z\right)\boldsymbol{\mathrm{q}}_{j}\left(\frac{x}{z},\mu^{2}\right)\,.\label{eq:sect22-PDF-convolution-defn}
\end{equation}

The last things we need to introduce in our resummation formula, eq.~\ref{eq:sect22-NNLL_sigma-differential-xsecn-i}-\ref{eq:sect22-NNLL_sigma-differential-xsecn-ii},
are the $\mathcal{H}_{1}$ and $\mathcal{C}_{1}$ terms. To this end
we first define the cumulant, $\Sigma_{{\scriptscriptstyle \mathcal{R}}}$,
of the $\mathrm{NNLL}_{\sigma}$ resummed spectrum as 
\begin{eqnarray}
\frac{d\Sigma_{{\scriptscriptstyle \mathcal{R}}}\left(L\right)}{d\Phi}\, & = & \int_{\infty}^{L}dL^{\prime}\,\frac{d\sigma_{{\scriptscriptstyle \mathcal{R}}}}{d\Phi dL^{\prime}}\,.\label{eq:sect22-dSigmaR}
\end{eqnarray}
Since $d\sigma_{{\scriptscriptstyle \mathcal{R}}}$ is expressed as
a total derivative we quickly find the following approximation to
the NLO $\mathcal{B}$/$\mathcal{B}$\noun{j }production cross section:
\begin{eqnarray}
\left.\frac{d\Sigma_{{\scriptscriptstyle \mathcal{R},1}}\left(L\right)}{d\Phi}\right|_{{\scriptscriptstyle \mathcal{H}_{1},\mathcal{C}_{1}\rightarrow0}} & = & \frac{d\sigma_{0}}{d\Phi}\,\left[1+\bar{\alpha}_{{\scriptscriptstyle \mathrm{S}}}\, G_{12}\, L^{2}+\bar{\alpha}_{{\scriptscriptstyle \mathrm{S}}}\,\left[G_{11}+2S_{1}-\sum_{\ell=1}^{n_{i}}\,\frac{\left[\boldsymbol{\mathrm{P}}\otimes\boldsymbol{\mathrm{q}}^{\left(\ell\right)}\right]_{i}}{q^{\left(\ell\right)}}\right]\, L\right]\,.\label{eq:sect22-dSigmaR-with-chibar-set-to-zero-expanded-to-NLO}
\end{eqnarray}
The cross section $\left.d\Sigma_{{\scriptscriptstyle \mathcal{R},1}}/d\Phi\right|_{{\scriptscriptstyle \mathcal{H}_{1},\mathcal{C}_{1}\rightarrow0}}$,
essentially by definition, contains all of the logarithmically enhanced
contributions to the exact NLO $\mathcal{B}$/$\mathcal{B}$\noun{j
}production cross section. The only parts of the exact NLO $\mathcal{B}$/$\mathcal{B}$\noun{j
}production cross section not accounted for by $\left.d\Sigma_{{\scriptscriptstyle \mathcal{R},1}}/d\Phi\right|_{{\scriptscriptstyle \mathcal{H}_{1},\mathcal{C}_{1}\rightarrow0}}$
are finite, unenhanced, parts for $v\rightarrow0$. These unenhanced
parts of the NLO contribution have two sources: finite virtual corrections
and contributions to the real emission part of the cross section which
are regular as $v\rightarrow0$ (terms of collinear origin). Thus
we can write 
\begin{eqnarray}
\frac{d\Sigma_{{\scriptscriptstyle \mathrm{NLO}}}\left(L\right)}{d\Phi} & = & \left.\frac{d\Sigma_{{\scriptscriptstyle \mathcal{R},1}}\left(L\right)}{d\Phi}\right|_{{\scriptscriptstyle \mathcal{H}_{1},\mathcal{C}_{1}\rightarrow0}}+\int_{\infty}^{L}dL^{\prime}\,\left[\bar{\alpha}_{{\scriptscriptstyle \mathrm{S}}}\bar{\chi}_{1}\left(\Phi\right)\,\frac{d\sigma_{0}}{d\Phi}\,\delta\left(L^{\prime}-\infty\right)+\frac{d\sigma_{{\scriptscriptstyle \mathcal{F},1}}}{d\Phi dL^{\prime}}\,\right]\,,\label{eq:sect22-dSigmaNLO}
\end{eqnarray}
where $\bar{\chi}_{1}\left(\Phi\right)$,%
\footnote{What we have denoted $\bar{\chi}_{1}\left(\Phi\right)$ ref.~\cite{Banfi:2010xy}
denotes as $C_{1,{\scriptscriptstyle \mathcal{B}}}^{\left(\delta\right)}$. %
} being localized at $v=0$, encodes the regular virtual and the hard
collinear contributions, with $d\sigma_{{\scriptscriptstyle \mathcal{F},1}}$
being the real emission contribution to $d\Sigma_{{\scriptscriptstyle \mathrm{NLO}}}\left(L\right)$,
with its $v\rightarrow0$ end-point subtracted and included in $\bar{\chi}_{1}$.
From eq.~\ref{eq:sect22-dSigmaNLO} we obtain directly 
\begin{equation}
\bar{\alpha}_{{\scriptscriptstyle \mathrm{S}}}\bar{\chi}_{1}\left(\Phi\right)=\lim_{L\rightarrow\infty}\,\left(\frac{d\Sigma_{{\scriptscriptstyle \mathrm{NLO}}}\left(L\right)}{d\Phi}-\left.\frac{d\Sigma_{{\scriptscriptstyle \mathcal{R},1}}\left(L\right)}{d\Phi}\right|_{{\scriptscriptstyle \bar{\chi}_{1}\rightarrow0}}\right)\,/\,\frac{d\sigma_{0}}{d\Phi}\,.\label{eq:sect22-chibar-defn}
\end{equation}
We separate hard-virtual and hard-collinear corrections in $\bar{\chi}_{1}$
as follows: 
\begin{eqnarray}
\bar{\chi}_{1}\left(\Phi\right) & = & \mathcal{H}_{1}\left(\Phi,\mu_{{\scriptscriptstyle R}}^{2},Q^{2}\right)+\sum_{\ell=1}^{n_{i}}\,\frac{\left[\boldsymbol{\mathcal{C}}_{1}\otimes\boldsymbol{\mathrm{q}}^{\left(\ell\right)}\right]_{i}\left(x_{\ell},\mu_{{\scriptscriptstyle F}}^{2}\right)}{q^{\left(\ell\right)}\left(x_{\ell},\mu_{{\scriptscriptstyle F}}^{2}\right)}\,,\label{eq:sect22-chibar-decomposed-into-H-and-C}
\end{eqnarray}
where the $\mathcal{C}_{1}$ terms represent the contribution due
to hard-collinear splitting in the initial-state and $\mathcal{H}_{1}$
is the remainder, including the hard-virtual component. $\mathcal{H}_{1}$
contains terms canceling the $\mu_{{\scriptscriptstyle R}}$ dependence,
while $\mathcal{C}_{1}$ has terms which correspondingly compensate
the $\mu_{{\scriptscriptstyle F}}$ dependence, of $d\sigma_{0}/d\Phi_{{\scriptscriptstyle \mathcal{B}}}$.
The precise details of these terms are irrelevant for the implementation
of the method being proposed (this can be considered one of its advantages),
so we can safely leave further specification of them to appendix \ref{sub:Derivation-of-NNLLsigma-jet-resolution-spectra}. We only stress
that in the $\mathcal{C}_{1}$ function, in eq.~\ref{eq:sect22-NNLL_sigma-differential-xsecn-ii},  which is convoluted
with a PDF evaluated at scale $\mu_{{\scriptscriptstyle F}}\sqrt{v}$,
the explicit factorization scale is $\mu_{{\scriptscriptstyle F}}$,
not $\mu_{{\scriptscriptstyle F}}\sqrt{v}$, i.e.~$\mathcal{C}_{1}$
in eq.~\ref{eq:sect22-NNLL_sigma-differential-xsecn-ii} is precisely
as it is written in eq.~\ref{eq:app1-curly-C-decomposition}, so
the derivative with respect to $L$ in eq.~\ref{eq:sect22-NNLL_sigma-differential-xsecn-i}
passes through $\mathcal{C}_{1}$, only acting on whatever follows
it.

The structure of eq.~\ref{eq:sect22-NNLL_sigma-differential-xsecn-i}
with regard to the inclusion of the $\bar{\chi}_{1}$ term, can be
intuitively understood by considering that the Sudakov and PDF factors
in eq.~\ref{eq:sect22-NNLL_sigma-differential-xsecn-i} are resumming
the effects of all orders soft/collinear radiation around the hard
scattering, described by $d\sigma_{0}$ in the case of the leading
term, and, that the same pattern of radiation also occurs with respect
to the hard scattering including hard radiative effects $d\sigma_{0}\,\bar{\chi}_{1}$.
In other words, one can view the resummation as being taken with respect
to essentially two separate processes and then adding these two resummations
together, one process being the higher order analogue of the other.
The fact that the resummation should be identical with respect to
either process can be understood by considering that the soft long-wavelength
radiative corrections --- encoded by the Sudakov form factor, running
coupling and PDFs --- will not be able to probe the internal details
of the hard scatterings they attach to.

\subsection{Comparison to \noun{Ckkw}/\noun{Minlo}\label{subsec:comparison-to-ckkw-minlo}}

Before continuing, it is worth comparing the formulae presented
here to those of \noun{Ckkw}/\noun{Minlo} \cite{Catani:2001cc,Hamilton:2012np},
in particular the Sudakov form factors. The Sudakov form factor exponent
in the latter articles%
\footnote{Specifically eqs.~2.8 and 2.9 of ref.~\cite{Catani:2001cc}, and
eqs.~A.1-A.3 of ref.~\cite{Hamilton:2012np}.%
} for a collection of (pseudo-)partons (indexed by $\ell$) evolving
from some scale $Q$ down to a resolution scale $y$, is given by
\begin{eqnarray}
\sum_{\ell}\ln\Delta_{\ell}\left(y,Q^{2}\right) & = & -R\left(v\right)-\int_{0}^{L}dL^{\prime}\,\left[\bar{\alpha}_{{\scriptscriptstyle \mathrm{S}}}\left(y^{\prime}\right)\,\left[2S_{1}\right]+\bar{\alpha}_{{\scriptscriptstyle \mathrm{S}}}^{2}\left(y^{\prime}\right)\,2G_{12}\,\left[4\mathcal{F}_{2}G_{12}\right]\, L^{\prime}\right]\,.\label{eq:sect23-comparing-Suds-for-old-minlo-and-this-minlo}
\end{eqnarray}
The integral on the right-hand side of eq.~\ref{eq:sect23-comparing-Suds-for-old-minlo-and-this-minlo}
is the difference between the total Sudakov form factor exponents
used in the \noun{Ckkw}/\noun{Minlo }prescription and
that proposed here based on \noun{Caesar} $\left(-R\left(v\right)\right)$.

For $\mathcal{B}$ production $\mathcal{F}_{2}=0$ and $S_{1}\propto\ln\frac{m_{{\scriptscriptstyle \mathcal{B}}}}{Q_{{\scriptscriptstyle \mathcal{B}}}}$,
with $Q_{{\scriptscriptstyle \mathcal{B}}}$ set equal to $m_{{\scriptscriptstyle \mathcal{B}}}$
in the original \noun{Minlo }proposal, hence, in this case, the second
term on the right of eq.~\ref{eq:sect23-comparing-Suds-for-old-minlo-and-this-minlo}
vanishes. Thus, in $\mathcal{B}$ production the Sudakov form factor
of the original \noun{Minlo }procedure is
fully consistent with that prescribed by \noun{Caesar}.

For the $\mathcal{B}$\noun{j }case, not forgetting that here in section
\ref{sec:Merging-without-a-merging-scale-two-units}
we are restricting ourselves to considering
the region $y_{01}\gtrsim\mathcal{O}(m_{{\scriptscriptstyle \mathcal{B}}}^{2})$,
$\mathcal{F}_{2}$ is not zero, and $S_{1}$ has non-trivial dependence
on the underlying $\mathcal{B}$\noun{j }kinematics. Therefore, in
the region where our \noun{Caesar}-based formula is strictly valid
we have a discrepancy between what is suggested by it and by \noun{Minlo}.
In particular the original \noun{Minlo }proposal has omitted $\mathrm{NNLL}_{\sigma}$
terms due to multiple emission corrections $\left(\mathcal{F}_{2}\right)$
and, more importantly, $\mathrm{NLL}_{\sigma}$ contributions due
to soft-wide-angle radiation $\left(S_{1}\right)$. Thus, in the region
$y_{01}\gtrsim\mathcal{O}(m_{{\scriptscriptstyle \mathcal{B}}}^{2})$
$\mathcal{B}$\noun{jj-Minlo}, implemented according to the original
proposal in ref.~\cite{Hamilton:2012np}, would formally not be LO
accurate in the description of $\mathcal{B}$\noun{j-}inclusive quantities,
with ambiguities arising between it and conventional LO of order $\sqrt{\bar{\alpha}_{{\scriptscriptstyle \mathrm{S}}}}$
times the leading order term. With the benefit of hindsight it is
perhaps obvious that the original \noun{Minlo} procedure would have
this problem in this region, since we know that its Sudakov form factors
contain only soft-collinear and collinear terms, yet soft-wide-angle
radiation from a $\mathcal{B}$\noun{j }state will be logarithmically
enhanced too, even if the underlying Born partons are widely separated.

In section \ref{sec:Merging-without-a-merging-scale-three-units}
we also consider this comparison (for $\mathcal{B}$\noun{jj-Minlo})
in the region $y_{01}<\mathcal{O}(m_{{\scriptscriptstyle \mathcal{B}}}^{2})$.

\subsection{\noun{Minlo} jet resolution spectra\label{sub:MiNLO-jet-resolution-spectrum}}

In the \noun{Minlo }framework, in all cases, we start with an NLO
cross section: for the $v_{01}$ resummation in $\mathcal{B}$-production
our fundamental ingredient is the NLO $\mathcal{B}$\noun{j} cross
section, while for $v_{12}$ resummation in $\mathcal{B}$\noun{j}-production
it is that for NLO $\mathcal{B}$\noun{jj}. We write these cross sections
as a sum of a part which is finite as $v\rightarrow0$, $d\sigma_{{\scriptscriptstyle \mathcal{F}}}$,
plus a singular part obtained by expanding the $\mathrm{NNLL}_{\sigma}$
resummation formula (eq.~\ref{eq:sect22-NNLL_sigma-differential-xsecn-i})
$d\sigma_{{\scriptscriptstyle \mathcal{S}}}$, and a further singular-remainder
piece, $d\sigma_{{\scriptscriptstyle \mathcal{SR}}}$, which is defined
as all singular terms which were not already contained in $d\sigma_{{\scriptscriptstyle \mathcal{S}}}$:
\begin{equation}
d\sigma_{{\scriptscriptstyle \mathcal{\phantom{X}}}}=d\sigma_{{\scriptscriptstyle \mathcal{S}}}+d\sigma_{{\scriptscriptstyle \mathcal{SR}}}+d\sigma_{{\scriptscriptstyle \mathcal{F}}}\,.\label{eq:sect24-NLO-decomposed-as-S-plus-F}
\end{equation}
Expanding the resummed differential cross section up to and including
$\mathcal{O}\left(\bar{\alpha}_{{\scriptscriptstyle \mathrm{S}}}^{2}\right)$
terms, we obtain 
\begin{eqnarray}
\frac{d\sigma_{{\scriptscriptstyle \mathcal{S}}}}{d\Phi dL} & = & \frac{d\sigma_{0}}{d\Phi}\,\sum_{n=1}^{2}\sum_{m=0}^{2n-1}\, H_{nm}\bar{\alpha}_{{\scriptscriptstyle \mathrm{S}}}^{n}\left(\mu_{{\scriptscriptstyle R}}^{2}\right)L^{m}\,,\label{eq:sect24-dsigmaS-FO-expansion-master-formula}
\end{eqnarray}
where the explicit $H_{nm}$ coefficients are documented in the appendix
\ref{sub:Fixed-order-expansion}. Since the resummation formula we
used to derive this fixed order expansion was $\mathrm{NNLL}_{\sigma}$
accurate, it only predicts part of the full $\mathrm{N}^{3}\mathrm{LL}_{\sigma}$
coefficient, $\sim\bar{\alpha}_{{\scriptscriptstyle \mathrm{S}}}^{2}$,
thus we have a singular remainder term, 
\begin{equation}
\frac{d\sigma_{{\scriptscriptstyle \mathcal{SR}}}}{d\Phi dL}=\frac{d\sigma_{0}}{d\Phi}\,\bar{\alpha}_{{\scriptscriptstyle \mathrm{S}}}^{2}\left(\mu_{{\scriptscriptstyle R}}^{2}\right)\left[\, L\,\widetilde{R}_{21}+\widetilde{R}_{20}\,\right]\,,\label{eq:sect24-dsigmaSR-FO-expansion}
\end{equation}
where $\widetilde{R}_{21}=0$ and we proceed under the assumption
that the coefficient $\widetilde{R}_{20}$ is generally unknown to
us. We introduce the strange $\widetilde{R}_{21}=0$ term here in
order to make the transition to the discussion on merging by three
units of multiplicity, in sect.~\ref{sec:Merging-without-a-merging-scale-three-units},
a little bit cleaner; there our formulae are applied in regions where
they lose $\mathrm{NNLL}_{\sigma}$ accuracy. The $d\sigma_{{\scriptscriptstyle \mathcal{SR}}}$
term can be considered as a valid parametrization of our ignorance
of the $v\rightarrow0$ singular part of the NLO cross section. Importantly,
since $d\sigma_{{\scriptscriptstyle \mathcal{S}}}$ alone is invariant
under $\mu_{{\scriptscriptstyle R}}$/$\mu_{{\scriptscriptstyle F}}$ shifts,
up to NNLO terms, $\widetilde{R}_{21}$
and $\widetilde{R}_{20}$ have no $\mu_{{\scriptscriptstyle R}}$
or $\mu_{{\scriptscriptstyle F}}$ dependence.

In practice, the \noun{Minlo} prescription consists of a series of
clearly defined, straightforward, operations on the fully differential input
NLO calculations. These can be summarized as renormalization and factorization
scale setting, together with matching to the Sudakov form factor
($\exp\left[\,-R\left(v\right)\,\right]$,
eq.~(\ref{eq:sect22-Big-Sudakov-minus-Ry})). To ease readibility, we have
deferred the precise details of these steps to the appendix
(\ref{sub:minlo-steps-for-two-units-of-multiplicity}). We suffice to say that
if one carefully traces the effects of the latter operations on the NLO cross
section, in particular on the singular parts,
$d\sigma_{{\scriptscriptstyle \mathcal{S}}}$ and
$d\sigma_{{\scriptscriptstyle \mathcal{SR}}}$, neglecting
$\mathcal{O}\left(\mathrm{N}^{4}\mathrm{LL}_{\sigma}\right)$ terms, one
finds the resulting \noun{Minlo} cross section can be written as
\begin{eqnarray}
d\sigma_{{\scriptscriptstyle \mathcal{M}}} & = & d\sigma_{{\scriptscriptstyle \mathcal{R}}}+d\sigma_{{\scriptscriptstyle \mathcal{MR}}}+d\sigma_{{\scriptscriptstyle \mathcal{F}}}\,,\label{eq:sect24-digmaM-up-to-N4LL-ambiguities-proc-dep}
\end{eqnarray}
where $d\sigma_{{\scriptscriptstyle \mathcal{R}}}$ is the resummation cross section, eq.~\ref{eq:sect22-NNLL_sigma-differential-xsecn-i}, 
a total derivative,
and $d\sigma_{{\scriptscriptstyle \mathcal{MR}}}$
holds all remaining large logs: 
\begin{eqnarray}
\frac{d\sigma_{{\scriptscriptstyle \mathcal{MR}}}}{d\Phi dL} & = & \frac{d\sigma_{0}}{d\Phi}\,\exp\left[\,-R\left(v\right)\,\right]\,\prod_{\ell=1}^{n_{i}}\frac{q^{\left(\ell\right)}\left(x_{\ell},\mu_{{\scriptscriptstyle F}}^{2}v\right)}{q^{\left(\ell\right)}\left(x_{\ell},\mu_{{\scriptscriptstyle F}}^{2}\right)}\,\left[\bar{\alpha}_{{\scriptscriptstyle \mathrm{S}}}^{2}\left(K_{{\scriptscriptstyle R}}^{2}\, y\right)\,\left[\widetilde{R}_{21}\, L+\widetilde{R}_{20}\right]+\bar{\alpha}_{s}^{3}\left(K_{{\scriptscriptstyle R}}^{2}\, y\right)L^{2}\,\widetilde{R}_{32}\right]\,,\nonumber \\
\widetilde{R}_{32} & = & 2G_{12}\bar{\beta}_{0}\mathcal{H}_{1}\left(\mu_{{\scriptscriptstyle R}}^{2}\right)\,.\label{eq:sect24-digmaMR-up-to-N4LL-ambiguities-proc-dep}
\end{eqnarray}
In eq.~\ref{eq:sect24-digmaMR-up-to-N4LL-ambiguities-proc-dep} the 
$K_{{\scriptscriptstyle R/F}}\in\left[\frac{1}{2},2\right]$ denote rescaling
factors applied to the renormalization and factorization scales,
$\mu_{{\scriptscriptstyle R/F}}$, defined at the start of the \noun{Minlo}
procedure (see \ref{sub:minlo-steps-for-two-units-of-multiplicity} for details),
for the purposes of assessing scale uncertainties.
The last term in eq.~\ref{eq:sect24-digmaM-up-to-N4LL-ambiguities-proc-dep},
$d\sigma_{{\scriptscriptstyle \mathcal{F}}}$, is more precisely $d\sigma_{{\scriptscriptstyle \mathcal{MF}}}$,
the replacement $d\sigma_{{\scriptscriptstyle \mathcal{MF}}}\rightarrow d\sigma_{{\scriptscriptstyle \mathcal{F}}}$ 
being made on the grounds that the \noun{Minlo} operations preserve
the fixed order expansion up to and including NLO terms, as well as
the fact that $d\sigma_{{\scriptscriptstyle \mathcal{F}}}$ (and $d\sigma_{{\scriptscriptstyle \mathcal{MF}}}$)
is finite for $v\rightarrow0$.

Since $\widetilde{R}_{21}=0$, the \noun{Minlo }jet resolution spectra
in eqs.~\ref{eq:sect24-digmaM-up-to-N4LL-ambiguities-proc-dep}
are equal to the $\mathrm{NNLL}_{\sigma}$
jet resolution spectrum in sect.~\ref{sub:NNLLsigma-jet-resolution-spectrum}
$\left(d\sigma_{{\scriptscriptstyle \mathcal{R}}}\right)$ up to $\mathrm{N^{3}LL}_{\sigma}$
differences.

\subsection{Integrated\noun{ Minlo} jet resolution spectra\label{sub:MiNLO-integrated-jet-resolution-spectrum}}

Making use of the fact that $d\sigma_{{\scriptscriptstyle \mathcal{R}}}$
is a total derivative with respect to $L$ (eq.~\ref{eq:sect24-digmaM-up-to-N4LL-ambiguities-proc-dep}),
and the definitions of $\bar{\chi}$ in terms of $\mathcal{H}_{1}$
and $\mathcal{C}_{1}$, it is fairly straightforward to show%
\footnote{For more details see appendix \ref{sub:Integral-of-Minlo-v-spectrum}.%
} that on integrating over all $v$
\begin{eqnarray}
\frac{d\sigma_{{\scriptscriptstyle \mathcal{M}}}}{d\Phi} & = & \frac{d\sigma_{{\scriptscriptstyle \mathrm{NLO}}}}{d\Phi}+\int dL^{\prime}\,\frac{d\sigma_{{\scriptscriptstyle \mathcal{MR}}}}{d\Phi dL^{\prime}}+\mathcal{O}\left(\bar{\alpha}_{{\scriptscriptstyle \mathrm{S}}}^{2}\right)\,.\label{eq:sect25-integral-of-Minlo-spectrum-dSigmaM}
\end{eqnarray}
The contaminating $\int d\sigma_{{\scriptscriptstyle \mathcal{MR}}}$
term consists of a $\mathrm{N^{2}LL}_{\sigma}$ piece, $\propto\widetilde{R}_{21}$,
and $\mathrm{N^{3}LL}$ piece $\propto\widetilde{R}_{20}-\bar{\beta}_{0}\mathcal{H}_{1}$.
For the regions in which the \noun{Caesar} formalism holds $\widetilde{R}_{21}=0$,
as discussed under eq.~\ref{eq:sect24-dsigmaSR-FO-expansion}.

If we assume that we were ignorant of the value of $\widetilde{R}_{21}$,
dropping terms over which we have no control, i.e.~beyond $\mathrm{NNLL}_{\sigma}$ order, we can neglect
the $L$ dependence of $\bar{\alpha}_{{\scriptscriptstyle \mathrm{S}}}$
and PDFs in $d\sigma_{{\scriptscriptstyle \mathcal{MR}}}$,
and all but the leading term in the Sudakov form factor exponent $\propto G_{12}$.
With these approximations the $d\sigma_{{\scriptscriptstyle \mathcal{MR}}}$
integral becomes: 
\begin{eqnarray}
\int dL^{\prime}\,\frac{d\sigma_{{\scriptscriptstyle \mathcal{MR}}}}{d\Phi dL^{\prime}} & = & -\frac{d\sigma_{0}}{d\Phi}\,\widetilde{R}_{21}\,\frac{1}{\left|2G_{12}\right|}\,\bar{\alpha}_{{\scriptscriptstyle \mathrm{S}}}\left(1+\mathcal{O}\left(\sqrt{\bar{\alpha}_{{\scriptscriptstyle \mathrm{S}}}}\right)\right)\,.\label{eq:sect25-dSigmaMR-no-NNNLL-integrated}
\end{eqnarray}
The $\mathcal{O}(\bar{\alpha}_{{\scriptscriptstyle \mathrm{S}}}^{3/2})$
ambiguity in eq.~\ref{eq:sect25-dSigmaMR-no-NNNLL-integrated} attributes
to neglect of $\mathrm{N}^{3}\mathrm{LL}_{\sigma}$ terms. So, if
our knowledge of $\widetilde{R}_{21}$ be wrong, for whatever reason,
the \noun{Minlo }inclusive cross section would deviate from the exact
NLO one by terms of order $\mathcal{O}\left(\bar{\alpha}_{{\scriptscriptstyle \mathrm{S}}}\right)$
relative to the LO contribution $\left(d\sigma_{0}\right)$.

Sticking to the regions for which the \noun{Caesar }formalism holds,
our starting resummation formula and the \noun{Minlo }cross section
formulated with it is $\mathrm{NNLL}_{\sigma}$ accurate, i.e.~$\widetilde{R}_{21}=0$
and our ignorance is located downstream in the $\mathrm{N^{3}LL}_{\sigma}$
terms $\propto\widetilde{R}_{20}-\bar{\beta}_{0}\mathcal{H}_{1}$.
Dropping terms now only of $\mathrm{N}^{4}\mathrm{LL}_{\sigma}$ accuracy
we can again neglect the $L$ dependence of the coupling constant
and PDF terms, and all but the leading double log term in the Sudakov
form factor, giving 
\begin{eqnarray}
\int dL^{\prime}\,\frac{d\sigma_{{\scriptscriptstyle \mathcal{MR}}}}{d\Phi dL^{\prime}} & = & -\frac{d\sigma_{0}}{d\Phi}\,\left[\widetilde{R}_{20}-\bar{\beta}_{0}\mathcal{H}_{1}\left(\mu_{{\scriptscriptstyle R}}^{2}\right)\right]\,\sqrt{\frac{\pi}{2}}\,\frac{1}{\left|2G_{12}\right|^{1/2}}\,\bar{\alpha}_{{\scriptscriptstyle \mathrm{S}}}^{3/2}\left(1+\mathcal{O}\left(\sqrt{\bar{\alpha}_{{\scriptscriptstyle \mathrm{S}}}}\right)\right)\,.\label{eq:sect25-dSigmaMR-no-NNNNLL-integrated}
\end{eqnarray}
Now the \noun{Minlo }inclusive cross section and that of the exact
NLO calculation differ by terms of order $\mathcal{O}(\bar{\alpha}_{{\scriptscriptstyle \mathrm{S}}}^{3/2})$
relative to the LO contribution; for the \noun{Minlo }cumulant cross
section to be certified NLO accurate it needs to agree with conventional
NLO up to relative $\mathcal{O}(\bar{\alpha}_{{\scriptscriptstyle \mathrm{S}}}^{2})$
(NNLO) ambiguities.

\subsection{Removal of spurious terms in the \noun{Minlo }integrated cross section\label{sub:Removal-of-spurious-terms}}

\subsubsection*{Original\noun{ Minlo}$^{\prime}$ approach }

If we replace the \noun{Minlo }Sudakov form factor exponent in step
\ref{enu:Minlo-procedure-last-step} according to 
\begin{equation}
-R\left(v\right)\rightarrow-R\left(v\right)-\Delta R\left(v\right)\,,\qquad-\Delta R\left(v\right)=\int_{0}^{L}dL^{\prime}\,\bar{\alpha}_{{\scriptscriptstyle \mathrm{S}}}^{2}\left(y^{\prime}\right)\,\left[\widetilde{R}_{21}\, L^{\prime}+\widetilde{R}_{20}-\bar{\beta}_{0}\mathcal{H}_{1}\left(\mu_{{\scriptscriptstyle R}}^{2}\right)\right]\,,\label{eq:sect26-Rv-goes-to-Rv-plus-DeltaRv}
\end{equation}
we find, neglecting $\mathrm{N^{4}LL}_{\sigma}$ terms, 
\begin{equation}
d\sigma_{{\scriptscriptstyle \mathcal{M}}}\rightarrow\frac{d\sigma_{0}}{d\Phi}\,\frac{d}{dL}\,\left[\left[1+\bar{\alpha}_{{\scriptscriptstyle \mathrm{S}}}\left(\mu_{{\scriptscriptstyle R}}^{2}v\right)\,\mathcal{H}_{1}\left(\mu_{{\scriptscriptstyle R}}^{2}\right)\right]\,\exp\left[\,-R\left(v\right)\,-\Delta R\left(v\right)\right]\,\mathcal{L}\left(\left\{ x_{\ell}\right\} ,\mu_{{\scriptscriptstyle F}},v\right)\right]+\frac{d\sigma_{{\scriptscriptstyle \mathcal{F}}}}{d\Phi dL}\,.\label{eq:sect26-dsigmaM-with-Rv-plus-DeltaRv-Sudakov}
\end{equation}
All large logarithms in this modified \noun{Minlo }spectrum, eq.~\ref{eq:sect26-dsigmaM-with-Rv-plus-DeltaRv-Sudakov},
are wrapped up as a total derivative and it's trivial to verify (more-or-less
exactly as in appendix \ref{sub:Integral-of-Minlo-v-spectrum}) that
the integral over all $v$ $\left(L\right)$ gives the conventional
NLO cross section without any spurious terms, as were examined in
sect.~\ref{sub:MiNLO-integrated-jet-resolution-spectrum}.

Thus we interpret the spurious terms that arise on integration in
sect.~\ref{sub:MiNLO-integrated-jet-resolution-spectrum}, as being
due to neglect of any $\mathrm{NNLL}_{\sigma}$ (for the scenario
$\widetilde{R}_{21}\ne0$) and $\mathrm{N^{3}LL}_{\sigma}$ terms
in our \noun{Minlo} Sudakov form factor (eq.~\ref{eq:sect22-Big-Sudakov-minus-Ry}).
To remove the spurious terms and recover NLO accuracy on integration
over $L$ we should try and include these terms in the latter. The
\noun{Minlo}$^{\prime}$ approach of ref.~\cite{Hamilton:2012rf}
does this explicitly, as in eq.~\ref{eq:sect26-Rv-goes-to-Rv-plus-DeltaRv},
extracting all relevant ingredients from known analytic results for
the full NLO singular behaviour of the Higgs/vector-boson transverse
momentum spectrum.

Neglecting $\mathrm{N^{4}LL}_{\sigma}$ terms, the modification to
the Sudakov form factor in eq.~\ref{eq:sect26-Rv-goes-to-Rv-plus-DeltaRv},
to be used in step \ref{enu:Minlo-procedure-last-step} of sect.~\ref{sub:MiNLO-jet-resolution-spectrum},
can be equivalently written as 
\begin{equation}
\exp\left[-R\left(v\right)\right]\rightarrow\exp\left[-R\left(v\right)\right]\,\left(1-\Delta R\left(v\right)\right)\,,\label{eq:sect26-expRv-goes-to-expRv-times-one-minus-DeltaRv}
\end{equation}
with $-\Delta R\left(v\right)$ exactly as in eq.~\ref{eq:sect26-Rv-goes-to-Rv-plus-DeltaRv},
leading to 
\begin{equation}
d\sigma_{{\scriptscriptstyle \mathcal{M}}}\rightarrow\frac{d\sigma_{0}}{d\Phi}\,\frac{d}{dL}\,\left[\left[1+\bar{\alpha}_{{\scriptscriptstyle \mathrm{S}}}\left(\mu_{{\scriptscriptstyle R}}^{2}v\right)\,\mathcal{H}_{1}\left(\mu_{{\scriptscriptstyle R}}^{2}\right)\right]\,\exp\left[-R\left(v\right)\right]\,\left(1-\Delta R\left(v\right)\right)\,\mathcal{L}\left(\left\{ x_{\ell}\right\} ,\mu_{{\scriptscriptstyle F}},v\right)\right]+\frac{d\sigma_{{\scriptscriptstyle \mathcal{F}}}}{d\Phi dL}\,.\label{eq:sect26-dsigmaM-with-expRv-times-one-minus-DeltaRv-Sudakov}
\end{equation}
The modification to the Sudakov form factor in eq.~\ref{eq:sect26-Rv-goes-to-Rv-plus-DeltaRv}
is equal to that in eq.~\ref{eq:sect26-expRv-goes-to-expRv-times-one-minus-DeltaRv}
with differences only starting at the $\mathrm{N^{4}LL}_{\sigma}$
level. Thus, in this modified \noun{Minlo }spectrum, eq.~\ref{eq:sect26-dsigmaM-with-expRv-times-one-minus-DeltaRv-Sudakov},
the integral over all $v$ $\left(L\right)$ gives the conventional
NLO cross section without any spurious terms.

\subsubsection*{Alternative approach to \noun{Minlo$^{\prime}$\label{sub:Variation-on-MinloPrime}}}

The message from eqs.~\ref{eq:sect26-Rv-goes-to-Rv-plus-DeltaRv}-\ref{eq:sect26-dsigmaM-with-Rv-plus-DeltaRv-Sudakov}
and eqs.~\ref{eq:sect26-expRv-goes-to-expRv-times-one-minus-DeltaRv}-\ref{eq:sect26-dsigmaM-with-expRv-times-one-minus-DeltaRv-Sudakov}
is the same: including the appropriate corrective factor on top of
the default \noun{Minlo} Sudakov form factor, $\exp\left[-R\left(v\right)\right]$,
we recover from $\mathcal{B}n$\noun{j-Minlo }NLO accuracy also for
$\mathcal{B}m$\noun{j }inclusive observables ($m=n-1$). We now suggest
to turn around the latter fact and use unitarity to effectively determine
the missing piece of the Sudakov factor, $\exp\left[-\Delta R\left(v\right)\right]$,
at a level of accuracy sufficient for our aims.

Minded by the equivalence between the \noun{Minlo$^{\prime}$ }formulations
in eqs.~\ref{eq:sect26-Rv-goes-to-Rv-plus-DeltaRv}-\ref{eq:sect26-dsigmaM-with-Rv-plus-DeltaRv-Sudakov}
and eqs.~\ref{eq:sect26-expRv-goes-to-expRv-times-one-minus-DeltaRv}-\ref{eq:sect26-dsigmaM-with-expRv-times-one-minus-DeltaRv-Sudakov},
we describe now how to implement, approximately, the $1-\Delta R\left(v\right)$
factor of the latter, without explicit knowledge of the $\widetilde{R}_{21}$,
$\widetilde{R}_{20}$ and $\mathcal{H}_{1}$ terms. We define what
we consider to be the discrepancy in the Sudakov form factor at $\mathrm{NNLL}_{\sigma}$
as: 
\begin{eqnarray}
\delta(\Phi) & = & \left(\frac{d\sigma_{{\scriptscriptstyle \mathcal{M}}}}{d\Phi}-\frac{d\sigma_{{\scriptscriptstyle \mathrm{NLO}}}}{d\Phi}\right)\,/\,\int dL\, h\left(L\right)\,\frac{d\sigma_{{\scriptscriptstyle \mathcal{M}}}}{d\Phi dL}\,,\label{eq:sect26-first-delta-defn}\\
\nonumber \\
h\left(L\right) & = & \bar{\alpha}_{{\scriptscriptstyle \mathrm{S}}}\left[\,\bar{\alpha}_{{\scriptscriptstyle \mathrm{S}}}L^{2}\Theta\left(\rho-\bar{\alpha}_{{\scriptscriptstyle \mathrm{S}}}L^{2}\right)+\rho\,\Theta\left(\bar{\alpha}_{{\scriptscriptstyle \mathrm{S}}}L^{2}-\rho\right)\,\right]\,.\label{eq:sect26-freezing-out-the-weight-factor-at-aSL2-gtrsim-1}
\end{eqnarray}
In eqs.~\ref{eq:sect26-first-delta-defn}-\ref{eq:sect26-freezing-out-the-weight-factor-at-aSL2-gtrsim-1}
we abbreviated $\bar{\alpha}_{{\scriptscriptstyle \mathrm{S}}}\left(Q^{2}\right)\rightarrow\bar{\alpha}_{{\scriptscriptstyle \mathrm{S}}}$.
The leading term in the integrand of the denominator of $\delta(\Phi)$
is $\sim\bar{\alpha}_{{\scriptscriptstyle \mathrm{S}}}^{3}L^{3}$
($\mathrm{NNLL}_{\sigma}$). The choice of the $h\left(L\right)$
function in eq.~\ref{eq:sect26-first-delta-defn} is not a rigid
one, as we comment on later. The freezing parameter $\rho$ in eq.~\ref{eq:sect26-freezing-out-the-weight-factor-at-aSL2-gtrsim-1}
is taken to be $\sim$1 by default. From eqs.~\ref{eq:sect25-integral-of-Minlo-spectrum-dSigmaM}-\ref{eq:sect25-dSigmaMR-no-NNNLL-integrated}
we get an expression for the numerator of $\delta(\Phi)$, and by
similar approximations to those used for the latter, an expression
for the denominator (appendix \ref{sub:Expression-for-denominator-of-delta-Phi}),
giving overall 
\begin{equation}
\delta(\Phi)=-\frac{1}{2}\,\widetilde{R}_{21}\,\left(1-\exp\left[G_{12}\rho\right]\right)^{-1}\,\left(1+\mathcal{O}\left(\sqrt{\bar{\alpha}_{{\scriptscriptstyle \mathrm{S}}}}\right)\right)\,.\label{eq:sect26-deltaPhi-eq-order-1-plus-order-root-aS}
\end{equation}
Since $\delta(\Phi)$ is an order one quantity (provided $\rho\gtrsim1$)
we can safely define the following modification of the original \noun{Minlo
}distribution 
\begin{equation}
\frac{d\sigma_{{\scriptscriptstyle \mathcal{M}}}^{\prime}}{d\Phi dL}=\frac{d\sigma_{{\scriptscriptstyle \mathcal{M}}}}{d\Phi dL}\,\left(1-\Delta R\left(v\right)^{{\scriptscriptstyle \mathrm{approx}}}\right)\,,\quad\mathrm{with}\quad-\Delta R\left(v\right)^{{\scriptscriptstyle \mathrm{approx}}}=-h\left(L\right)\,\delta(\Phi)\,.\label{eq:sect26-dsigmaMcorr-eq-dsigmaM-times-1-minus-DeltaRapprox}
\end{equation}
To help appreciate the correspondence between eq.~\ref{eq:sect26-dsigmaMcorr-eq-dsigmaM-times-1-minus-DeltaRapprox}
and eqs.~\ref{eq:sect26-expRv-goes-to-expRv-times-one-minus-DeltaRv}-\ref{eq:sect26-dsigmaM-with-expRv-times-one-minus-DeltaRv-Sudakov}
(and hence also back to the original \noun{Minlo$^{\prime}$ }approach
of eqs.~\ref{eq:sect26-Rv-goes-to-Rv-plus-DeltaRv}-\ref{eq:sect26-dsigmaM-with-Rv-plus-DeltaRv-Sudakov})
setting $\rho=\infty$ we point out that 
\begin{equation}
\Delta R\left(v\right)^{{\scriptscriptstyle \mathrm{approx}}}=\Delta R\left(v\right)+\mathrm{N^{3}LL}_{\sigma}\,.\label{eq:sect26-DeltaRv-approx-eq-DeltaRv-plus-N3LLsigma}
\end{equation}
Inserting our definition for $\delta(\Phi)$, eq.~\ref{eq:sect26-first-delta-defn},
into eq.~\ref{eq:sect26-dsigmaMcorr-eq-dsigmaM-times-1-minus-DeltaRapprox}
we find the identity 
\begin{eqnarray}
\frac{d\sigma_{{\scriptscriptstyle \mathcal{M}}}^{\prime}}{d\Phi} & = & \frac{d\sigma_{{\scriptscriptstyle \mathrm{NLO}}}}{d\Phi}\,,\label{eq:sect26-dSigmaMcorr-eq-dSigmaNLO-i}
\end{eqnarray}
i.e.~the corrected \noun{Minlo} distribution precisely returns the
true NLO inclusive cross section on integrating out the radiation,
unambiguously. This correction is achieved while leaving the NLO accuracy
of the input cross section intact; the weighting factor in square
brackets in eq.~\ref{eq:sect26-dsigmaMcorr-eq-dsigmaM-times-1-minus-DeltaRapprox}
being $\lesssim1+\mathcal{O}(\bar{\alpha}_{{\scriptscriptstyle \mathrm{S}}}^{2})$.
The modification in eq.~\ref{eq:sect26-dsigmaMcorr-eq-dsigmaM-times-1-minus-DeltaRapprox}
also does not interfere with the \noun{Minlo }cross section at $\mathrm{NLL}_{\sigma}$.%
\footnote{To see this consider re-expressing the square bracket term in eq.~\ref{eq:sect26-dsigmaMcorr-eq-dsigmaM-times-1-minus-DeltaRapprox}
as an exponential.%
}

The $\rho$ parameter guards against the $1-\Delta R\left(v\right)^{{\scriptscriptstyle \mathrm{approx}}}$
factor in eq.~\ref{eq:sect26-dsigmaMcorr-eq-dsigmaM-times-1-minus-DeltaRapprox}
becoming negative, which can happen in the region $\bar{\alpha}_{{\scriptscriptstyle \mathrm{S}}}L\gtrsim1$,
if $\delta\left(\Phi\right)$ is positive, leading to an unphysical
spectrum at $v\rightarrow0$. We remind that the region $\bar{\alpha}_{{\scriptscriptstyle \mathrm{S}}}L\gtrsim1$
has been anyway, from the beginning, outside the control of our calculational
setup, which is only adequate down to the region $\bar{\alpha}_{{\scriptscriptstyle \mathrm{S}}}L^{2}\sim1$.
Introducing $\rho$ also tames the integrand in the denominator of
$\delta\left(\Phi\right)$, so it can be determined/applied by simply
weighting events appropriately in analysis of the original $d\sigma_{{\scriptscriptstyle \mathcal{M}}}$
distribution, without issues of numerical convergence.

To provide some advance reassurance, should any be needed, in our
feasibility study in section \ref{sec:Feasibility-study} we carry
out what we consider to be broad variations of the $\rho$ parameter,
finding our results exhibit marginal sensitivity to it, in regions
of practical interest.

In the region of applicability of \noun{Caesar}, our knowledge of
the spectrum is complete at $\mathrm{NNLL}_{\sigma}$, i.e.~we know
that if the Sudakov form factor in eq.~\ref{eq:sect22-Big-Sudakov-minus-Ry}
is implemented in \noun{Minlo}, $\widetilde{R}_{21}=0$. All equations
and analysis above remain valid for $\widetilde{R}_{21}=0$ though.
The latter implies $\delta\left(\Phi\right)$ (eq.~\ref{eq:sect26-deltaPhi-eq-order-1-plus-order-root-aS})
is merely $\mathcal{O}(\sqrt{\bar{\alpha}_{{\scriptscriptstyle \mathrm{S}}}})$
instead of $\mathcal{O}(1)$, meaning the correction factor, eq.~\ref{eq:sect26-dsigmaMcorr-eq-dsigmaM-times-1-minus-DeltaRapprox},
has effectively less work to do. For $\widetilde{R}_{21}\ne0$ the
latter correction factor clearly affects the spectrum at $\mathrm{NNLL}_{\sigma}$.
However, we argue that since, for $\widetilde{R}_{21}=0$, the coefficient
of the correction in eq.~\ref{eq:sect26-dsigmaMcorr-eq-dsigmaM-times-1-minus-DeltaRapprox}
is $\mathcal{O}(\sqrt{\bar{\alpha}_{{\scriptscriptstyle \mathrm{S}}}})$,
not $\mathcal{O}(1)$, the effect is really $\mathcal{O}(\mathrm{N^{3}LL}_{\sigma})$,
i.e.~it would not spoil $\mathrm{NNLL}_{\sigma}$ were it already
in place. Since $\widetilde{R}_{21}=0$ for the domain of validity
of the \noun{Caesar }formalism, it may have seemed more natural to
have made $h\left(L\right)\sim\bar{\alpha}_{{\scriptscriptstyle \mathrm{S}}}^{2}L$
in our discussion here, instead of $\sim\bar{\alpha}_{{\scriptscriptstyle \mathrm{S}}}^{2}L^{2}$,
however, for the formal reasons just discussed, we see no great advantage
in doing so. Furthermore, we plan to employ the method also in the
region where \noun{Caesar }is not valid, in the next section, so having
a more widely applicable $h\left(L\right)$ function, which nominally
assumes the distribution it is correcting is $\mathrm{NLL}_{\sigma}$
accurate, is preferable.

It may be tempting to think that one can also apply this procedure
even if the initial input \noun{Minlo} distribution was only
$\mathrm{LL}_{\sigma}$ accurate, supplying, in that case, the
$h\left(L\right)$ function with one more power of $L$, in order to
keep $\delta\left(\Phi\right)\sim\mathcal{O}\left(1\right)$.
While this appears compatible with the recovery of 
NLO Born kinematics, maintaining also NLO accuracy of the
initial \noun{Minlo} simulation, the expansion of the product of
the latter factor and the initial $\mathrm{LL}_{\sigma}$
Sudakov form factor has a different functional form to that
of a $\mathrm{NLL}_{\sigma}$ Sudakov form factor. In other words,
one cannot then view the resulting correction (eq.~\ref{eq:sect26-dsigmaMcorr-eq-dsigmaM-times-1-minus-DeltaRapprox}) as approximating missing
higher order pieces of the Sudakov form factor, which was has
been our guiding principle throughout. This conflict can only
be resolved by making
$h\left(L\right)\sim\bar{\alpha}_{{\scriptscriptstyle \mathrm{S}}}L$,
however, in that case the correction factor
(eq.~\ref{eq:sect26-dsigmaMcorr-eq-dsigmaM-times-1-minus-DeltaRapprox})
will clearly violate NLO accuracy of the initial \noun{Minlo} program.
We therefore consider it a requirement that the relevant resummation in
the initial, uncorrected, \noun{Minlo} program be at least
$\mathrm{NLL}_{\sigma}$. Fortunately, this is a rather low theoretical
threshold to cross by today's standards, and really the only non-trivial
$\mathrm{NLL}_{\sigma}$ ingredients required are the (soft-wide-angle)
$S_1$ Sudakov coefficients
(eqs.~\ref{eq:sect22-Big-Sudakov-minus-Ry}-\ref{eq:sect22-n-eq-2-S1-defn}).
It is well understood how to obtain the latter soft-wide-angle
pieces, and it is not a particularly onerous task to do so nowadays.
Indeed there is much publicly available, automated, machinery which
can be straightforwardly adapted to this end, e.g. in \noun{Powheg-Box}
\cite{Alioli:2010xd} and $\tt{Madgraph5\_aMC@NLO}$ \cite{Alwall:2014hca}.
Finally, it is also the case that the aforementioned $S_1$ terms are
trivial for processes where the underlying Born comprises only
two or three coloured partons (e.g. $\mathcal{B}$\noun{j}- and 
$\mathcal{B}$\noun{jj}-\noun{Minlo}).

There are numerous possible variations, tangents and refinements one
can explore along the lines presented here, all leading to eq.~\ref{eq:sect26-dSigmaMcorr-eq-dSigmaNLO-i},
with or without ambiguities. For example, one can easily enough conceive
of modifications which avoid the introduction of the parameter $\rho$.
Equally, there are other ways to view the formulae in this section,
most of which are obvious. We do not want to digress, to avoid diluting
the basic idea and straying too far from the goals in the introduction.
In particular, we choose not to discuss to what extent we have formally
improved the description of the resummation region, but rather we now
get on with demonstrating the practicality of the above, and its
extension beyond the merging of two units of multiplicity.

\section{Merging three units of multiplicity\label{sec:Merging-without-a-merging-scale-three-units}}

We now turn to address the problem of getting the $\mathcal{B}$\noun{jj}-\noun{Minlo
}calculation to return NLO predictions for inclusive $\mathcal{B}$-production
observables, as well as $\mathcal{B}$\noun{j} and $\mathcal{B}$\noun{jj}
inclusive quantities.

Since $\mathcal{B}$\noun{j-Minlo }contains at most one final state
parton with NLO accuracy, from now on we discuss modifications to
$\mathcal{B}$\noun{jj-Minlo }only. Furthermore, we focus on modifications
needed to address the remaining problematic region $y_{12}\ll y_{01}\ll m_{{\scriptscriptstyle \mathcal{B}}}^{2}$,
with $y_{01}\gtrsim m_{{\scriptscriptstyle \mathcal{B}}}^{2}$ having
been covered in sect.~\ref{sec:Merging-without-a-merging-scale-two-units}.

Where necessary, we use a superscript $\left[01\right]$/$\left[12\right]$
on quantities $G_{ij}$, $S_{1}$ etc to distinguish those associated
to the $y_{01}$ resummation, from those associated to that of $y_{12}$.

The basic idea here is simple and can easily be improved; we briefly
discuss such refinements later on, with a view to future work. We
require that a lower multiplicity $\mathcal{B}$\noun{j-Minlo$^{\prime}$/Nnlops
}simulation has already been built, e.g.~with the procedure of sect.~\ref{sub:Removal-of-spurious-terms},
or along the lines of ref.~\cite{Hamilton:2012rf}, recovering NLO
accuracy for $\mathcal{B}$\noun{j-} and (NNLO) $\mathcal{B}$-inclusive
observables. We propose to apply method of sect.~\ref{sub:Removal-of-spurious-terms}
to the $\mathcal{B}$\noun{jj-Minlo }simulation, with the obvious
replacement $d\sigma_{{\scriptscriptstyle \mathcal{M}}}\rightarrow d\sigma_{{\scriptscriptstyle \mathcal{M}}}^{{\scriptscriptstyle \mathcal{B}\mathrm{JJ}}}$
therein, but also with the conventional fixed order distribution $d\sigma_{{\scriptscriptstyle \mathrm{NLO}}}$
replaced by $d\sigma_{{\scriptscriptstyle \mathcal{M}}}^{{\scriptscriptstyle \mathcal{B}\mathrm{J}}\,\prime}$.
It then follows, trivially, that 
\begin{eqnarray}
\frac{d\sigma_{{\scriptscriptstyle \mathcal{M}}}^{{\scriptscriptstyle \mathcal{B}\mathrm{JJ}}\,\prime}}{d\Phi_{{\scriptscriptstyle \mathcal{B}\mathrm{J}}}} & = & \frac{d\sigma_{{\scriptscriptstyle \mathcal{M}}}^{{\scriptscriptstyle \mathcal{B}\mathrm{J}}\,\prime}}{d\Phi_{{\scriptscriptstyle \mathcal{B}\mathrm{J}}}}\,.\label{eq:sect3-formula-from-intro-blurb}
\end{eqnarray}
Thus, the resulting $\mathcal{B}$\noun{jj}-\noun{Minlo$^{\prime}$
}distribution is targeted onto the $\mathcal{B}$\noun{j}-\noun{Minlo$^{\prime}$}
inclusive $\Phi_{{\scriptscriptstyle \mathcal{B}\mathrm{J}}}$ distribution,
without diminishing its own NLO accuracy. In this way the $\mathcal{B}$\noun{jj}-\noun{Minlo$^{\prime}$}
simulation can be made NLO accurate for $\mathcal{B}$\noun{j-} and
(NNLO) $\mathcal{B}$-inclusive observables.

The essential point one needs to prove for the self-consistency of
the method in this context is the same one as in sect.~\ref{sub:Removal-of-spurious-terms},
i.e.~that the $\delta(\Phi_{{\scriptscriptstyle \mathcal{B}\mathrm{J}}})$
that gets extracted does not blow up and risk breaking the NLO accuracy
of the initial uncorrected $\mathcal{B}$\noun{jj-Minlo}. This basically
boils down to saying that the existing \noun{Minlo }procedure resums
$v_{12}$ and $v_{01}$ both with $\mathrm{NLL}_{\sigma}$ accuracy.
As discussed at the end of sect.~\ref{sub:Removal-of-spurious-terms},
if all we cared about was unitarizing the cross section, this requirement
could be loosened to that of having just $\mathrm{LL}_{\sigma}$ accuracy
in place, including a further power of $L$ in $h\left(L\right)$. The
price of that ignorance would be that the correction can no longer be
interpreted as an approximation to missing higher order contributions
in the Sudakov form factor, i.e. one essentially gives up on a physical
interpretation of the mechanism of unitarity violation and, correspondingly,
one begins to warp the spectrum by higher order ambiguities that bear 
no relation to any kind of resummation. However, as we go on to explain,
we understand the $\mathcal{B}$\noun{jj-Minlo} cross section meets
already the above $\mathrm{NLL}_{\sigma}$ specification, with the
exception of a sub-leading kinematic region, which should not be
difficult to accommodate.

\subsection{\texorpdfstring{$\mathrm{NLL}_{\sigma}$}{NLLo} resummation\label{sub:NLL-jet-resolution}}

It is a general underlying assumption of the \noun{Caesar} formalism
that the Born configurations ($\Phi_{{\scriptscriptstyle \mathcal{B}\mathrm{J}}}$
in this case) consist of hard, well-separated, partons. So the $\mathrm{NLL}$/$\mathrm{NNLL}_{\sigma}$
theoretical framework from which we derived the resummation formula,
that was the starting point for section \ref{sec:Merging-without-a-merging-scale-two-units},
is not guaranteed to hold here, where we also need control $v_{12}$
resummation in the region $\sqrt{y_{01}}\ll m_{{\scriptscriptstyle \mathcal{B}}}$.

In section \ref{sub:The-validity-of-CAESAR-for-y12-when-y01-is-small}
we argue that the \noun{Caesar} resummation formula for $v_{12}=y_{12}/y_{01}$
resums large logarithms $\bar{\alpha}_{{\scriptscriptstyle \mathrm{S}}}^{n}L_{12}^{m}$,
$m\ge2n-1$, independently of the value of $v_{01}=y_{01}/m_{{\scriptscriptstyle \mathcal{B}}}^{2}$,
i.e.~even in the region $y_{01}\ll m_{{\scriptscriptstyle \mathcal{B}}}^{2}$.
This is based on the following two considerations:

\renewcommand{\labelenumi}{\roman{enumi}.)}
\begin{enumerate}
\item it is straightforward to show that the \noun{Caesar} Sudakov form
factor for $v_{12}$ resummation is equivalent to that prescribed
by the coherent parton branching formalism at $\mathrm{NLL}_{\sigma}$,
except for a sub-dominant subset of soft wide-angle radiation contributions,
beyond the accuracy of the latter formalism;%
\footnote{In other words the \noun{Caesar }Sudakov form factors capture the
same leading soft wide-angle terms as those in the coherent branching
formalism, as well as sub-leading ones which the latter discards. %
} 
\item the leading $\mathrm{NLL}_{\sigma}$ terms in the expansion of the
\noun{Caesar} $v_{12}$ cumulant distribution, are determined by integrating
the radiation pattern of a single soft/collinear emission relative
to an emitting $\mathcal{B}$J state, over the ordered region $y_{12}<y_{01}$,
and this pattern/integral derives independently of whether $y_{01}\lesssim m_{{\scriptscriptstyle \mathcal{B}}}^{2}$
or not. 
\end{enumerate}
\renewcommand{\labelenumi}{\arabic{enumi}.}
\noindent In i.) we are saying that the \noun{Caesar }Sudakov form
factor must be at least $\mathrm{LL}$ accurate for $v_{12}$,
since it agrees with the analogous expression derived from the coherent
branching formalism, which is understood to have at least that accuracy,
regardless of the value of the underlying $\Phi_{{\scriptscriptstyle \mathcal{B}\mathrm{J}}}$
configuration.%
\footnote{This statement also holds regardless of PDF considerations, since
$\mathrm{LL}$ effects only pertain to soft-collinear emissions.%
} Accepting i.) and ii.) together then implies that the \noun{Caesar
}resummation formula is $\mathrm{NLL}_{\sigma}$ regardless of the
value of the underlying $\Phi_{{\scriptscriptstyle \mathcal{B}\mathrm{J}}}$
configuration: since the Sudakov form factor is present in the resummation
formula as an overall factor, if the expansion of the formula generates
just the leading $\mathrm{NLL}_{\sigma}$ terms in the cross section
correctly, it generates all of them correctly.

For the reader who is willing to accept the statements above without
detailed explanation (the first of which is not obvious) we recommend
skipping \ref{sub:The-validity-of-CAESAR-for-y12-when-y01-is-small}.

In section \ref{sub:y01-resummation} we go on to include $v_{01}$
resummation at $\mathrm{NLL}_{\sigma}$. To this end we notice how,
if we include on top of the \noun{Caesar} $v_{12}$ resummation formula,
matched to \emph{leading} \emph{order} \noun{$\mathcal{B}$jj}, also
the $v_{01}$ Sudakov form factor, on integrating out $y_{12}$ we
obtain the \noun{$\mathcal{B}$j-Minlo }distribution to $\mathrm{NLL}_{\sigma}$.
The analysis in sect.~\ref{sub:MiNLO-jet-resolution-spectrum}
has made it clear already that this \noun{$\mathcal{B}$j-Minlo }distribution
recovers the \noun{Caesar $v_{01}$ }resummation formula on further
integration over the rapidity, $\mathrm{y}_{{\scriptscriptstyle \mathrm{J}}}$, and azimuth, $\phi_{{\scriptscriptstyle \mathrm{J}}}$, of the remaining pseudoparton.

With the latter modification we come full-circle: by concatenating
the two \noun{Caesar }resummations we get the same resummation as
the original \noun{Ckkw} \cite{Catani:2001cc,Krauss:2002up} and \noun{Minlo
}articles \cite{Hamilton:2012np}, to $\mathrm{NLL}_{\sigma}$, modulo
the terms in the $v_{12}$ Sudakov form factor mentioned overhead
in item i. If we restrict ourselves to the same accuracy remit as
the coherent branching formalism aims at, the only part of our prescription
not already specified in the original \noun{Ckkw} paper \cite{Catani:2001cc},
is the inclusion of the PDFs. Again, our argument to extend the prescription
to include PDFs (and also the aforementioned wide angle terms) is
based on the idea that if the resummation formula carries an overall
$\mathrm{LL}$ accurate Sudakov form factor and reproduces
just the leading $\mathrm{NLL}_{\sigma}$ terms correctly, it surely
reproduces all of the $\mathrm{NLL}_{\sigma}$ terms in the cross
section. It is reassuring then that our extension, in that respect,
to take into account the PDF effects, is also consistent with that
of the \noun{Ckkw} paper on hadronic collisions \cite{Krauss:2002up},
and the original \noun{Minlo} prescription \cite{Hamilton:2012np}.

While the arguments behind our nested resummation are strong enough
to convince us of its correctness, we do not consider that we have
definitively proven it.

Since it can lead to confusion, in reading this subsection \ref{sub:NLL-jet-resolution},
the sole goal of which is to give a conjectured $\mathrm{NLL}_{\sigma}$
resummation formula, we advise the reader to temporarily abandon all
thoughts about matching to NLO, and imagine instead just matching
to LO $\mathcal{B}$\noun{jj} cross sections.

\subsubsection{\texorpdfstring{$y_{12}$}{y12} resummation when \texorpdfstring{$y_{01}\lesssim m_{{\scriptscriptstyle \mathcal{B}}}^{2}$}{y01<mB2}\label{sub:The-validity-of-CAESAR-for-y12-when-y01-is-small}}

For what concerns this subsection we focus on large $L_{12}$ logarithms
relative to a given $\mathcal{B}$\noun{j }state. Large $L_{01}$
logarithms are discussed next in sect.~\ref{sub:y01-resummation}.
The statements we make regarding $L_{12}$ resummation should hold
independently of the behaviour of the $d\sigma/d\Phi_{{\scriptscriptstyle \mathcal{B}\mathrm{J}}}$
underlying Born cross section, be it pathological or otherwise. However,
should reassurance be needed already, the divergent $y_{01}\rightarrow0$
behaviour of $d\sigma/d\Phi_{{\scriptscriptstyle \mathcal{B}\mathrm{J}}}$
is ultimately tamed by inclusion of a Sudakov form factor consistent
with \noun{Caesar} and the coherent parton branching formalism.

We now specify the connection between the Sudakov form factors in
the \noun{Caesar} approach and those used for the coherent parton
branching formalism/\noun{Ckkw} \cite{Marchesini:1983bm,Webber:1983if,Catani:1990rr,Catani:1991hj,Catani:1992rm,Catani:1992ua,Catani:1992zp,Catani:1993hr,Catani:1993yx,Gerwick:2012fw,Gerwick:2013haa,Gerwick:2014koa}.
The latter formalism is capable of resumming $v_{12}$ logarithms
also in the small $v_{01}$ region. The key point that comes out of
this analysis, in regards to making the case for the nested resummation,
is that (ignoring potentially enhanced $\ln z$ terms%
\footnote{The \noun{Caesar }framework (like many other works) neglects the potential
small $x$ problems anyway. %
}) the Sudakov form factors associated with the $y_{12}$ resummation
in both approaches are the same to $\mathrm{NLL}_{\sigma}$.

For processes with $n=3$ hard legs, all $S_{1}^{\left[12\right]}$
coefficients (eq.~\ref{eq:sect22-n-eq-3-S1-defn}) can be written,
without approximations, as a piece containing a logarithm of $y_{01}$
plus a remainder term, $\Delta S_{1}$, which, crucially, for $Q_{{\scriptscriptstyle \mathcal{B}}}=m_{{\scriptscriptstyle \mathcal{B}}}$,
$Q_{{\scriptscriptstyle \mathcal{B}\mathrm{J}}}=\sqrt{y_{01}}$, has
no large-logarithmic dependence on $y_{01}$: 
\begin{equation}
S_{1}^{\left[12\right]}=G_{12}^{\left[01\right]}L_{01}+\Delta S_{1}\,.\label{eq:sect31-S1-decomposed-as-L01-plus-DeltaS1}
\end{equation}
Rewriting $S_{1}^{{\scriptscriptstyle \left[12\right]}}$ as in eq.~\ref{eq:sect31-S1-decomposed-as-L01-plus-DeltaS1}
is the key to understanding the connection between \noun{Ckkw} and
\noun{Caesar} here. In eq.~\ref{eq:sect31-S1-decomposed-as-L01-plus-DeltaS1}
the $G_{12}^{\left[01\right]}$ coefficient is that which one would
write down for the $n=2$ process underlying the $n=3$ one; $qq^{\prime}\rightarrow W/Z$
and $gg\rightarrow H$ for jet-associated $W/Z$ and Higgs boson production
processes. Explicit expressions for $\Delta S_{1}$ are given in appendix~\ref{sub:Expressions-for-n-eq-3-soft-wide-angle-coeffs-S1}.

While $\Delta S_{1}$ is free of large $y_{01}$ logarithms, it is
not zero. In the $n=3$, $2\rightarrow2$, hard configurations with
a gluon in the final-state, $\Delta S_{1}$ contains terms proportional
to $\ln z$, where $z=m_{{\scriptscriptstyle \mathcal{B}}}^{2}/\hat{s}$,
with $\hat{s}$ the invariant mass of the $2\rightarrow2$ collision.
In the $n=3$, $2\rightarrow2$, hard configurations with a fermion
emitted in the final-state, also terms proportional to $\ln\left(1-z\right)$
are present in $\Delta S_{1}$. Such terms are thrown out in the coherent
parton branching formalism as being beyond the accuracy aimed at there
for exclusive quantities, namely, control of all terms $\bar{\alpha}_{{\scriptscriptstyle \mathrm{S}}}^{n}L_{01}^{p}L_{12}^{q}$,
$p+q\ge2n-1$; heuristically, that accuracy implies a resummation
of an infinite number of soft and collinear emissions with, in addition,
up to one soft-wide-angle, or hard-collinear emission. Thus, in order
for soft-wide-angle emissions, which the $S_{1}^{{\scriptscriptstyle \left[12\right]}}$
terms are to account for, to be within the accuracy remit they must
have been emitted from an underlying $\mathcal{B}$\noun{j }state
for which $z\rightarrow1$. Equally, in the case of the $n=3$ reactions
with a fermion in final-state of the underlying Born, the fact that
the $n=3$ state is arrived at by fermion emission, means that further
radiation must be soft-collinear to register within the formalism's
accuracy. Thus the $\Delta S_{1}$ terms are (so far) outside the
scope of the coherent parton branching framework, indeed, they would
appear to exactly the kind of ``large angle soft gluon contributions
of order $\alpha_{{\scriptscriptstyle \mathrm{S}}}^{n}L^{2n-2}$''
which the formalism neglects in the case of multi-jet distributions
(pg.~11, ref.~\cite{Catani:1992ua}).

In comparing the \noun{Caesar} formulas to those of the coherent parton
branching framework we therefore now drop $\Delta S_{1}$ terms, and
those contributing beyond $\mathrm{NLL}_{\sigma}$, in the Sudakov
form factors, leading to the replacements 
\begin{eqnarray}
-R\left(v_{01}\right)\rightarrow-\bar{R}\left(v_{01}\right) & = & \int_{y_{01}}^{Q_{{\scriptscriptstyle \mathcal{B}}}^{2}}\frac{dy^{\prime}}{y^{\prime}}\,\bar{\alpha}_{{\scriptscriptstyle \mathrm{S}}}\left(y^{\prime}\right)\,\left[\,2G_{12}^{{\scriptscriptstyle \left[01\right]}}\ln\frac{m_{{\scriptscriptstyle \mathcal{B}}}^{2}}{y^{\prime}}+G_{11}^{{\scriptscriptstyle \left[01\right]}}\,\right]\,,\label{eq:sect31-Rv01-NLLsigma-approx}\\
\nonumber \\
-R\left(v_{12}\right)\rightarrow-\bar{R}\left(v_{12}\right) & = & \int_{y_{12}}^{y_{01}}\frac{dy^{\prime}}{y^{\prime}}\,\bar{\alpha}_{{\scriptscriptstyle \mathrm{S}}}\left(y^{\prime}\right)\,\left[\,2G_{12}^{{\scriptscriptstyle \left[01\right]}}\ln\frac{m_{{\scriptscriptstyle \mathcal{B}}}^{2}}{y^{\prime}}+G_{11}^{{\scriptscriptstyle \left[01\right]}}\,\right]\label{eq:sect31-Rv12-NLLsigma-approx}\\
 & + & \int_{y_{12}}^{y_{01}}\frac{dy^{\prime}}{y^{\prime}}\,\bar{\alpha}_{{\scriptscriptstyle \mathrm{S}}}\left(y^{\prime}\right)\,\left[\,2\left(G_{12}^{{\scriptscriptstyle \left[12\right]}}-G_{12}^{{\scriptscriptstyle \left[01\right]}}\right)\ln\frac{y_{01}}{y^{\prime}}+\left(G_{11}^{{\scriptscriptstyle \left[12\right]}}-G_{11}^{{\scriptscriptstyle \left[01\right]}}\right)\,\right]\,.\nonumber 
\end{eqnarray}
Translating eqs.~\ref{eq:sect31-Rv01-NLLsigma-approx}-\ref{eq:sect31-Rv12-NLLsigma-approx}
in terms of the notation of the coherent parton branching formalism
we get, without approximations, 
\begin{equation}
\mathrm{e}^{-\bar{R}\left(v_{01}\right)}=\prod_{\ell\in{\scriptscriptstyle \left[01\right]}}\Delta_{\ell}\left(\sqrt{y_{01}},m_{{\scriptscriptstyle \mathcal{B}}}\right)\,,\qquad\mathrm{e}^{-\bar{R}\left(v_{12}\right)}=\frac{\prod_{\ell\in{\scriptscriptstyle \left[01\right]}}\Delta_{\ell}\left(\sqrt{y_{12}},m_{{\scriptscriptstyle \mathcal{B}}}\right)}{\prod_{\ell\in{\scriptscriptstyle \left[01\right]}}\Delta_{\ell}\left(\sqrt{y_{01}},m_{{\scriptscriptstyle \mathcal{B}}}\right)}\,\,\frac{\prod_{\ell\in{\scriptscriptstyle \left[12\right]}}\Delta_{\ell}\left(\sqrt{y_{12}},\sqrt{y_{01}}\right)}{\prod_{\ell\in{\scriptscriptstyle \left[01\right]}}\Delta_{\ell}\left(\sqrt{y_{12}},\sqrt{y_{01}}\right)}\,.\label{eq:sect31-CAESAR-Sudakovs-as-CKKW-Sudakovs}
\end{equation}
where $\ell\in\left[01\right]$ means one of the two coloured legs
$\ell$ which directly attaches itself to $\mathcal{B}$,%
\footnote{In the cases at hand $\ell\in\left[01\right]$ then means $\ell$
is always a quark if $\mathcal{B}$ is a vector boson, it is always
a gluon if $\mathcal{B}$ is the Higgs boson.%
} while $\ell\in\left[12\right]$ means any of the three coloured legs
external to the $\mathcal{B}$\noun{j} state. Definitions of the Sudakov
form factors $\Delta_{\ell}$ are given in appendix \ref{sub:NLLsigma-resummation-in-CKKW-notation},
they are the same as those used widely in the literature on the coherent
parton branching formalism/\noun{Ckkw} (e.g.~ref.~\cite{Catani:2001cc}).
Observe how the form of the product of the two \noun{Caesar}-style
Sudakov form factors gives the breakdown one expects in terms of the
Sudakov form factors employed by the coherent parton branching formalism/\noun{Ckkw}
method. Continuing to neglect $\Delta S_{1}$ terms,
the $y_{12}$ Sudakov form factor can be rewritten without further
approximation as 
\begin{equation}
\mathrm{e}^{-\bar{R}\left(v_{12}\right)}=\exp\left[\int_{0}^{L_{12}}dL_{12}^{\prime}\,\bar{\alpha}_{{\scriptscriptstyle \mathrm{S}}}\left(y^{\prime}\right)\,2S_{1}^{\left[12\right]}\right]\,\prod_{\ell\in{\scriptscriptstyle \left[12\right]}}\Delta_{\ell}\left(\sqrt{y_{12}},\sqrt{y_{01}}\right)\,,\qquad(y^{\prime}=y_{01}\exp\left[-L_{12}^{\prime}\right])\,,\label{eq:sect31-CAESAR-y12-Sudakov-as-naive-CKKW-Sudakovs-times-soft-wide-angle-Sudakov}
\end{equation}
making clear the difference between it and what one might have expected
based on a naive transverse momentum ordering, i.e.~the same expression
without the first exponential accounting for coherent soft-wide-angle
emission.

Finally, we have that the coherent parton branching formalism and
the \noun{Caesar }$y_{12}$ resummation formula are consistent in
regards to the Sudakov form factors they would assign for the $y_{12}$
(and $y_{01}$) resummations, at the level to which the former is
accurate. \noun{Caesar}'s accounting for soft-wide angle resummation,
via the leading part of its $S_{1}^{{\scriptscriptstyle \left[12\right]}}$
term (eq.~\ref{eq:sect31-S1-decomposed-as-L01-plus-DeltaS1}) is
essential for this non-trivial agreement. Beyond the domain of validity
of the coherent branching formalism we only have $\mathrm{LL}$ agreement
with the coherent parton branching formalism in the \noun{Caesar}
Sudakov form factor for $y_{12}$ resummation.

Following the argument laid out surrounding bullets i.) and ii.) in
sect.~\ref{sub:NLL-jet-resolution}, we therefore consider the \noun{Caesar
$y_{12}$ }resummation formula to be $\mathrm{NLL}_{\sigma}$ accurate
also in the region $y_{01}\lesssim m_{{\scriptscriptstyle \mathcal{B}}}^{2}$.
For this to be false requires either: i.) the statement that the $y_{12}$
resummation formula is not $\mathrm{LL}$ accurate for arbitrary
$\Phi_{{\scriptscriptstyle \mathcal{B}\mathrm{J}}}$ to be false,
which conflicts with the coherent parton branching formalism; ii.)
the leading $\mathrm{NLL}_{\sigma}$ terms in the expansion of the
\noun{Caesar} $y_{12}$ cumulant in the region $y_{01}\lesssim m_{{\scriptscriptstyle \mathcal{B}}}^{2}$
do not follow directly from integrating the soft/collinear radiation
pattern of a single emission with respect to the emitting $\mathcal{B}$\noun{j
}configuration\noun{, }over the region $y_{12}<y_{01}$.

\subsubsection{\texorpdfstring{$y_{01}$}{y01} resummation\label{sub:y01-resummation}}

Going back to our initial resummation formula of section \ref{sec:Merging-without-a-merging-scale-two-units},
neglecting higher order terms, we now understand the following resummation
formula to be $\mathrm{NLL}_{\sigma}$ independently of $\Phi_{{\scriptscriptstyle \mathcal{B}\mathrm{J}}}$,

\begin{eqnarray}
\frac{d\sigma_{{\scriptscriptstyle \mathcal{R}}}^{{\scriptscriptstyle \mathcal{B}\mathrm{JJ}}}}{d\Phi_{{\scriptscriptstyle \mathcal{B}\mathrm{J}}}dL_{12}} & = & \frac{d\sigma_{0}^{{\scriptscriptstyle \mathcal{B}\mathrm{J}}}}{d\Phi_{{\scriptscriptstyle \mathcal{B}\mathrm{J}}}}\,\frac{d}{dL_{12}}\,\left[\mathrm{e}^{-R\left(v_{12}\right)}\,\prod_{\ell=1}^{n_{i}}\frac{q^{\left(\ell\right)}\left(x_{\ell}^{{\scriptscriptstyle \left[12\right]}},y_{12}\right)}{q^{\left(\ell\right)}\left(x_{\ell}^{{\scriptscriptstyle \left[12\right]}},y_{01}\right)}\right]\,,\label{eq:sect31-dsigmaRBJJ-NLL}
\end{eqnarray}
where $x_{\ell}^{{\scriptscriptstyle \left[12\right]}}$ refers to
the momentum fractions of the incoming partons colliding to make the
$\mathcal{B}$\noun{j }system. Integrating this formula over $y_{12}$
we obtain the leading order $\Phi_{{\scriptscriptstyle \mathcal{B}\mathrm{J}}}$
distribution, $d\sigma_{0}^{{\scriptscriptstyle \mathcal{B}\mathrm{J}}}/d\Phi_{{\scriptscriptstyle \mathcal{B}\mathrm{J}}}$,
up to NLO-sized ambiguities. The renormalization scale in the coupling
constants in $d\sigma_{0}^{{\scriptscriptstyle \mathcal{B}\mathrm{J}}}$
is $\mu_{{\scriptscriptstyle R}}$ and the factorization scale in
the PDFs is $\mu_{{\scriptscriptstyle F}}$. It then follows directly
(given the correspondence between the \noun{Minlo }procedure in sect.~\ref{sub:MiNLO-jet-resolution-spectrum}
and the initial resummation formula eq.~\ref{eq:sect22-NNLL_sigma-differential-xsecn-i})
that for $\mu_{{\scriptscriptstyle F}}=\sqrt{y_{01}}$ in $d\sigma_{0}^{{\scriptscriptstyle \mathcal{B}\mathrm{J}}}$,
if we include a factor $\mathcal{W}^{{\scriptscriptstyle \left[01\right]}}$
in the form 
\begin{equation}
\frac{d\bar{\sigma}_{{\scriptscriptstyle \mathcal{R}}}^{{\scriptscriptstyle \mathcal{B}\mathrm{JJ}}}}{d\Phi_{{\scriptscriptstyle \mathcal{B}\mathrm{J}}}dL_{12}}=\mathcal{W}^{{\scriptscriptstyle \left[01\right]}}\,\frac{d\sigma_{{\scriptscriptstyle \mathcal{R}}}^{{\scriptscriptstyle \mathcal{B}\mathrm{JJ}}}}{d\Phi_{{\scriptscriptstyle \mathcal{B}\mathrm{J}}}dL_{12}}\,,\qquad\mathcal{W}^{{\scriptscriptstyle \left[01\right]}}=\exp\left[-R\left(v_{01}\right)\right]\,\frac{\bar{\alpha}_{{\scriptscriptstyle \mathrm{S}}}\left(y_{01}\right)}{\bar{\alpha}_{{\scriptscriptstyle \mathrm{S}}}\left(\mu_{{\scriptscriptstyle R}}\right)}\,,\label{eq:sect31-W01-factor}
\end{equation}
we reproduce the $\mathcal{B}$\noun{j-Minlo }distribution, and hence
also the \noun{Caesar }$y_{01}$ resummation formula, to $\mathrm{NLL}_{\sigma}$
accuracy. We conclude that $d\bar{\sigma}_{{\scriptscriptstyle \mathcal{R}}}^{{\scriptscriptstyle \mathcal{B}\mathrm{JJ}}}$
in eq.~\ref{eq:sect31-W01-factor} above, is $\mathrm{NLL}_{\sigma}$
accurate in the resummation of $L_{12}$ for arbitrary given $\Phi_{{\scriptscriptstyle \mathcal{B}\mathrm{J}}}$,
and that it reproduces, on integration, the $L_{01}$ resummation
to the same precision.

\subsection{\texorpdfstring{$\mathcal{B}$\noun{jj-Minlo}}{BJJ} jet resolution spectra\noun{\label{sub:Extended-Bjj-Minlo}}}

Expanding the conjectured resummation formula in $\bar{\alpha}_{{\scriptscriptstyle \mathrm{S}}}$
to give the associated NLO approximation for the $\mathcal{B}$\noun{jj}
cross section, we get 
\begin{eqnarray}
\frac{d\bar{\sigma}_{{\scriptscriptstyle \mathcal{S}}}^{{\scriptscriptstyle \mathcal{B}\mathrm{JJ}}}}{d\Phi_{{\scriptscriptstyle \mathcal{B}\mathrm{J}}}dL_{12}} & = & \frac{d\sigma_{0}^{{\scriptscriptstyle \mathcal{B}\mathrm{J}}}}{d\Phi_{{\scriptscriptstyle \mathcal{B}\mathrm{J}}}}\,\left[1+\sum_{m=1}^{2}\, R_{1m}^{{\scriptscriptstyle \left[01\right]}}\bar{\alpha}_{{\scriptscriptstyle \mathrm{S}}}L_{01}^{m}\right]\,\left[\sum_{n=1}^{2}\sum_{m=2n-2}^{2n-1}\, H_{nm}^{{\scriptscriptstyle \left[12\right]}}\bar{\alpha}_{{\scriptscriptstyle \mathrm{S}}}^{n}L_{12}^{m}\right]\,,\label{eq:sect-32-expanded-resummation-formula}
\end{eqnarray}
where the coefficients $H_{nm}^{{\scriptscriptstyle \left[12\right]}}$
have the same form as those introduced in sect.~\ref{sub:MiNLO-jet-resolution-spectrum},
with the renormalization and factorization scales $\mu_{{\scriptscriptstyle R}}=m_{{\scriptscriptstyle \mathcal{B}}}$
and $\mu_{{\scriptscriptstyle F}}=\sqrt{y_{01}}$ throughout;%
\footnote{Including inside the PDF factors of the $d\sigma_{0}^{{\scriptscriptstyle \mathcal{B}\mathrm{J}}}$
term. %
} explicit expressions for these can be found in appendix \ref{sub:Fixed-order-expansion-of-conjectured-resummation-formula}.
If we now trace the effects of the \noun{Minlo }procedure on this
fixed order expansion, by analogy to the exercise of sect.~\ref{sub:MiNLO-jet-resolution-spectrum},
we find that the final \noun{Minlo }cross section is in agreement
with the resummation formula, eq.~\ref{eq:sect31-W01-factor}, up
to sub-leading terms outside the control of the latter. Specifically,
here the \noun{Minlo }procedure for the NLO $\mathcal{B}$\noun{jj
}cross section is: 
\begin{enumerate}
\item Set $\mu_{{\scriptscriptstyle R}}$ and $\mu_{{\scriptscriptstyle F}}$
according to, 
\[
d\sigma\rightarrow d\sigma^{\prime}=d\sigma\,\left(\mu_{{\scriptscriptstyle R}}\rightarrow m_{{\scriptscriptstyle \mathcal{B}}},\,\mu_{{\scriptscriptstyle F}}\rightarrow\sqrt{y_{12}}\right)\,.
\]

\item Multiply the LO component by the $\mathcal{O}\left(\bar{\alpha}_{{\scriptscriptstyle \mathrm{S}}}\right)$
expansion of the inverse of the product of $y_{01}$ and $y_{12}$
Sudakov form factors times $\bar{\alpha}_{s}\left(y_{01}\right)/\bar{\alpha}_{s}\left(\mu_{{\scriptscriptstyle R}}^{2}\right)$
and $\bar{\alpha}_{s}\left(y_{12}\right)/\bar{\alpha}_{s}\left(\mu_{{\scriptscriptstyle R}}^{2}\right)$
(terms beyond $\mathrm{NLL}_{\sigma}$ accuracy): 
\begin{eqnarray*}
d\sigma^{\prime}\rightarrow d\sigma^{\prime\prime} & = & d\sigma^{\prime}-\left.d\sigma^{\prime}\right|_{{\scriptscriptstyle \mathrm{LO}}}\bar{\alpha}_{{\scriptscriptstyle \mathrm{S}}}\left(\mu_{{\scriptscriptstyle R}}^{2}\right)\left(G_{12}^{{\scriptscriptstyle \left[01\right]}}\, L_{01}^{2}+\left(G_{11}^{{\scriptscriptstyle \left[01\right]}}+2S_{1}^{{\scriptscriptstyle \left[01\right]}}+2\bar{\beta}_{0}\right)L_{01}\right)\\
 &  & \phantom{d\sigma^{\prime\prime}}-\left.d\sigma^{\prime}\right|_{{\scriptscriptstyle \mathrm{LO}}}\bar{\alpha}_{{\scriptscriptstyle \mathrm{S}}}\left(\mu_{{\scriptscriptstyle R}}^{2}\right)\left(G_{12}^{{\scriptscriptstyle \left[12\right]}}\, L_{12}^{2}+\left(G_{11}^{{\scriptscriptstyle \left[12\right]}}+2S_{1}^{{\scriptscriptstyle \left[12\right]}}+\bar{\beta}_{0}\right)L_{12}\right)\,.
\end{eqnarray*}

\item Multiply by the \noun{Minlo }Sudakov form factors and $\bar{\alpha}_{{\scriptscriptstyle \mathrm{S}}}$
ratios: 
\begin{eqnarray}
d\sigma^{\prime\prime}\rightarrow d\sigma_{{\scriptscriptstyle \mathcal{M}}}^{{\scriptscriptstyle \mathcal{B}\mathrm{JJ}}} & = & \mathrm{e}^{-R\left(v_{01}\right)}\,\frac{\bar{\alpha}_{s}\left(y_{01}\right)}{\bar{\alpha}_{s}\left(\mu_{{\scriptscriptstyle R}}^{2}\right)}\,\mathrm{e}^{-R\left(v_{12}\right)}\,\frac{\bar{\alpha}_{s}\left(y_{12}\right)}{\bar{\alpha}_{s}\left(\mu_{{\scriptscriptstyle R}}^{2}\right)}\, d\sigma^{\prime\prime}\,.\label{eq:sect32-minlo-step-4-sudakov-aS-ratio-times-subtracted-xsecn}
\end{eqnarray}

\end{enumerate}
With these operations we find we can write $d\sigma_{{\scriptscriptstyle \mathcal{M}}}^{{\scriptscriptstyle \mathcal{B}\mathrm{JJ}}}=d\bar{\sigma}_{{\scriptscriptstyle \mathcal{R}}}^{{\scriptscriptstyle \mathcal{B}\mathrm{JJ}}}$,
as in eq.~\ref{eq:sect31-W01-factor}, neglecting sub-leading terms
unaccounted for by the $d\bar{\sigma}_{{\scriptscriptstyle \mathcal{R}}}^{{\scriptscriptstyle \mathcal{B}\mathrm{JJ}}}$
formula.

Recalling that the product of the \noun{Caesar }Sudakov form factors
is equivalent at $\mathrm{NLL}_{\sigma}$ accuracy to the product
of those prescribed in the \noun{Ckkw} method and the original \noun{Minlo
}procedure \cite{Hamilton:2012np}, modulo the $\Delta S_{1}^{\left[{\scriptscriptstyle 12}\right]}$
soft-wide angle contributions, already elaborated on. The only other
difference between the original \noun{Minlo }procedure and that enumerated
above is the prescription for the scale to use in the addition factor
of $\bar{\alpha}_{{\scriptscriptstyle \mathrm{S}}}$ accompanying
the NLO corrections --- the original \noun{Minlo }procedure suggests
to use the arithmetic mean of all other $\bar{\alpha}_{{\scriptscriptstyle \mathrm{S}}}$
factors, on an event-by-event basis --- a difference affecting terms
beyond level of accuracy needed here. In conclusion, then the \noun{Minlo
}procedure outlined above, deriving from joining the \noun{Caesar
$y_{01}$ }and $y_{12}$ resummations, boils down to the original
\noun{Minlo }prescription at the $\mathrm{NLL}_{\sigma}$ level specified
at the end of sect.~\ref{sub:y01-resummation}, excepting the sub-dominant
wide-angle $\Delta S_{1}^{\left[{\scriptscriptstyle 12}\right]}$
Sudakov form factor terms. As indicated already in the introduction
to this section, the `product' of the two \noun{Caesar }resummations
has returned us, somewhat remarkably, almost exactly back to the \noun{Ckkw/Minlo}
recipe.

\subsection{Integrated \texorpdfstring{$\mathcal{B}$\noun{jj-Minlo}}{BJJ-MINLO} jet resolution spectra\label{sub:MiNLO-integrated-jet-resolution-spectrum-3-units}}

Granted that the $\mathcal{B}$\noun{jj-Minlo }procedure is $\mathrm{NLL}_{\sigma}$
accurate for the $v_{12}$ resummation and $\mathrm{NLL}_{\sigma}$
for $v_{01}$ when $y_{12}$ is integrated out, it follows that 
\begin{equation}
\delta(\Phi_{{\scriptscriptstyle \mathcal{B}\mathrm{J}}})=\left(\frac{d\sigma_{{\scriptscriptstyle \mathcal{M}}}^{{\scriptscriptstyle \mathcal{B}\mathrm{JJ}}}}{d\Phi_{{\scriptscriptstyle \mathcal{B}\mathrm{J}}}}-\frac{d\sigma_{{\scriptscriptstyle \mathcal{M}}}^{{\scriptscriptstyle \mathcal{B}\mathrm{J}}\,\prime}}{d\Phi_{{\scriptscriptstyle \mathcal{B}\mathrm{J}}}}\right)\,/\,\int dL_{12}\, h\left(L_{12}\right)\,\frac{d\sigma_{{\scriptscriptstyle \mathcal{M}}}^{{\scriptscriptstyle \mathcal{B}\mathrm{JJ}}}}{d\Phi_{{\scriptscriptstyle \mathcal{B}\mathrm{J}}}dL_{12}}=\sum_{n=0}^{\infty}e_{n}(\Phi_{{\scriptscriptstyle \mathcal{B}\mathrm{J}}})\,\bar{\alpha}_{{\scriptscriptstyle \mathrm{S}}}^{n/2}L_{01}^{n}\quad,\label{eq:sect33-delta-Phi-BJ}
\end{equation}
where the $e_{n}$ coefficients are $\mathcal{O}\left(1\right)$.
The $e_{n}$ coefficients carry no divergent $1/y_{01}$ factors ---
these cancel between the numerator and denominator of $\delta$%
\footnote{Both terms in the numerator of $\delta\left(\Phi_{{\scriptscriptstyle \mathcal{B}\mathrm{J}}}\right)$,
and the denominator, are proportional to the Born $d\sigma_{{\scriptscriptstyle 0}}^{{\scriptscriptstyle \mathcal{B}\mathrm{J}}}/d\Phi_{{\scriptscriptstyle \mathcal{B}\mathrm{J}}}$%
.} --- equally, they contain no large $L_{01}$ factors.\footnote{The $e_{n}$ coefficients do contain powers of $\bar{\alpha}_{{\scriptscriptstyle \mathrm{S}}}L_{01}$ and other subleading contributions.}
Thus $\delta(\Phi_{{\scriptscriptstyle \mathcal{B}\mathrm{J}}})$
is formally also $\mathcal{O}\left(1\right)$, neglecting deep Sudakov
regions where $\bar{\alpha}_{{\scriptscriptstyle \mathrm{S}}}L_{01}^{2}\gtrsim 1$.
This means that we are justified in applying the procedure of sect.~\ref{sub:Variation-on-MinloPrime}
with the replacements $d\sigma_{{\scriptscriptstyle \mathcal{M}}}\rightarrow d\sigma_{{\scriptscriptstyle \mathcal{M}}}^{{\scriptscriptstyle \mathcal{B}\mathrm{JJ}}}$,
$d\sigma_{{\scriptscriptstyle \mathrm{NLO}}}\rightarrow d\sigma_{{\scriptscriptstyle \mathcal{M}}}^{{\scriptscriptstyle \mathcal{B}\mathrm{J}}\,\prime}$,
leading to eq.~\ref{eq:sect3-formula-from-intro-blurb}, and hence
recover also NLO accuracy for $\mathcal{B}$-inclusive quantities, or NNLO
accuracy, should the $\mathcal{B}$\noun{j-Minlo$^{\prime}$ }distribution
have been reweighted to NNLO \cite{Hamilton:2013fea}.

\section{Feasibility study\label{sec:Feasibility-study}}

In the following we show how the above merging of three units of multiplicity
works in a practical implementation. For this we consider Higgs production
at the LHC, with a collision energy of 8 TeV. We `merge' the \noun{Hjj-Minlo}
simulation to an existing\noun{ Hj-Minlo$^{\prime}$} simulation reweighted
to NNLO according to the prescription of ref.~\cite{Hamilton:2013fea}.
In the following we therefore make predictions that are NNLO accurate
for inclusive Higgs boson production, and NLO accurate for \noun{Hj}
and \noun{Hjj }observables. The inclusive matrix element predictions
are matched to the parton shower using the \noun{Powheg }method.

\subsection{Implementation\label{sub:Implementation}}

In order to simplify the implementation and require no changes to
the existing \noun{Hjj} and \noun{Hnnlops} processes in the \noun{Powheg-Box},
we have chosen to work at the level of the Les Houches events (LHE).
The distributions formed from LHE in \noun{Powheg} are NLO accurate,
i.e.~differences with a fixed order NLO computation are beyond NLO
accuracy. Relatedly, they respect the $\mathrm{NLL}_{\sigma}$ accuracy
of \noun{Minlo}: the difference in phase space w.r.t.~the matrix
elements is beyond $\mathrm{LL}$ and the matching to NLO is then
enough to preserve distributions at the $\mathrm{NLL}_{\sigma}$ level.
We consider that working at the level of the LHE also simplifies the
generation of the final results: we have written an independent code
that reads in \noun{Hjj} and \noun{Hnnlops} LHE files and writes out
a reweighted \noun{Hjj} LHE file to achieve the results of the three
units of multiplicity merging.

As described in sect.~\ref{sub:Variation-on-MinloPrime}, we need
to correct the $\frac{d\sigma_{{\scriptscriptstyle \mathcal{M}}}^{{\scriptscriptstyle \mathrm{HJJ}}}}{d\Phi_{{\scriptscriptstyle \mathrm{HJ}}}dL_{12}}$
in such a way that when integrated over $L_{12}$ it returns the \noun{Hj}-\noun{Minlo$^{\prime}$/Hnnlops}
$\Phi_{{\scriptscriptstyle \mathrm{HJ}}}$ distribution. This can
be done by multiplying the fully differential \noun{Hjj} calculation
by the $\left(1-\Delta R(v_{12})^{{\scriptscriptstyle \mathrm{approx}}}\right)$
factor as described in eq.~\ref{eq:sect26-dsigmaMcorr-eq-dsigmaM-times-1-minus-DeltaRapprox}.
This factor can only be computed \emph{after} integration over $L_{12}$,
as is clear from eq.~\ref{eq:sect26-first-delta-defn}. To avoid
performing the complete $L_{12}$ integration for every $\Phi_{{\scriptscriptstyle \mathrm{HJ}}}$
phase-space point, and given that this integral is too complicated
to perform analytically, we instead have chosen to setup three three-dimensional
interpolation grids for the three contributions to $\delta(\Phi_{{\scriptscriptstyle \mathrm{HJ}}})$:
the two terms in the numerator and the term in the denominator, respectively.
The three dimensions are the rapidity of the Higgs boson, the rapidity of
the hardest jet and the transverse momentum of the hardest jet. These
being the dimensions making up the non-trivial part of the \noun{Hj} phase space $\Phi_{{\scriptscriptstyle \mathrm{HJ}}}$; the dependence on the azimuthal
angle, $\phi_{{\scriptscriptstyle \mathrm{J}}}$, is completely flat.
Indeed these interpolation grids can be filled quickly with the LHE
from the existing \noun{Hjj} and \noun{Hnnlops} implementations in
the \noun{Powheg-Box} framework. We have generated the interpolation
grids using rigid binning as well as a method based upon \noun{Parni}~\cite{vanHameren:2007pt}
to dynamically create hypercubes in the three dimensions; we did not
see appreciable improvements using the more involved \noun{Parni}
method and the results we present here are therefore based on the
implementation using the simpler fixed interpolation grid bins.

In implementing the $h\left(L\right)$ function of eq.~\ref{eq:sect26-freezing-out-the-weight-factor-at-aSL2-gtrsim-1},
we have softened the abrupt transition at the freezing scale, manifested
by the step functions. Specifically we implement $h\left(L\right)$
as 
\begin{eqnarray}
h_{\phantom{0}}\left(L\right) & = & \bar{\alpha}_{{\scriptscriptstyle \mathrm{S}}}\left[\,\bar{\alpha}_{{\scriptscriptstyle \mathrm{S}}}L^{2}h_{0}\left(L\right)+\frac{\rho}{2\left|G_{12}\right|}\,\left[1-h_{0}\left(L\right)\right]\,\right]\,,\label{eq:sect41-implemented-hL-fn}\\
h_{0}\left(L\right) & = & \rho^{\gamma}/\left[\rho^{\gamma}-\left(2\left|G_{12}\right|\bar{\alpha}_{{\scriptscriptstyle \mathrm{S}}}L^{2}\right)^{\gamma}\right]\,,\label{eq:sect41-implemented-h0L-fn}
\end{eqnarray}
taking $\gamma=5$. As $\gamma\rightarrow\infty$ the $h\left(L\right)$
function becomes exactly that of eq.~\ref{eq:sect26-freezing-out-the-weight-factor-at-aSL2-gtrsim-1},
but for a rescaling $\rho\rightarrow\rho/(2\left|G_{12}\right|)$.
Thus, $h\left(L\right)$ becomes frozen in the region where the leading
double log term in the Sudakov exponent is $\approx\rho$. We probe
the sensitivity of our results to $\rho$ (and therefore, indirectly,
also $\gamma$) by computing predictions with $\rho=1$, 3, 9, 18
and 27, with the central renormalization and factorization scale choices.
To avoid confusion, we already remark that the results in the next
section prove to be quite robust against variations of $\rho$: for
quite a number of observables it appears there is no visible variation
at all, although, for sufficiently inclusive observables, that is
not unexpected.

Because we have chosen not to change the existing \noun{Minlo} implementation
of the \noun{Hjj} process, the $\Delta S_{1}$ terms, as introduced
in eq.~\ref{eq:sect31-S1-decomposed-as-L01-plus-DeltaS1}, are not
included in our Sudakov exponents. Recall that these $\mathrm{NLL}_{\sigma}$
terms only become relevant for $\Phi_{{\scriptscriptstyle \mathcal{B}\mathrm{J}}}$
configurations where the leading pseudoparton is hard-collinear. Furthermore,
the region where $y_{01}\gtrsim\mathcal{O}(m_{{\scriptscriptstyle \mathrm{H}}}^{2})$,
is beyond the scope of the coherent parton branching formalism, because
the first emission, i.e.~the one entering $y_{01}$, is not enhanced
by any large logarithm in that case. As clarified in sect.~\ref{subsec:comparison-to-ckkw-minlo},
it follows that \noun{Ckkw}/\noun{Minlo} does not lead to the correct
Sudakov factors at $\mathrm{NLL}_{\sigma}$ accuracy in the $y_{12}$
variable, in this region of phase-space: they miss the $S_{1}$ contribution
due to soft-wide-angle radiation. In this feasibility study we chose
to ignore these facts, with the expectation that technical issues
might well instead present us with more serious, immovable obstacles.
Formally, if these missing contributions would turn out to be important,
$\delta(\Phi_{{\scriptscriptstyle \mathrm{HJ}}})$, with the definition
of $h(L_{12})$ as in eq.~\ref{eq:sect26-first-delta-defn}, would
no longer be an order one quantity, c.f.~eq.~\ref{eq:sect26-deltaPhi-eq-order-1-plus-order-root-aS}.
It is not difficult to include these terms in the \noun{Minlo }framework,
indeed one of the results of this work has been to pinpoint these
and other terms, which can improve the quality of the resummation.
We leave the implementation of such terms to future work, although
all indications from the following results suggest that this
stands to be, fortunately and unfortunately, a null exercise. We have
checked that in all of phase-space $\delta(\Phi_{{\scriptscriptstyle \mathrm{HJ}}})$
remains within the range of values associated to resummation constants
used in Higgs boson transverse momentum resummation.

Lastly we comment that we work in the SM theory in which the top quark
is integrated out. This results in a well-known Higgs effective theory
with tree-level interactions between the Higgs boson and gluons. This
approximation breaks down if the Higgs or the gluons carry enough
energy to resolve the shrunk top quark loop, e.g., when the leading
jet transverse momentum exceeds the top quark mass. We also do not
include b quark mass effects, setting the b quark mass and Yukawa
coupling to zero. Accounting for these finite-quark mass effects at
the Born level in the \noun{Hjj Powheg }generator could be done in
an analogous way to ref.~\cite{Hamilton:2015nsa}.

\subsection{Results and testing\label{sub:Results-and-testing}}

In the hard matrix elements we set the Higgs mass to $m_{{\scriptscriptstyle \mathrm{H}}}=125$
GeV and keep it stable throughout the simulation. The LHE are showered
with \noun{Pythia6}~\cite{Sjostrand:2006za}, using the \noun{Perugia~0}
tune~\cite{Skands:2010ak} but with hadronization and multiple-parton
interactions turned off. The central renormalisation and factorization
scales are set according to the \noun{Minlo} procedure. To assess
the scale dependence in \noun{Hjj-Minlo} we vary the renormalisation
and factorization scales independently, by a factor two around their
central values, omitting the two values where the scales are changed
oppositely. This results in 7 curves, whose envelope gives the uncertainty
band. For the \noun{Nnlops} we use procedure advocated in ref.~\cite{Hamilton:2013fea},
resulting into 21 curves. In the merging of the new improved \noun{Hjj-Minlo} 
results we keep the scales used in the input \noun{Hjj} and \noun{Hj}
(which on its own is an input to \noun{Nnlops}) calculations correlated.
Hence, this also results in 21 curves, the envelope of which defines
the uncertainty band. We employ the \noun{MSTW2008nnlo} PDF set~\cite{Martin:2009iq}
for all contributions and refrain from showing uncertainties of PDF
origin.

All figures that we present here have the same layout. They contain
a main panel on the left and three ratio plots on the right-hand sides.
In the main panel, we show the central values for the \noun{Nnlops}
predictions for inclusive Higgs boson production in green (\noun{Nnlops}),
the pre-existing \noun{Hjj-Minlo }ones in blue (\noun{Hjj}), and the
predictions of our new improved \noun{Hjj-Minlo} procedure in red
($\noun{Hjj}^{\star}$), together with its scale uncertainty band.
The right-hand plots display the ratio of these predictions, from
top to bottom, with respect to the $\noun{Hjj}^{\star}$, \noun{Nnlops}
and \noun{Hjj} results. The coloured band in each of the latter plots
shows the scale uncertainty associated to the prediction in the denominator
of the corresponding ratio.

In the upper right-hand panel we also show, in all cases, superimposed
on top of the light-red scale uncertainty band, a much darker red
uncertainty band, formed by varying the $\rho$ parameter of the correction
procedure (see again sects.~\ref{sub:Variation-on-MinloPrime} and
\ref{sub:Implementation}). The precise implementation of the $h\left(L\right)$
function, through which dependence on this parameter enters, was described
in the previous section, surrounding eq.~\ref{eq:sect41-implemented-hL-fn}.
We re-iterate that the dark-red band, depicting uncertainty due to
this $\rho$ parameter, was formed by taking the envelope of predictions
made with $\rho=1$, 3, 9, 18 and 27, using the central renormalization
and factorization scale choices.

We remind that the correction procedure, as described in sects.~\ref{sub:Variation-on-MinloPrime}
and \ref{sec:Merging-without-a-merging-scale-three-units}, should
function such that quantities which are fully inclusive with respect
to the $y_{12}$ variable have no sensitivity to $\rho$ at all. Thus,
for at least the first few figures we look at in this section, focusing
on fully inclusive and $\noun{Hj}$-inclusive observables, the aforementioned
dark-red band should be (and is) invisible, being obscured by the
horizontal black reference line. Moving on to more interesting observables,
particularly probing the behaviour of the second jet/second pseudoparton
in the event, the dark-red $\rho$-parameter band starts to emerge,
but it is generally quite elusive.

We do not claim that variation of $\rho$, together with the renormalization
and factorization scales, gives a realistic estimate of theoretical
uncertainties in regions where large Sudakov logarithms occur. We
content ourselves to say that $\rho$ is an unphysical technical parameter
introduced in our procedure, with systematics associated to it. We
believe our variation of $\rho$, as described above, is a conservative
estimate of these systematics, and we find them to be very much negligible.

Finally, statistical uncertainties are shown as vertical lines, however,
for the most part these are negligible to the point of being invisible.

\subsubsection*{Inclusive quantities}

\begin{figure}[htbp]
\centering{}\includegraphics[scale=0.9]{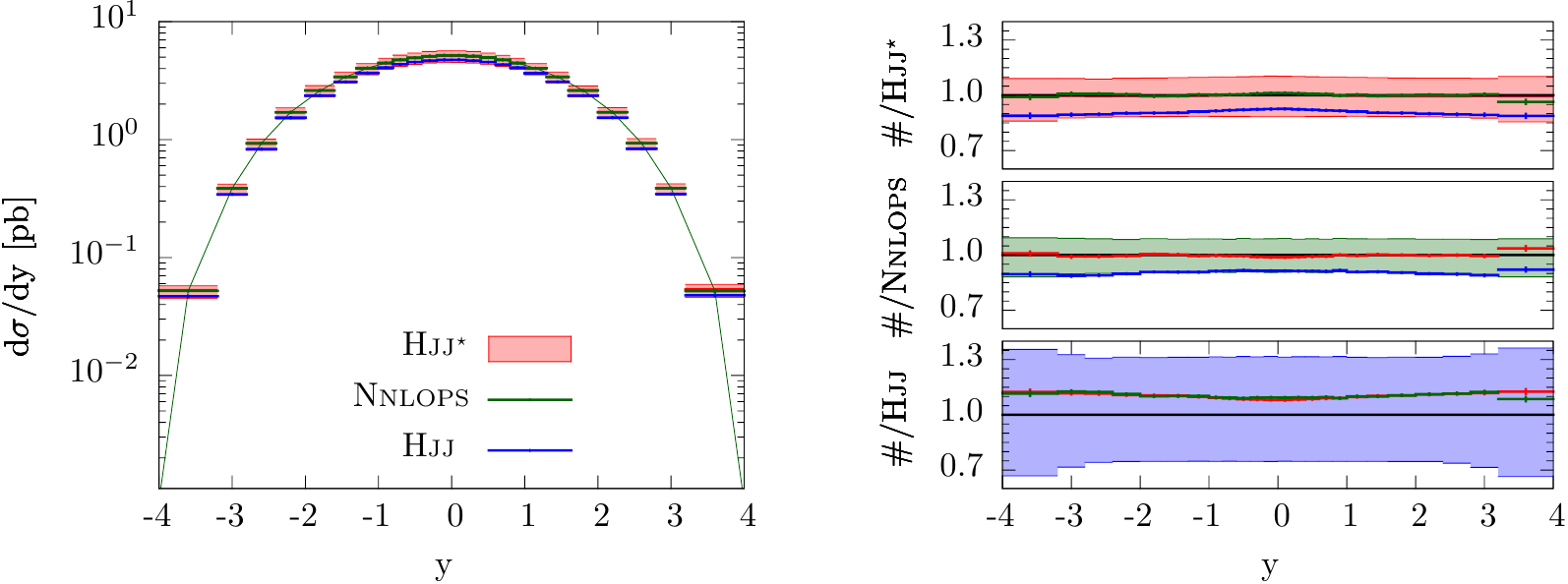}\protect\caption{\label{fig:yH-fully-incl}Rapidity of the Higgs boson as predicted
by the \noun{Hjj-Minlo} (\noun{$\noun{Hjj}$}, blue)\noun{,} \noun{$\noun{Nnlops}$
}(dark green) and improved \noun{Hjj-Minlo} ($\noun{Hjj}^{\star}$, red) generators. }
\end{figure}

In figure \ref{fig:yH-fully-incl} we plot the rapidity of the Higgs
boson; no cuts have been applied to the final state. The\noun{ $\noun{Hjj}{}^{\star}$
}and $\noun{Nnlops}$ central predictions agree with one another to
within 2\%, with their uncertainty bands exhibiting a similar level
of agreement. This indicates that the method and its implementation
are performing as expected (eqs.~\ref{eq:sect26-dSigmaMcorr-eq-dSigmaNLO-i}-\ref{eq:sect3-formula-from-intro-blurb}).
The uncorrected \noun{Hjj-Minlo }prediction in blue is 10\% away from
the central \noun{Nnlops }results, but this is fortuitous given that
the scale uncertainty on the former is $\sim30\%$. Moreover, given
our theoretical analysis in the preceding sections of this paper,
neglecting the sub-leading $\mathrm{NLL}_{\sigma}$ $\Delta S_{1}$
terms, we expect the \noun{Hjj-Minlo} prediction here is only LO accurate,
so the $\sim30\%$ uncertainty assigned to it is arguably too small.
The uncertainty band associated to varying the $\rho$ parameter as
described at the beginning of this subsection \ref{sub:Results-and-testing}
is so small that it is concealed within thickness of the black reference
line in the upper right plot; indeed since this quantity is fully
inclusive in $L_{12}$, by construction of the procedure (sect.~\ref{sub:Variation-on-MinloPrime}),
the only way any such uncertainty could manifest here is as a result
of technical problems and/or some statistical issues.

\begin{figure}[htbp]
\centering{}\includegraphics[scale=0.9]{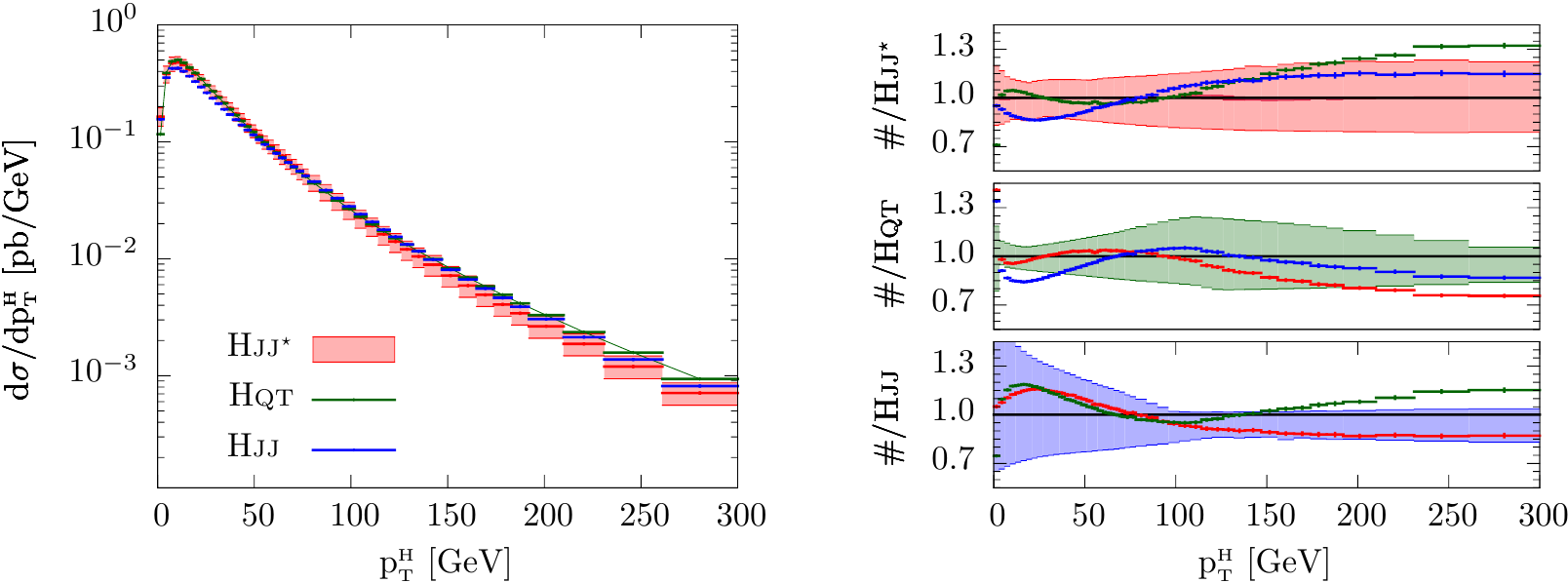}\protect\caption{\label{fig:pTH-HQT-vs-HJJ-vs-HJJstar}Transverse momentum of the Higgs
boson as obtained from the \noun{Hjj-Minlo} (\noun{$\noun{Hjj}$}, blue) and improved \noun{Hjj-Minlo} ($\noun{Hjj}^{\star}$, red) generators, together with the associated NNLL+NNLO computation
from the \noun{$\noun{Hqt}$ }program (dark green) \cite{Bozzi:2003jy,Bozzi:2005wk,deFlorian:2011xf}.}
\end{figure}

In figure \ref{fig:pTH-HQT-vs-HJJ-vs-HJJstar} we plot the Higgs boson
transverse momentum spectrum. As with the Higgs boson rapidity distribution
no cuts have been applied to the final state. Exceptionally, in this
figure we compare \noun{$\noun{Hjj}{}^{\star}$ }and \noun{$\noun{Hjj}$
}to the \noun{NNLL+NNLO }predictions of the $\noun{Hqt}$ program
\cite{Bozzi:2003jy,Bozzi:2005wk,deFlorian:2011xf,deFlorian:2012mx,Grazzini:2013mca},
instead of $\noun{Nnlops}$. Comparing $\noun{Nnlops }$ (not shown)
and $\noun{Hjj}^{\star}$ we find the two generators agree with one
another to within $3\%$ throughout the spectrum, except for the region
$p_{{\scriptscriptstyle \mathrm{T}}}\lesssim5\,\mathrm{GeV}$, where
the difference rises up to $15\%$ in the $p_{{\scriptscriptstyle \mathrm{T}}}<2\,\mathrm{GeV}$
region. The latter differences owe to the finite size of the bins
in our interpolation grids, coupled with the fact that the distribution
is changing very rapidly for $p_{{\scriptscriptstyle \mathrm{T}}}\lesssim5\,\mathrm{GeV}$.
Given this technicality, and the fact that this region is under poor
theoretical control anyway, the conclusion, again, is that the method
and its implementation work well. Turning then to the comparison with
$\noun{Hqt}$ in figure \ref{fig:pTH-HQT-vs-HJJ-vs-HJJstar}, we see,
pleasingly, that the method substantially corrects the shape of the
pre-existing \noun{Hjj-Minlo} simulation, with the resulting \noun{$\noun{Hjj}{}^{\star}$}
prediction agreeing very well with $\noun{Hqt}$ in the region where
the latter is undeniably the superior calculation ($p_{{\scriptscriptstyle \mathrm{T}}}\lesssim100\,\mathrm{GeV}$).%
\footnote{In $\noun{Hqt}$ we have used the `switched' mode and taken the central
renormalization, factorization and resummation scales to be $\frac{1}{2}m_{{\scriptscriptstyle \mathrm{H}}}$.
The uncertainty band comprises the envelope of a 7-point variation
of the first two scales: $\mu_{{\scriptscriptstyle R}}\rightarrow K_{{\scriptscriptstyle R}}\mu_{{\scriptscriptstyle R}}$,
$\mu_{{\scriptscriptstyle F}}\rightarrow K_{{\scriptscriptstyle F}}\mu_{{\scriptscriptstyle F}}$,
with $K_{{\scriptscriptstyle R/F}}=\frac{1}{2}\,,\,1,\,2,$ omitting
the two combinations for which $K_{{\scriptscriptstyle R}}$ and $K_{{\scriptscriptstyle F}}$
differ by more than a factor of two.%
} In the high transverse momentum tail both\noun{ $\noun{Hjj}{}^{\star}$}
and $\noun{Hqt}$ computations have the same NLO accuracy for this
distribution. Differences between \noun{$\noun{Hjj}{}^{\star}$} and
$\noun{Hqt}$ occur there due to the different choice of scales in
each code, roughly, $p_{{\scriptscriptstyle \mathrm{T}}}^{{\scriptscriptstyle \mathrm{H}}}$
in the case of \noun{$\noun{Hjj}{}^{\star}$}, compared to $\frac{1}{2}m_{{\scriptscriptstyle \mathrm{H}}}$
in $\noun{Hqt}$.\textcolor{magenta}{{} }The same comments made above
for the Higgs boson rapidity distribution in regards to the uncertainty
associated with the $\rho$ parameter apply equally well again here.

\subsubsection*{Jet cross sections}

\begin{figure}[htbp]
\begin{centering}
\includegraphics[scale=0.9]{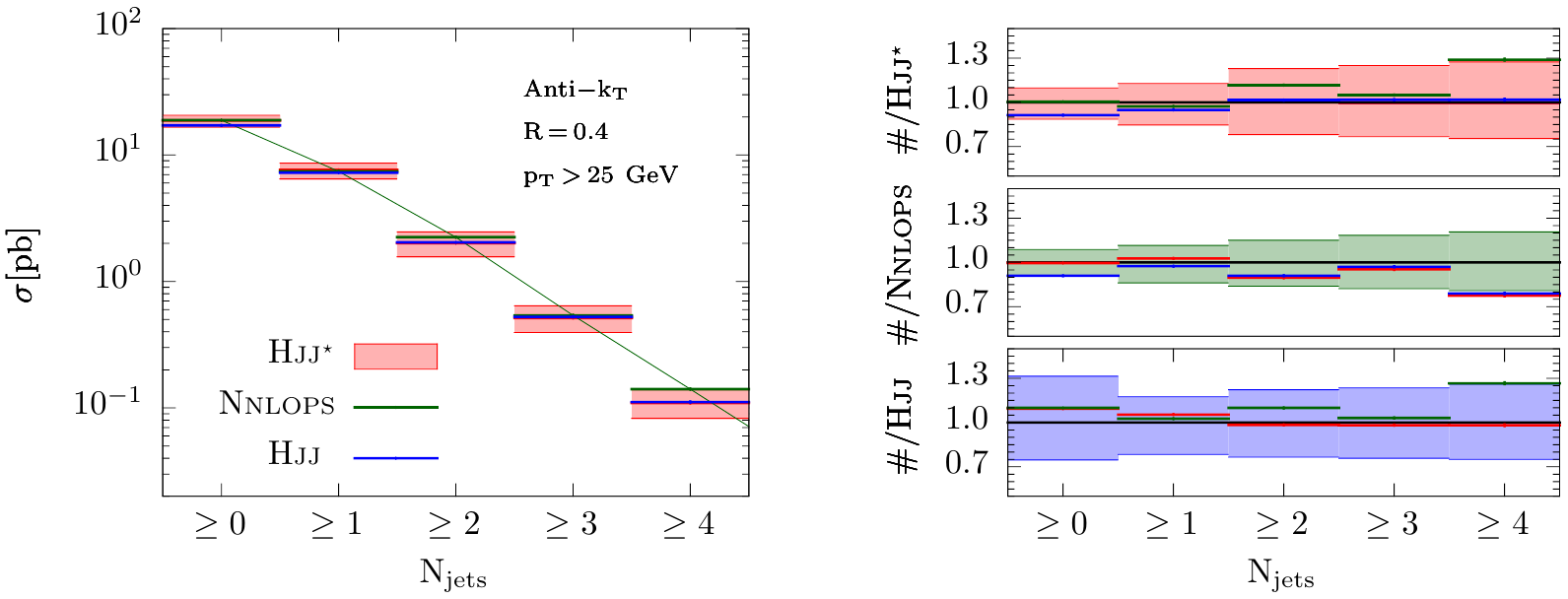} 
\par\end{centering}

\begin{centering}
~ 
\par\end{centering}

\begin{centering}
~ 
\par\end{centering}

\begin{centering}
\includegraphics[scale=0.9]{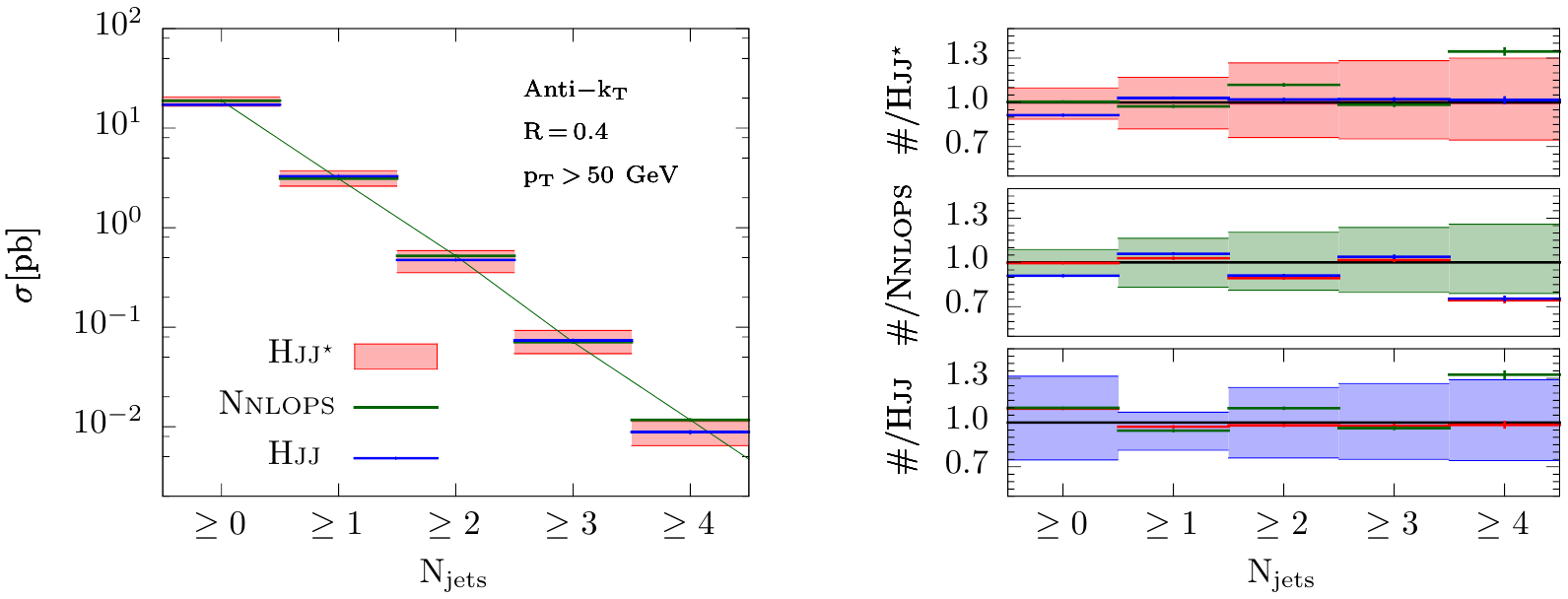} 
\par\end{centering}

\begin{centering}
~ 
\par\end{centering}

\begin{centering}
~ 
\par\end{centering}

\centering{}\includegraphics[scale=0.9]{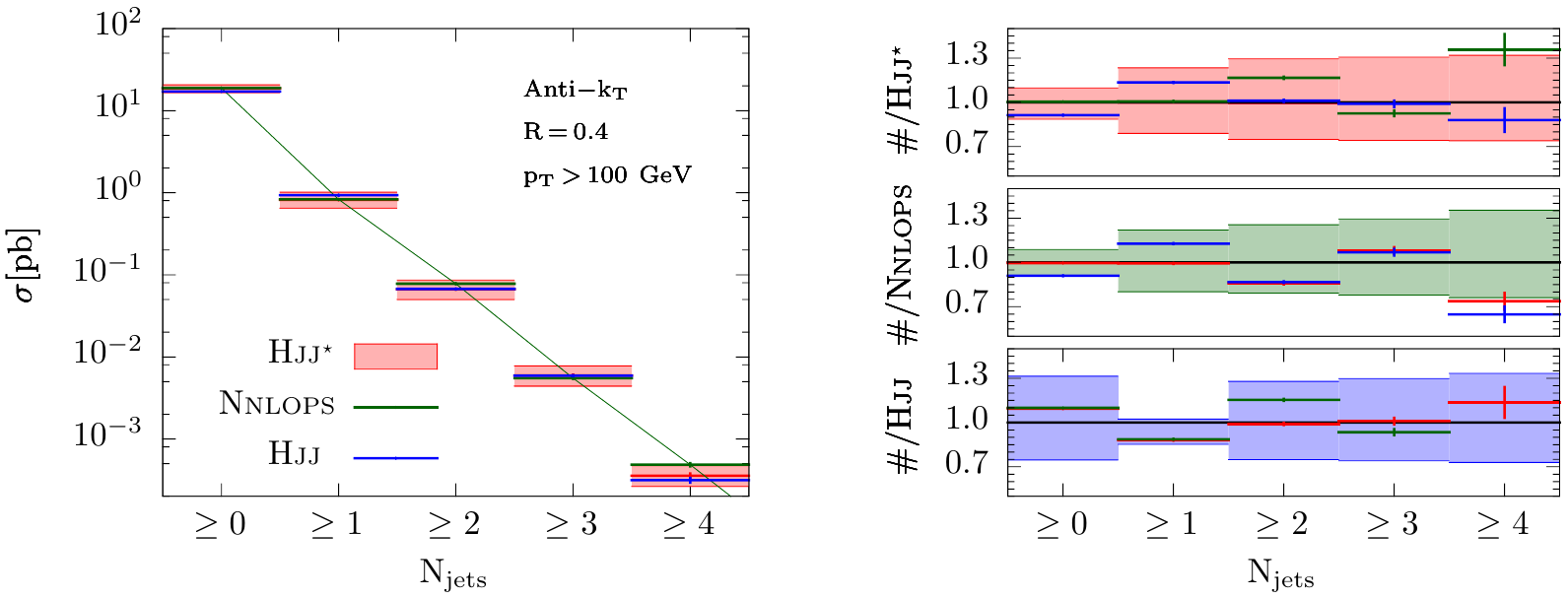}\protect\caption{\label{fig:Inclusive-jet-cross-section-akT-R-eq-0.4}Inclusive jet
cross sections, for jets defined according to the anti-$k_{t}$-jet
algorithm \cite{Cacciari:2008gp} with jet radius $R=0.4$. In the
upper, middle and lower plots jets are defined for transverse momentum
thresholds of $25$, $50$ and $100$ GeV, respectively. In each case
we compare output from the \noun{Hjj-Minlo} (\noun{$\noun{Hjj}$}, blue)\noun{,} \noun{$\noun{Nnlops}$
}(dark green) and improved \noun{Hjj-Minlo} ($\noun{Hjj}^{\star}$, red) generators. }
\end{figure}

\begin{figure}[htbp]
\begin{centering}
\includegraphics[scale=0.9]{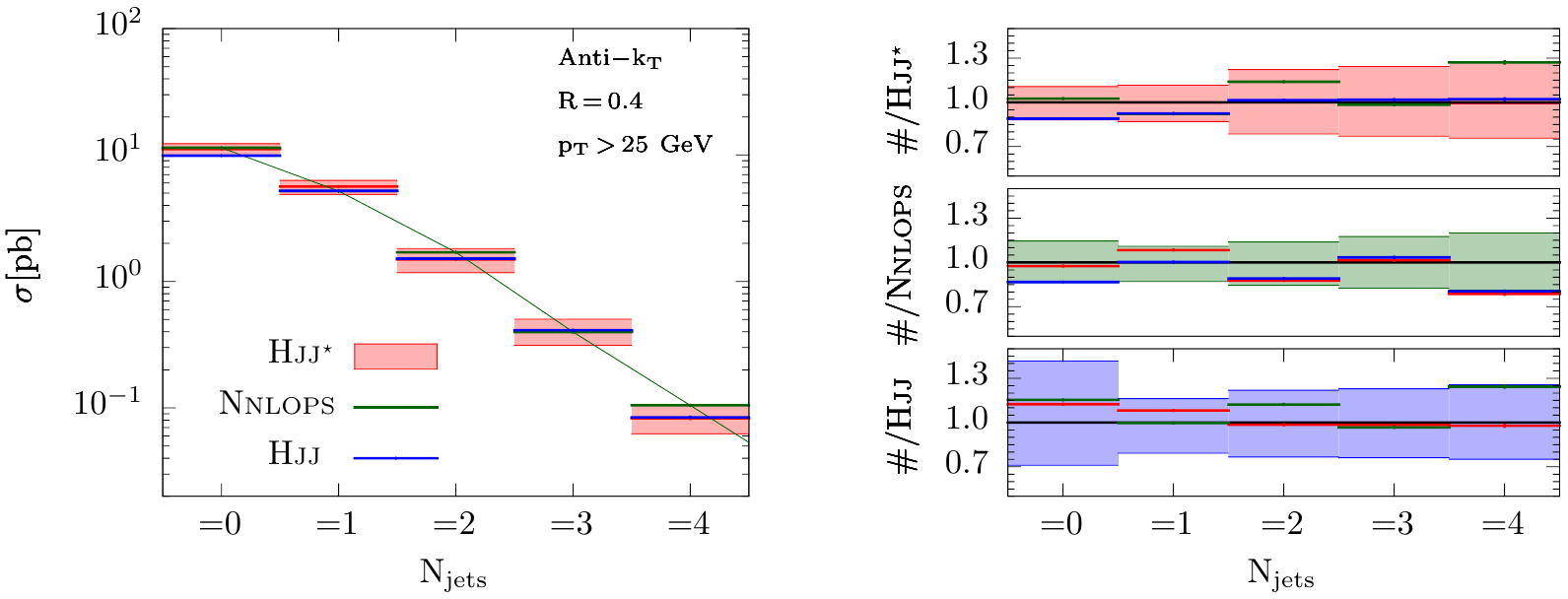} 
\par\end{centering}

\begin{centering}
~ 
\par\end{centering}

\begin{centering}
~ 
\par\end{centering}

\begin{centering}
\includegraphics[scale=0.9]{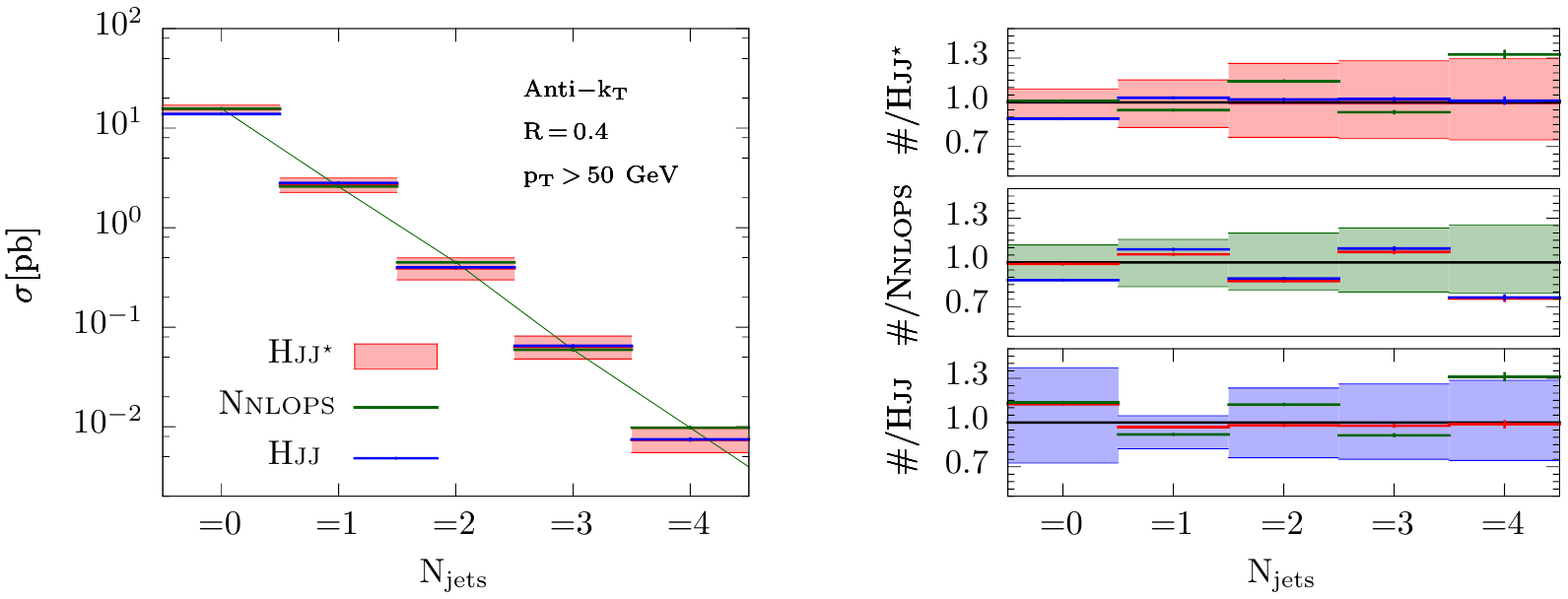} 
\par\end{centering}

\begin{centering}
~ 
\par\end{centering}

\begin{centering}
~ 
\par\end{centering}

\centering{}\includegraphics[scale=0.9]{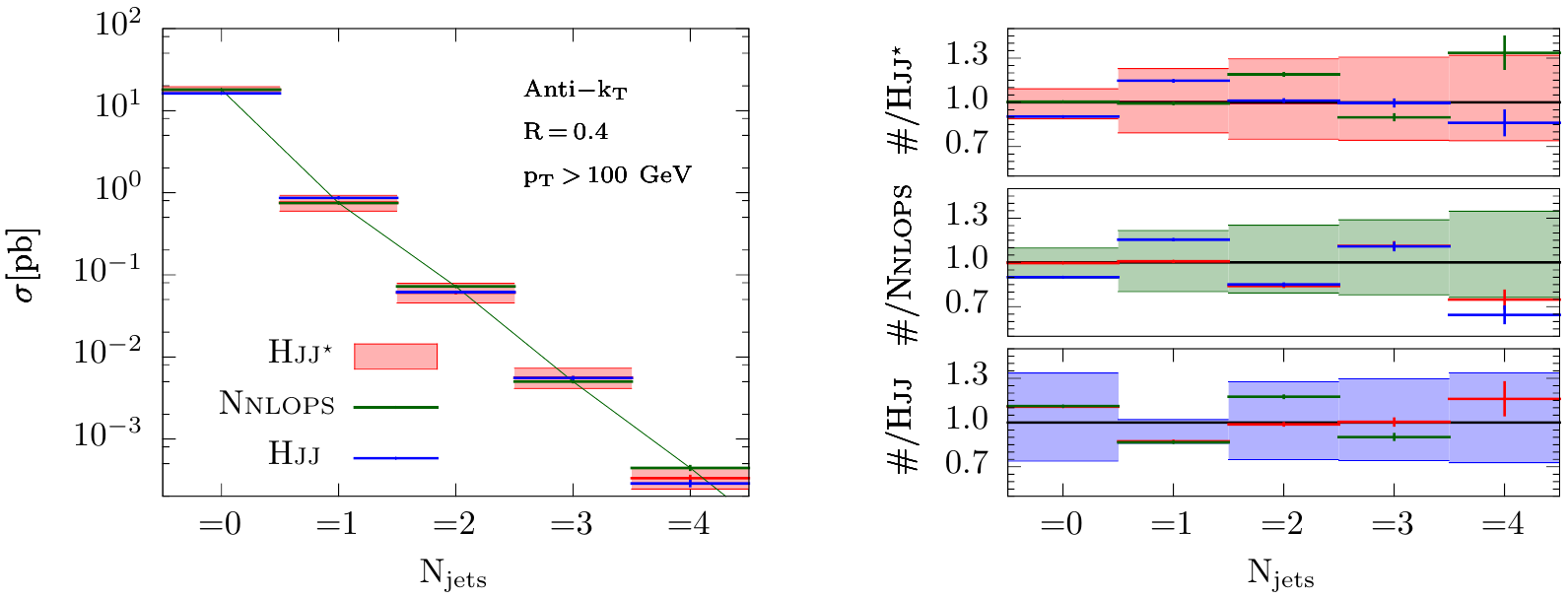}\protect\caption{\label{fig:Exclusive-jet-cross-section-akT-R-eq-0.4}Exclusive\emph{
}jet cross sections, for jets defined according to the anti-$k_{t}$-jet
algorithm \cite{Cacciari:2008gp} with jet radius $R=0.4$. In the
upper, middle and lower plots jets are defined for transverse momentum
thresholds of $25$, $50$ and $100$ GeV, respectively. In each case
we compare output from the \noun{Hjj-Minlo} (\noun{$\noun{Hjj}$}, blue)\noun{,} \noun{$\noun{Nnlops}$
}(dark green) and improved \noun{Hjj-Minlo} ($\noun{Hjj}^{\star}$, red) generators. }
\end{figure}

In figure \ref{fig:Inclusive-jet-cross-section-akT-R-eq-0.4} we compare
predictions for inclusive jet cross sections, between the \noun{$\noun{Hjj}$
}(blue)\noun{,} \noun{$\noun{Nnlops}$ }(dark green) and $\noun{Hjj}^{\star}$
(red) generators, defined according to the anti-$k_{t}$-jet algorithm
\cite{Cacciari:2008gp} with radius parameter $R=0.4$, for jet transverse
momentum thresholds of $25$, $50$ and $100$ GeV. In figure \ref{fig:Exclusive-jet-cross-section-akT-R-eq-0.4}
we show the analogous set of plots for the corresponding exclusive
jet cross sections. No rapidity cuts have been applied to the jets
in making these plots.

First we discuss the inclusive jet cross sections in fig.~\ref{fig:Inclusive-jet-cross-section-akT-R-eq-0.4}.
For the 0-jet inclusive cross sections, the improved $\noun{Hjj}^{\star}$
results are indistinguishable from the \noun{$\noun{Nnlops}$ }ones,
shifted upwards by 10\% with respect to the original \noun{Hjj-Minlo
}predictions (\noun{Hjj). }The 1-jet inclusive predictions show the
$\noun{Hjj}^{\star}$ results agreeing with the \noun{$\noun{Nnlops}$
}ones to within 2\%. Unlike the case of the 0-jet bin, in the 1-jet
bin, for 25 and 50 GeV jet $p_{{\scriptscriptstyle \mathrm{T}}}$
thresholds, the unimproved \noun{Hjj-Minlo} result was already in
agreement with the \noun{Nnlops }at the level of 5\% or better. So,
for the 1-jet inclusive cross sections the room for improvement is
very much smaller, with only a small amount visible in the case of
the 50 GeV $p_{{\scriptscriptstyle \mathrm{T}}}$ cut. For the case
of the 100 GeV jet $p_{{\scriptscriptstyle \mathrm{T}}}$ threshold
the unimproved \noun{Hjj-Minlo} prediction is 10-15\% away from the
\noun{Nnlops }one, whereas the improved $\noun{Hjj}^{\star}$ result
sits on top of it. Looking to the higher multiplicity bins, involving
at least two jets, we see, as desired, the $\noun{Hjj}^{\star}$ predictions
and those of the parent, unimproved, \noun{Hjj-Minlo} simulation are
in perfect agreement, but for a statistical fluctuation in the 4-jet
inclusive cross section with a 100 GeV jet $p_{{\scriptscriptstyle \mathrm{T}}}$
cut. We remind that the vertical error bars indicate statistical errors,
which are rarely visible, whereas the shaded bands indicate theoretical
uncertainties.

The behaviour seen in all of the inclusive jet cross sections of fig.~\ref{fig:Inclusive-jet-cross-section-akT-R-eq-0.4},
is as we would naively expect it to be. By construction, our improved
\noun{Minlo }method should reproduce \noun{Nnlops }results essentially
identically for 0- and 1-jet inclusive quantities (eqs.~\ref{eq:sect26-dSigmaMcorr-eq-dSigmaNLO-i}-\ref{eq:sect3-formula-from-intro-blurb}),
while observables that receive their leading contributions from higher
jet multiplicities are to be described as in the original \noun{Hjj-Minlo
}generator, which yields the more accurate predictions for those observables.

Given that our \noun{Minlo} improvement method is intended to return
the 0-jet and 1-jet inclusive results of its `target' \noun{Nnlops
}simulation, essentially without ambiguities, one might be tempted
to ask why we can see even 2\% differences between the \noun{Nnlops}
and $\noun{Hjj}^{\star}$ predictions for the 1-jet inclusive cross
sections. What the improvement procedure precisely does, without ambiguities,
assuming a perfect implementation, is to have the improved $\noun{Hjj}^{\star}$
result reproduce the \noun{Nnlops} underlying Born kinematics $\Phi_{{\scriptscriptstyle \mathcal{B}\mathrm{J}}}$
(eqs.~\ref{eq:sect26-dSigmaMcorr-eq-dSigmaNLO-i}-\ref{eq:sect3-formula-from-intro-blurb})
which are defined by clustering events with the exclusive $k_{t}$-jet
algorithm, with $R=1.0$. What is plotted in the 1-jet bins of fig.~\ref{fig:Inclusive-jet-cross-section-akT-R-eq-0.4}
is therefore not in one-to-one correspondence with the kinematics
$\Phi_{{\scriptscriptstyle \mathcal{B}\mathrm{J}}}$ (consisting of
a Higgs boson and a single pseudoparton in the final-state) but rather
it is something which is also sensitive to additional radiation. The
$\noun{Hjj}^{\star}$ and \noun{Nnlops }generators are further in
agreement as to the relative distribution of this additional radiation
at the level of $\bar{\alpha}_{{\scriptscriptstyle \mathrm{S}}}^{4}$
terms, i.e.~at the level of NLO corrections to \noun{Hj}, however,
at $\mathcal{O}(\bar{\alpha}_{{\scriptscriptstyle \mathrm{S}}}^{5})$
differences do enter. Hence, even if the implementation were a perfect
representation of our method, with infinite resolution in the $\Phi_{{\scriptscriptstyle \mathcal{B}\mathrm{J}}}$
grids, we can still expect to see differences between the \noun{Nnlops
}results and $\noun{Hjj}^{\star}$, for the 1-jet inclusive cross
section, which are formally NNLO-sized in the context of the inclusive
1-jet calculation. This being the case, one can be quite satisfied
with only 2\% differences between the \noun{Nnlops} and $\noun{Hjj}^{\star}$
predictions for the 1-jet inclusive cross sections. In fact, we examined
the 0- and 1-jet inclusive cross sections, with a 25 GeV jet $p_{{\scriptscriptstyle \mathrm{T}}}$
threshold, prior to interfacing with the parton shower, whereupon
we found the 0-jet and 1-jet $\noun{Hjj}^{\star}$ cross sections
to be indistinguishable from their \noun{Nnlops }counterparts, while
the 2- and 3-jet bins remained identical to those of \noun{Hjj-Minlo.}

Let us now turn our attention to the exclusive jet cross sections
of fig.~\ref{fig:Exclusive-jet-cross-section-akT-R-eq-0.4}. First,
for the high multiplicity bins, involving two or more jets, the $\noun{Hjj}^{\star}$
results are in complete agreement with those of its parent unimproved
\noun{Hjj-Minlo }generator (up to a single statistical fluctuation).
The \noun{$\noun{Nnlops}$ }predictions are nominally only LO accurate
for the $2$-jet bins, whereas for higher jet multiplicity bins the
simulation relies entirely on the parton shower approximation. The
\noun{$\noun{Hjj}$ }and $\noun{Hjj}^{\star}$ predictions, on the
other hand, are expected to be NLO accurate for the 2-jet bins, LO
for the 3-jet bins, only resorting to the parton shower approximation
in the 4-jet bins. $\noun{Hjj}^{\star}$ being in perfect agreement
with \noun{$\noun{Hjj}$-Minlo }for the latter cross sections is,
of course, the desired behaviour from our improved \noun{Minlo }prediction.

For the 0-jet exclusive cross sections in fig.~\ref{fig:Exclusive-jet-cross-section-akT-R-eq-0.4}
we see nice agreement between the $\noun{Hjj}^{\star}$ and \noun{Nnlops
}predictions at the 1-2\% level or better, as is to be expected by
construction of our method. To explain the 1-2\% differences that
can be seen we tender again the same theoretical explanation as above
(the \noun{$\noun{Nnlops}$} and $\noun{Hjj}^{\star}$ results differ,
by construction, at the level of $\mathcal{O}(\bar{\alpha}_{{\scriptscriptstyle \mathrm{S}}}^{5})$
terms), however, with such small differences we also cannot rule out
imperfections in the implementation, e.g.~artefacts due to the finite
granularity of the grids and grid interpolation. We suffice to say
that the differences between the $\noun{Hjj}^{\star}$ and \noun{Nnlops}
computations of the 0-jet exclusive cross sections are negligibly
small, while the unimproved \noun{Hjj-Minlo }result sits 10-15\% below
them.

Lastly, we look to the the 1-jet exclusive cross sections. The plots
in this case read that the $\noun{Hjj}^{\star}$ prediction is different
from the \noun{Nnlops }one by 7\% for the 25 GeV jet $p_{{\scriptscriptstyle \mathrm{T}}}$
threshold, 5\% for the 50 GeV threshold, and $\sim0\%$ for the 100
GeV threshold. Meanwhile, the unimproved \noun{Hjj-Minlo} prediction
is in agreement with the \noun{Nnlops} prediction at the level of
$\sim0$\%, 10\%, and 15\%, for the same $p_{{\scriptscriptstyle \mathrm{T}}}$
thresholds, respectively.

Since the \noun{Minlo} improvement method we propose works to correct
the inclusive 0- and 1-jet \noun{Hjj-Minlo }cross sections to be equal
to those of the target \noun{Nnlops }generator, while leaving inclusive
2-jet observables basically untouched, we consider it can be useful
to think of the exclusive 1-jet cross section as the difference of
the inclusive 1- and 2-jet cross sections: $\sigma\left(=1-\mathrm{jet}\right)=\sigma\left(\ge1-\mathrm{jet}\right)-\sigma\left(\ge2-\mathrm{jets}\right)$.
Clearly if $\sigma\left(\ge1-\mathrm{jet}\right)\gg\sigma\left(\ge2-\mathrm{jets}\right)$
differences in the latter will have limited impact on the exclusive
1-jet cross section. The latter scenario is enhanced by increasing
the jet $p_{{\scriptscriptstyle \mathrm{T}}}$ threshold and, sure
enough, the pattern of the exclusive jet-cross sections seen in the
case of the 100 GeV $p_{{\scriptscriptstyle \mathrm{T}}}$ threshold,
mirrors well what we see in the analogous inclusive jet cross section
case, discussed overhead. To explain then the differences seen between
the \noun{Nnlops }and\noun{ }$\noun{Hjj}^{\star}$ generators\noun{
}at the 25 GeV and 50 GeV jet $p_{{\scriptscriptstyle \mathrm{T}}}$
thresholds, we note that the \noun{Nnlops }exclusive 1-jet cross section
is given by its inclusive 1-jet cross section minus its inclusive
2-jet cross section, on the other hand, by design, as can be verified
in fig.~\ref{fig:Inclusive-jet-cross-section-akT-R-eq-0.4}, the
$\noun{Hjj}^{\star}$ exclusive 1-jet cross section is basically given
by the \noun{Nnlops }inclusive 1-jet cross section minus the \noun{Hjj-Minlo
}inclusive 2-jet cross section. Since the \noun{Hjj-Minlo} inclusive
2-jet cross section at the 25 GeV jet $p_{{\scriptscriptstyle \mathrm{T}}}$
thresholds is 10\% lower than the \noun{Nnlops }one, while the ratio
of the inclusive 1-jet to 2-jet cross sections is roughly two, it
follows that one can expect the $\noun{Hjj}^{\star}$ 1-jet exclusive
cross section to be 5\% higher than the corresponding \noun{Nnlops
}one. Adding in the fact that the $\noun{Hjj}^{\star}$ 1-jet inclusive
cross section was already 1-2\% above the corresponding \noun{Nnlops
}one, the 7\% excess is actually very much in line with expectations
based on how the method is intended to work, in particular, its preserving
of the inclusive cross sections. A similar explanation holds for the
50 GeV jet $p_{{\scriptscriptstyle \mathrm{T}}}$ threshold result,
however, there the fact that the 2-jet inclusive cross section of
\noun{Hjj-Minlo }is low does not imply as big an increase is needed
in the 1-jet exclusive bin to recover the 1-jet inclusive \noun{Nnlops
}result, since for that higher $p_{{\scriptscriptstyle \mathrm{T}}}$
threshold $\sigma\left(\ge1-\mathrm{jet}\right)\gg\sigma\left(\ge2-\mathrm{jets}\right)$.
We also remark that the $\noun{Hjj}^{\star}$ exclusive 1-jet cross
section results all agree with those of the \noun{Nnlops }generator
to within the thickness of the scale uncertainty bands.

We finally note that the dark-red scale uncertainty band associated
to variation of the $\rho$ parameter, as described in sects.~\ref{sub:Implementation}-\ref{sub:Results-and-testing},
is invisible here: the effect of varying this parameter on these distributions
is totally negligible.

To conclude the discussion on jet cross sections, we can say that
all of the results we find are very much in line with expectations
regarding how our method should function, all vindicating the method
and its implementation.

\subsubsection*{Leading jet}

\begin{figure}[htbp]
\centering{}\includegraphics[scale=0.9]{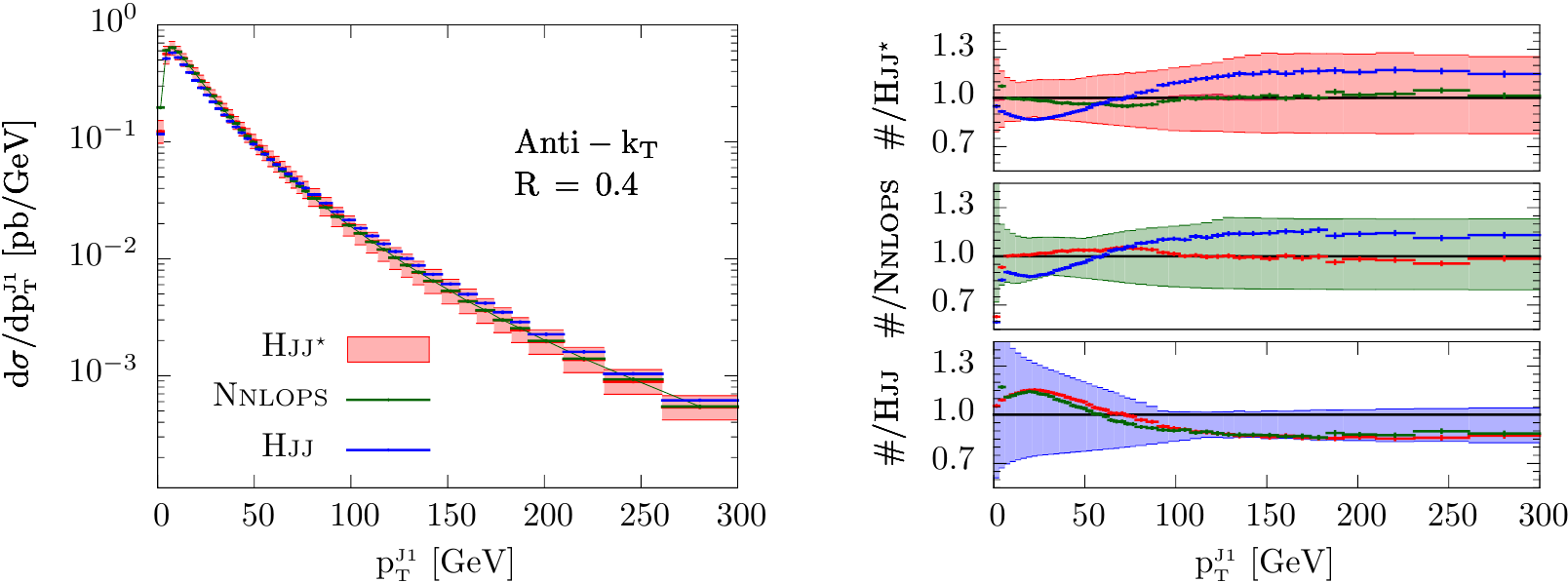}\protect\caption{\label{fig:Leading-jet-transverse-mom}Leading jet transverse momentum
spectrum, for anti-$k_{t}$-jets with radius parameter $R=0.4$.}
\end{figure}

\begin{figure}[htbp]
\centering{}~\includegraphics[width=0.4225\textwidth,height=0.22\textheight]{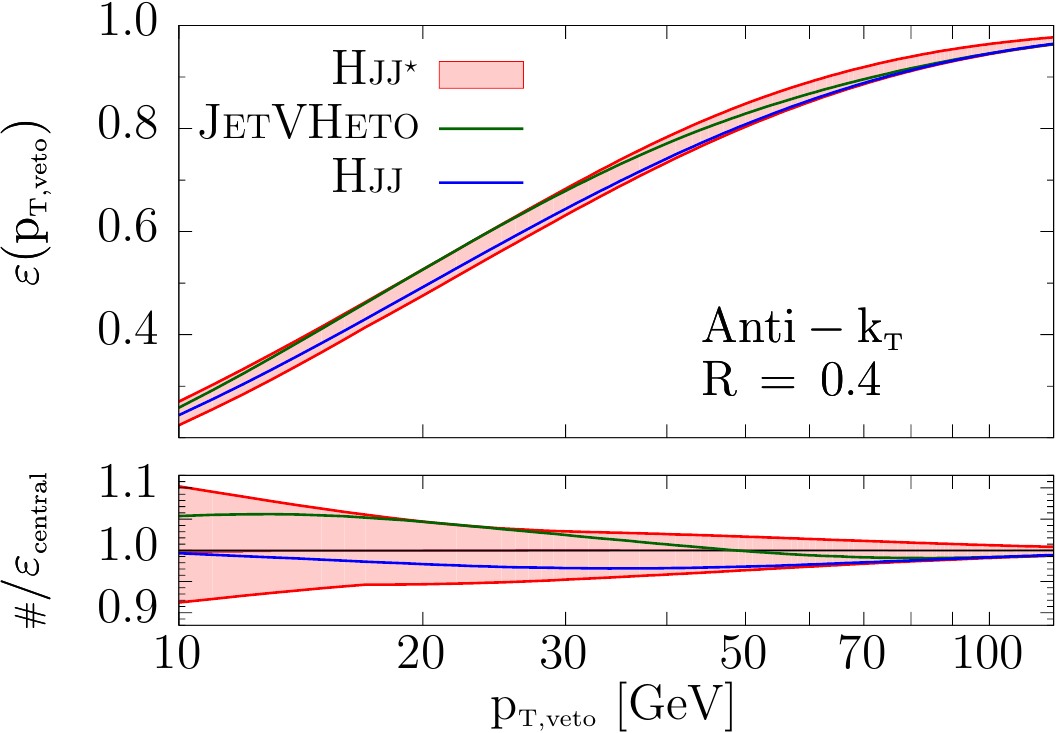}\hspace{10 mm}\includegraphics[width=0.4\textwidth,height=0.22\textheight]{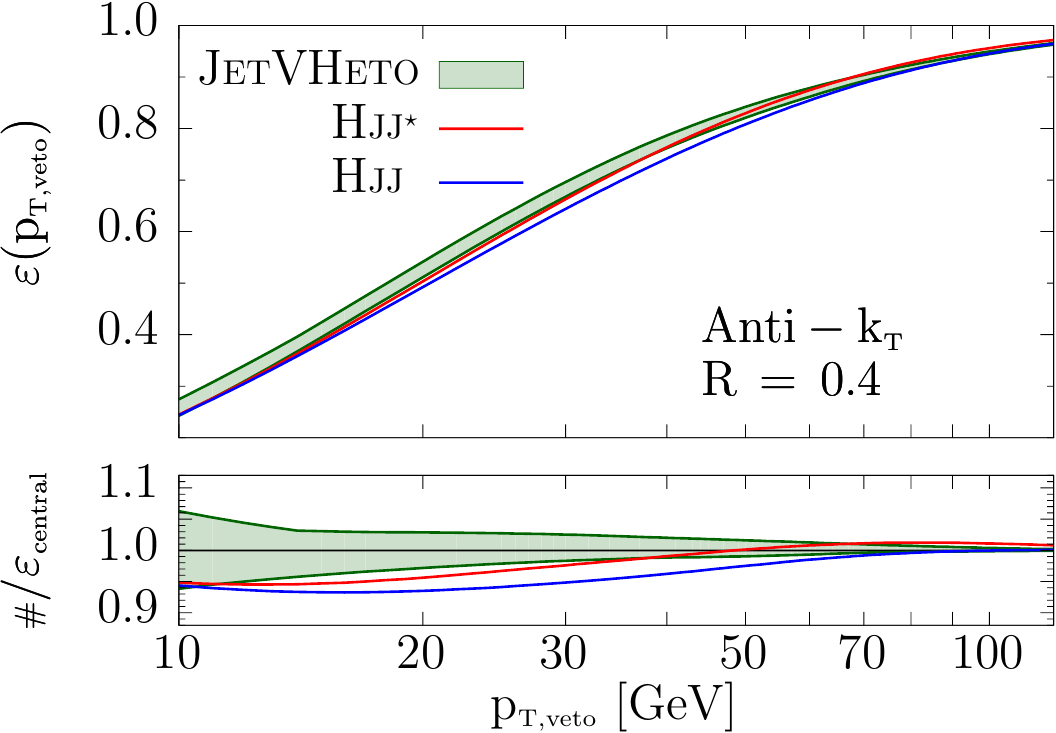}\protect\caption{\label{fig:jet-veto-effy}The jet veto efficiency, $\varepsilon(p_{{\scriptscriptstyle \mathrm{T,veto}}})$,
is defined as the cross section for Higgs boson production events
containing no jets with transverse momentum greater than $p_{{\scriptscriptstyle \mathrm{T,veto}}}$,
divided by the respective total inclusive cross section. Jets are
defined here according to the anti-$k_{t}$-jet algorithm with $R=0.4$.
On the left, in the red shaded area, one can see the scale uncertainty
band obtained from the improved \noun{Hjj-Minlo} ($\noun{Hjj}^{\star}$) simulation, with the
NNLL+NNLO prediction from the \noun{JetVHeto} program \cite{Banfi:2012yh,Banfi:2012jm}
overlaid in green, and that of the original \noun{Hjj-Minlo} program in
blue. The lower pane displays the same quantities as a ratio with
respect to the central $\noun{Hjj}^{\star}$ prediction. On the right
we display instead the corresponding uncertainty band obtained from
\noun{JetVHeto }(i.e.~renormalization and factorization scale variations
only), with the central value of the \noun{JetVHeto} prediction defining
the reference line in the associated ratio plot.}
\end{figure}

\begin{figure}[htbp]
\centering{}\includegraphics[scale=0.9]{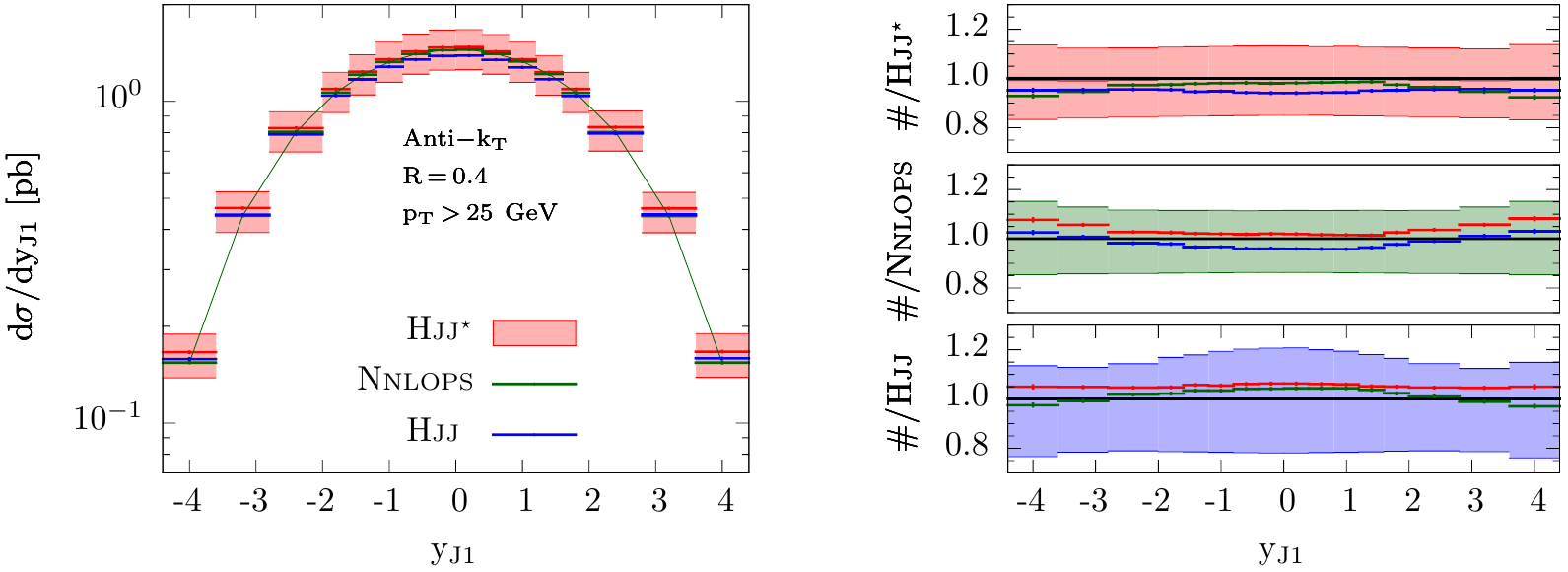}\protect\caption{\label{fig:yJ1-antikT-R-eq-0.4}Leading jet rapidity for anti-$k_{t}$-jets
with radius parameter $R=0.4$ and a transverse momentum threshold
of $25$ GeV.}
\end{figure}

\begin{figure}[htbp]
\centering{}\includegraphics[scale=0.9]{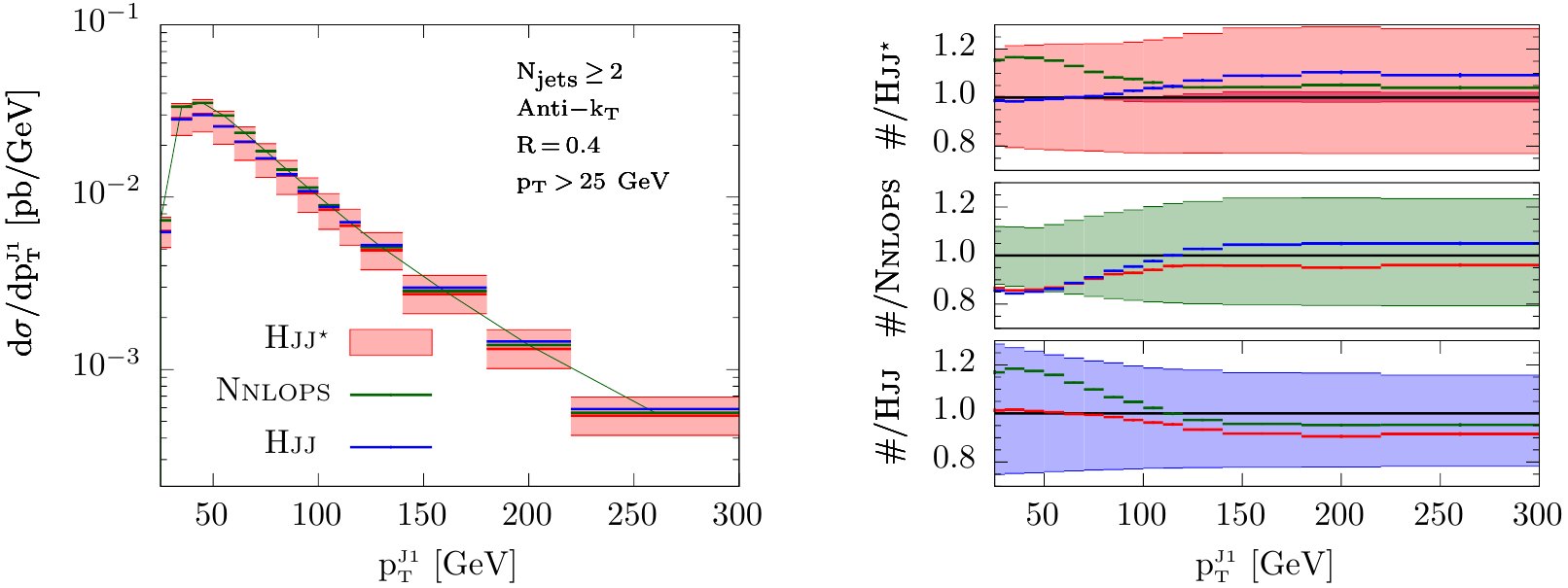}\protect\caption{\label{fig:Leading-jet-pT-Njet-ge-2}Leading jet transverse momentum
in events with two or more jets; jets are here defined according to
the anti-$k_{t}$ algorithm with radius $R=0.4$ and a transverse
momentum threshold of $25$ GeV.}
\end{figure}

\begin{figure}[htbp]
\centering{}\includegraphics[scale=0.9]{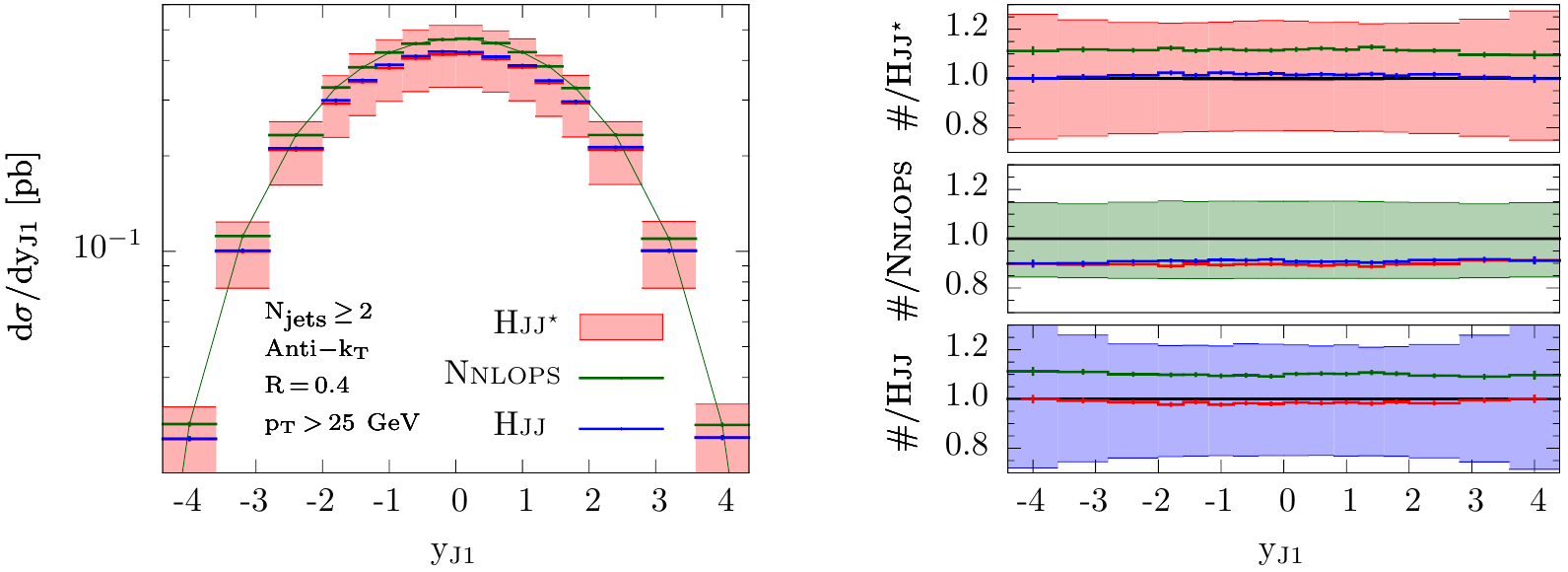}\protect\caption{\label{fig:yJ1-Njets-ge-2}Leading jet rapidity in events with two
or more anti-$k_{t}$, $R=0.4$, $p_{{\scriptscriptstyle \mathrm{T}}}\ge25\,\mathrm{GeV}$
jets.}
\end{figure}

In figures \ref{fig:Leading-jet-transverse-mom}-\ref{fig:yJ1-Njets-ge-2}
we plot various quantities relating to the kinematics of the leading
jet. In all of these figures jets have been defined according
to the anti-$k_{t}$ clustering algorithm, with jet radius $R=0.4$;
no rapidity cuts have been applied to the jets. For figs.~\ref{fig:yJ1-antikT-R-eq-0.4},
\ref{fig:Leading-jet-pT-Njet-ge-2}, \ref{fig:yJ1-Njets-ge-2} jets
were further defined as having a transverse momentum of at least 25
GeV.

The results for the leading jet transverse momentum spectrum in fig.~\ref{fig:Leading-jet-transverse-mom}
read similarly to those reported for the Higgs boson transverse momentum
spectrum (fig.~\ref{fig:pTH-HQT-vs-HJJ-vs-HJJstar}). The $\noun{Nnlops }$
and $\noun{Hjj}^{\star}$ predictions agree very well throughout the
spectrum, with the procedure correcting well for substantial ($\pm15\%$)
shape differences between the unimproved \noun{Hjj-Minlo }result and
the more accurate \noun{Nnlops }prediction. Regarding differences
between the $\noun{Nnlops }$ and $\noun{Hjj}^{\star}$ results in
the $p_{{\scriptscriptstyle \mathrm{T}}}\lesssim5\,\mathrm{GeV}$
region, the explanation here is the same as for the case of the Higgs
boson $p_{{\scriptscriptstyle \mathrm{T}}}$ spectrum, namely, that
the granularity in our discretized implementation of the $\Phi_{{\scriptscriptstyle \mathcal{B}\mathrm{J}}}$
phase space is not sufficiently fine to cope with the rapidly changing
distribution for $p_{{\scriptscriptstyle \mathrm{T}}}\lesssim5\,\mathrm{GeV}$.
We reiterate that this region is under limited theoretical control
anyway. Indeed, rather than seek improved agreement of $\noun{Nnlops }$
and $\noun{Hjj}^{\star}$ in the latter murky region, we might prefer
to lessen the 3-5\% deviation in the neighbourhood $60\le p_{{\scriptscriptstyle \mathrm{T}}}\le80$
GeV. This region, where the \noun{Hjj-Minlo} and \noun{Nnlops} lines
intersect, appears to be where the $p_{{\scriptscriptstyle \mathrm{T}}}$
derivative of the difference between the two predictions is changing
most rapidly, i.e.~the numerator of $\delta\left(\Phi_{{\scriptscriptstyle \mathcal{B}\mathrm{J}}}\right)$
in eq.~\ref{eq:sect26-first-delta-defn}/\ref{eq:sect33-delta-Phi-BJ}.
It should therefore be possible to improve agreement between the $\noun{Nnlops }$
and $\noun{Hjj}^{\star}$ results in this region by, for example,
making use of (irregular) optimized grids and interpolation methods
which can work on them. Overall, notwithstanding our unsophisticated
implementation, agreement between the $\noun{Nnlops }$ and $\noun{Hjj}^{\star}$
predictions is very satisfactory, providing significant improvement
across the whole $p_{{\scriptscriptstyle \mathrm{T}}}$ spectrum relative
to the original \noun{Hjj-Minlo }generator.

In fig.~\ref{fig:jet-veto-effy} we plot $\noun{Hjj}$, $\noun{Hjj}^{\star}$
and NNLL+NNLO \noun{JetVHeto} \cite{Banfi:2012yh,Banfi:2012jm} predictions
for the jet veto efficiency, $\varepsilon(p_{{\scriptscriptstyle \mathrm{T,veto}}})$,
defined as the cross section for Higgs boson production events containing
no jets with transverse momentum greater than $p_{{\scriptscriptstyle \mathrm{T,veto}}}$,
divided by the respective total inclusive cross section. In the left-hand
column, in the red shaded area, we show the scale uncertainty band
predicted by the $\noun{Hjj}^{\star}$ simulation, with the central
NNLL+NNLO resummed prediction of \noun{JetVHeto} superimposed in green
(matching scheme-(a), $\mu_{{\scriptscriptstyle \mathrm{R}}}=\mu_{{\scriptscriptstyle \mathrm{F}}}=\mu_{{\scriptscriptstyle \mathrm{Q}}}=m_{{\scriptscriptstyle \mathrm{H}}}$,
$\mu_{{\scriptscriptstyle \mathrm{Q}}}$ being the resummation scale).
The lower panel shows the ratio with respect to the $\noun{Hjj}^{\star}$
prediction obtained with its central scale choice. On the right we
have made the same plots as on the left but with the \noun{JetVHeto}
predictions replacing those of the $\noun{Hjj}^{\star}$ and vice-versa.
The uncertainty band in the \noun{JetVHeto} results is the envelope
of a seven point variation of $\mu_{{\scriptscriptstyle \mathrm{R}}}$
and $\mu_{{\scriptscriptstyle \mathrm{F}}}$ by a factor of two. This
is in contrast to the band associated with it in ref.~\cite{Banfi:2012jm},
where additionally resummation scale and matching scheme variations
were included in the envelope. Thus the \noun{JetVHeto} error band
here is considerably smaller than that shown in ref.~\cite{Banfi:2012jm}.
We restricted the \noun{JetVHeto} uncertainty estimate to the same
class of variations so as to have a more like-for-like comparison
to the $\noun{Hjj}^{\star}$ band.

The $\noun{Hjj}^{\star}$ and \noun{JetVHeto} predictions agree within
the $\noun{Hjj}^{\star}$ uncertainties, but not quite to within the
thickness of the restricted \noun{JetVHeto} band, in which case the
central $\noun{Hjj}^{\star}$ prediction is 1-2\% below the lower
edge of the uncertainty band. Nevertheless, considering the \noun{JetVHeto
}calculation has superior accuracy to both\noun{ }the $\noun{Hjj}^{\star}$
and \noun{Nnlops} predictions, through its high accuracy resummation,
the level of agreement we find should be understood as being, again,
quite satisfactory: the $\noun{Hjj}^{\star}$ prediction is always
within 5\% of the \noun{JetVHeto }result, moreover, for the region
$p_{{\scriptscriptstyle \mathrm{T},\mathrm{veto}}}>25$ GeV, it is
within 3\% of the \noun{JetVHeto }prediction. We also observed that
if we compare to the \noun{JetVHeto }results with the same uncertainty
prescription as ref.~\cite{Banfi:2012jm} (not shown), the central
$\noun{Hjj}^{\star}$ prediction lies within half the thickness of
the more conservative error band that results in that case.

In ref.~\cite{Hamilton:2013fea} we presented results showing the
\noun{Nnlops }prediction lying within 1-2\% of the \noun{JetVHeto
}prediction, over the full $p_{{\scriptscriptstyle \mathrm{T},\mathrm{veto}}}$
range. Some degree of that good agreement stemmed from exploiting
freedom in the distribution of the NNLO-to-NLO inclusive K-factor
across the leading jet $p_{{\scriptscriptstyle \mathrm{T}}}$ spectrum,
to `tune' the \noun{Nnlops} result. We expect that the slightly less
good agreement in the $\noun{Hjj}^{\star}$ result here is correlated
with the percent level differences seen above in our jet $p_{{\scriptscriptstyle \mathrm{T}}}$
spectrum, between $\noun{Hjj}^{\star}$ and \noun{Nnlops. }We remind
that these differences are technical in origin, and should be entirely
removable with a more refined implementation of our method.

Lastly, we remark that the unimproved \noun{Hjj-Minlo }results for
the jet veto efficiency are, somewhat surprisingly, also quite good.
This good agreement of unimproved \noun{Hjj-Minlo} and \noun{JetVHeto
}is, however, rather fortuitous. The 0-jet cross section in the numerator
of the definition of the jet veto efficiency, is equal to the $p_{{\scriptscriptstyle \mathrm{T}}}$
integral of the leading jet transverse momentum spectrum from $p_{{\scriptscriptstyle \mathrm{T}}}=0$
GeV up to $p_{{\scriptscriptstyle \mathrm{T}}}=p_{{\scriptscriptstyle \mathrm{T,veto}}}$.
One can clearly see from figure \ref{fig:Leading-jet-transverse-mom}
that the leading jet transverse momentum spectrum from the unimproved
\noun{Hjj-Minlo }generator is, in general, quite different with respect
to the \noun{Nnlops} and improved $\noun{Hjj}^{\star}$ results. For
the region $p_{{\scriptscriptstyle \mathrm{T},\mathrm{veto}}}\lesssim30$
GeV the $\noun{Hjj}^{\star}$ and unimproved \noun{Hjj-Minlo }jet
$p_{{\scriptscriptstyle \mathrm{T}}}$ spectra, while clearly different
in normalization, are actually not so different in shape. By definition,
the jet veto efficiency, $\varepsilon(p_{{\scriptscriptstyle \mathrm{T,veto}}})$,
divides out the respective total cross sections, and hence it is therefore
reasonable to expect $\varepsilon(p_{{\scriptscriptstyle \mathrm{T,veto}}})$
is not so different in the $\noun{Hjj}^{\star}$ and unimproved \noun{Hjj-Minlo
}predictions for the latter\noun{ }$p_{{\scriptscriptstyle \mathrm{T},\mathrm{veto}}}$
region. Moreover, since the numerator of $\varepsilon(p_{{\scriptscriptstyle \mathrm{T,veto}}})$
is the cumulant of the leading jet transverse momentum spectrum, which
receives, by far, its main contribution from the low $p_{{\scriptscriptstyle \mathrm{T}}}$
region, it follows that the behaviour of $\varepsilon(p_{{\scriptscriptstyle \mathrm{T,veto}}})$,
for $p_{{\scriptscriptstyle \mathrm{T},\mathrm{veto}}}\gtrsim30$
GeV, is less sensitive to differences in the latter spectrum in this
region, with all predictions converging steadily towards $\varepsilon(m_{{\scriptscriptstyle \mathrm{H}}})\approx1$.

Figure \ref{fig:yJ1-antikT-R-eq-0.4} shows the rapidity of the leading
$R=0.4$ anti-$k_{t}$ jet, with a 25 GeV cut on the jet transverse
momentum. Broadly speaking the structure of the results in this distribution,
in particular their normalization, can be explained in terms of the
inclusive 1-jet cross section with the same jet $p_{{\scriptscriptstyle \mathrm{T}}}$
threshold; the uppermost plot at the top of fig.~\ref{fig:Inclusive-jet-cross-section-akT-R-eq-0.4}.
We remind that the \noun{Hjj-Minlo }prediction for this observable
is nominally LO, whereas the \noun{Nnlops }and $\noun{Hjj}^{\star}$
results are NLO accurate. This being the case, it is a remarkable
coincidence that the unimproved \noun{Hjj-Minlo} result only exhibits
very small differences with respect to the other two predictions,
at the level of about 5\%.

The \noun{Nnlops }and $\noun{Hjj}^{\star}$ results are almost indistinguishable
in the central rapidity region, with the $\noun{Hjj}^{\star}$ prediction
improved in this aspect, relative to its parent \noun{Hjj-Minlo }simulation.
Towards the higher rapidity regions, differences in the \noun{Nnlops
}and $\noun{Hjj}^{\star}$ results, on the level of $\sim5\%$, become
visible. Generally speaking the $\mathrm{y}_{{\scriptscriptstyle \mathrm{J}1}}$
distributions of $\noun{Hjj}^{\star}$ and its parent \noun{Hjj-Minlo
}generator, exhibit very slight, and very similar, `smiles' with respect
to the \noun{Nnlops }distribution. In the \noun{Hjj-Minlo }case the
`smile' feature coupled with its smaller inclusive 1-jet cross section
conspires to make it agree very well with the \noun{Nnlops }prediction
in the high rapidity regions, where the improved $\noun{Hjj}^{\star}$
program is off by 5\%. 

We refer back to the discussion of the inclusive 1-jet cross section,
surrounding fig.~\ref{fig:Inclusive-jet-cross-section-akT-R-eq-0.4},
for comments on why one can expect to see small deviations between
the $\noun{Hjj}^{\star}$ result and the target \noun{Nnlops }distribution
for general inclusive 1-jet quantities, starting at the level of $\bar{\alpha}_{{\scriptscriptstyle \mathrm{S}}}^{5}$
terms. Our initial reaction, to seeing the difference in shape between
the $\mathrm{y}_{{\scriptscriptstyle \mathrm{J}1}}$ distributions
of the $\noun{Hjj}^{\star}$ and \noun{Nnlops} results, was to interpret
it as being due to a weakness in our implementation of our method.
Re-making this distribution at the level of the $\noun{Hjj}^{\star}$
and \noun{Nnlops} LHE events reveals, however, that the two are actually
indistinguishable from one another (the distributions agree at the
sub-percent level). Moreover, at the LHE level, the unimproved \noun{Hjj-Minlo
}code is more clearly out of agreement with both of the latter and,
in particular, it does no longer agree so well with and the \noun{Nnlops}
in the high rapidity region; the difference being at the level of
5\%.

It follows that our implementation of the method actually works perfectly
as intended, and that the small features above which were counter
to naive expectations, are actually fully attributable to the attachment
of the parton shower. The parton shower generates the 3rd hardest
radiation and beyond in the \noun{Nnlops }generator, while it starts
by generating the 4th hardest radiation in the case of \noun{Hjj-Minlo
}and $\noun{Hjj}^{\star}$. Naturally then the effect of the parton
shower on the $\mathrm{y}_{{\scriptscriptstyle \mathrm{J}1}}$ distribution
in the \noun{Nnlops }case is greater, acting to deplete the cross
section in the high rapidity side bands relative to the \noun{Hjj-Minlo
}and $\noun{Hjj}^{\star}$ results. Given that the difference between
the theoretically superior $\noun{Hjj}^{\star}$ and \noun{Nnlops
}results in these high rapidity regions has been traced to the effects
of the 3rd hardest emitted parton (i.e.~an $\bar{\alpha}_{{\scriptscriptstyle \mathrm{S}}}^{5}$
effect), we cannot say one result is better than the other. We suffice
to say that the difference is in any case small, in a region where
theoretical control is not as high as in other places, and it is very
much contained within the scale uncertainty bands.

Figure \ref{fig:Leading-jet-pT-Njet-ge-2} shows the transverse momentum
spectrum of the leading jet, in events containing at least two $R=0.4$
anti-$k_{t}$ jets, with transverse momenta above 25 GeV. Ostensibly,
this is a rather everyday observable, but it nevertheless probes Sudakov
effects on the $y_{12}$ distribution. So, it is really the first
distribution we have shown so far which is sensitive to non-trivial
workings of our method. Towards the low end of the spectrum, $p_{{\scriptscriptstyle \mathrm{T}}}^{{\scriptscriptstyle \mathrm{J}1}}\lesssim75$
GeV, there is essentially not enough phase space available to generate
large $L_{12}$ logarithms. By contrast, at the high $p_{{\scriptscriptstyle \mathrm{T}}}^{{\scriptscriptstyle \mathrm{J}1}}$
end, one can expect large $L_{12}$ logarithms, with a significant
contribution from events for which $\sqrt{y_{12}}\gtrsim25$ GeV and
$\sqrt{y_{01}}$ is of the order of $p_{{\scriptscriptstyle \mathrm{T}}}^{{\scriptscriptstyle \mathrm{J1}}}$.
So, even though the distribution is defined on 2-jet events, in the
high $p_{{\scriptscriptstyle \mathrm{T}}}^{{\scriptscriptstyle \mathrm{J1}}}$
limit, the second jet should generally be considered as secondary,
soft, radiation emitted from a hard, high-$p_{{\scriptscriptstyle \mathrm{T}}}$,
Higgs-plus-jet system.

By construction our method will only act to correct the \noun{Hjj-Minlo
}distribution for $y_{01}\gg y_{12}$, leaving regions where there
is no such strong scale hierarchy untouched. Thus, in fig.~\ref{fig:Leading-jet-pT-Njet-ge-2},
at low transverse momentum, we see the $\noun{Hjj}^{\star}$ distribution
agrees identically with \noun{Hjj-Minlo. }This\noun{ }is\noun{,} of
course, the desired behaviour, since in this region, for this (2-jet)
observable, \noun{Hjj-Minlo }is nominally NLO accurate, whereas \noun{Nnlops
}is only LO. We remind that, the analogous inclusive leading jet transverse
momentum spectrum, fig.~\ref{fig:Inclusive-jet-cross-section-akT-R-eq-0.4},
displays significant deviations in shape between \noun{Nnlops}/$\noun{Hjj}^{\star}$
and \noun{Hjj-Minlo }in this same $p_{{\scriptscriptstyle \mathrm{T}}}$
region, while \noun{Nnlops} and $\noun{Hjj}^{\star}$ are in near
perfect agreement. 

Turning instead to the high $p_{{\scriptscriptstyle \mathrm{T}}}^{{\scriptscriptstyle \mathrm{J}1}}$
region, the three predictions are in good agreement with one another.
In the high $p_{{\scriptscriptstyle \mathrm{T}}}^{{\scriptscriptstyle \mathrm{J}1}}$
region there is perhaps a faint hint of the $\noun{Hjj}^{\star}$
result tending to that of the \noun{Nnlops}. We assert that the latter
tendency would be the correct and desirable result there. Should the
transverse momentum of the leading jet enter a high enough $p_{{\scriptscriptstyle \mathrm{T}}}$
regime, a 25 GeV jet-defining $p_{{\scriptscriptstyle \mathrm{T}}}$
cut for the second jet will correspond to a cut deep in the Sudakov
region of the corresponding $\sqrt{y_{12}}$ distribution, in which
case, the leading jet $p_{{\scriptscriptstyle \mathrm{T}}}$ spectrum
in two-jet events increasingly corresponds to the inclusive leading
jet $p_{{\scriptscriptstyle \mathrm{T}}}$ spectrum.

Finally for fig.~\ref{fig:Leading-jet-pT-Njet-ge-2}, we notice that
the dark red band, depicting uncertainty due to variations of the
technical $\rho$ parameter, has become visible for the first time
in this section (in the upper-right ratio plot, at high transverse
momentum). This technical systematic is, however, seemingly limited
to a $\pm2\%$ uncertainty, which is dwarfed by the conventional theoretical
uncertainty coming from the renormalization and factorization scale
variations (the significantly larger light-red band). 

The last distribution we present showing the behaviour of the leading
jet is that of its rapidity in events with at least two-jets, fig.~\ref{fig:yJ1-Njets-ge-2}.
This distribution is rather unremarkable given what we have shown
immediately before, for the leading jet transverse momentum spectrum
in the same class of events (fig.~\ref{fig:Leading-jet-pT-Njet-ge-2}).
Here, as in fig.~\ref{fig:Leading-jet-pT-Njet-ge-2}, the distribution
shows that the $\noun{Hjj}^{\star}$ distribution overlaps the \noun{Hjj-Minlo}
prediction, which is NLO accurate in the descriptions of this observable,
while the \noun{Nnlops }result is only LO. This is the expected and,
of course, the desired behaviour of our improved $\noun{Hjj}^{\star}$
simulation.

\subsubsection*{Second jet and third hardest jets}

\begin{figure}[htbp]
\centering{}\includegraphics[scale=0.9]{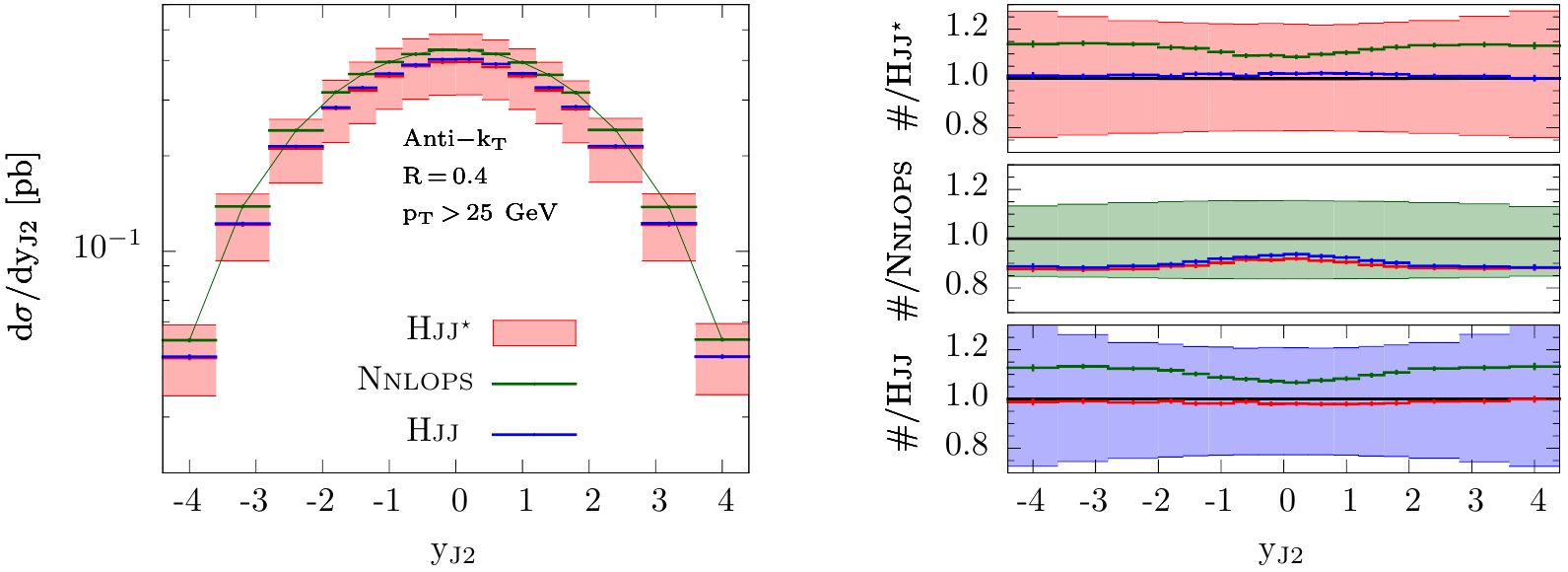}\protect\caption{\label{fig:yJ2-antikT-R-eq-0.4}Rapidity of the second hardest jet.}
\end{figure}

\begin{figure}[htbp]
\centering{}\includegraphics[scale=0.9]{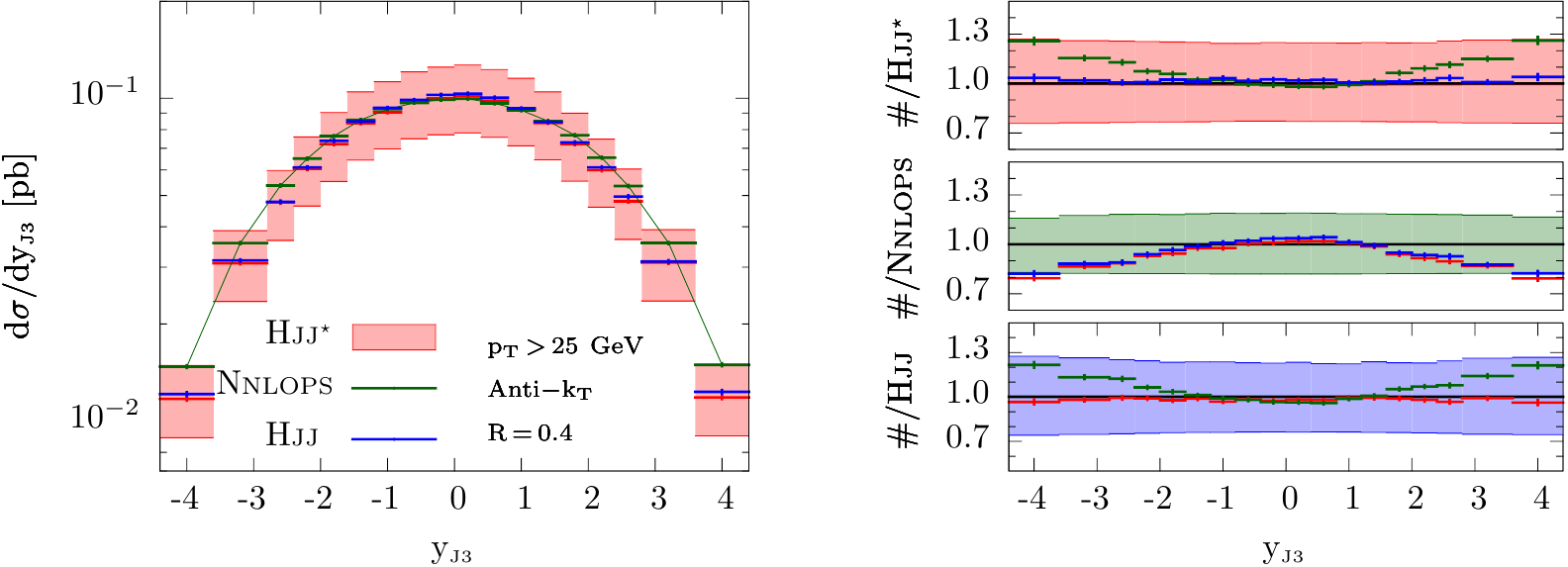}\protect\caption{\label{fig:yJ3-antikT-R-eq-0.4}Rapidity of the third hardest jet.}
\end{figure}

In this subsection we move to present plots of distributions probing
directly the behaviour of the second and third hardest jets produced
in association with the Higgs boson. As before, jets have been defined
according to the anti-$k_{t}$ clustering algorithm, with the jet
radius parameter $R=0.4$. Additionally, for the case of jet rapidity
distributions, in figures \ref{fig:yJ2-antikT-R-eq-0.4} and \ref{fig:yJ3-antikT-R-eq-0.4},
the jets are required to pass a transverse momentum threshold of 25
GeV.

\begin{figure}[htbp]
\centering{}\includegraphics[scale=0.9]{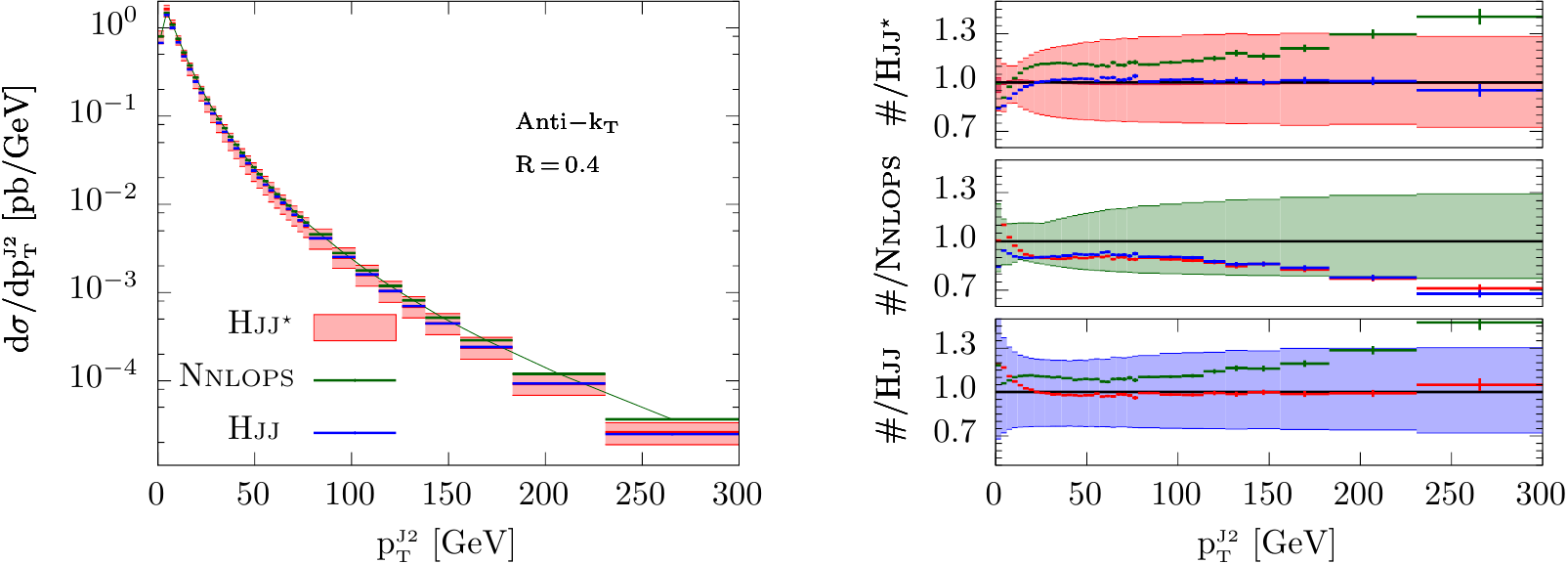}\protect\caption{\label{fig:pTJ2-antikT-R-eq-0.4}Transverse momentum spectrum of the
second jet.}
\end{figure}

The transverse momentum spectrum of the second hardest jet is plotted
in fig.~\ref{fig:pTJ2-antikT-R-eq-0.4}. In all simulations, before
(not shown) and after showering, the distribution peaks in the bin
at $3\,\mathrm{GeV}\le p_{{\scriptscriptstyle \mathrm{T}}}^{{\scriptscriptstyle \mathrm{J}2}}\le\,6\,\mathrm{GeV}$.
Moving upwards from the first bin at $p_{{\scriptscriptstyle \mathrm{T}}}^{{\scriptscriptstyle \mathrm{J}2}}=0$
GeV the $\noun{Hjj}^{\star}$ (red) and \noun{Hjj-Minlo }(blue) predictions
start off with a 20\% difference, which smoothly and monotonically
diminishes, with the two distributions coalescing at $p_{{\scriptscriptstyle \mathrm{T}}}^{{\scriptscriptstyle \mathrm{J}2}}\approx20$
GeV. For higher transverse momenta, the $\noun{Hjj}^{\star}$ and
\noun{Hjj-Minlo }histograms become indistinguishable from one another.
Meanwhile, in the same region, the \noun{Nnlops }result starts off
with a 15\% discrepancy between it and the latter simulations, which
rises with the transverse momentum. Nevertheless, the \noun{Nnlops
}prediction is within the margins set by all renormalization and factorization
scale uncertainty bands. 

The behaviour of the $\noun{Hjj}^{\star}$ and \noun{Hjj-Minlo }predictions
relative to one another is as intended. In general, the \noun{Hjj-Minlo
}prediction is NLO accurate in the description of $p_{{\scriptscriptstyle \mathrm{T}}}^{{\scriptscriptstyle \mathrm{J2}}}$,
and so it is of course desirable that the $\noun{Hjj}^{\star}$ tends
to that result in regions where Sudakov logarithms at higher orders
are not large, i.e.~away from the Sudakov peak.%
\footnote{In such regions where it is meaningful to quantify accuracy in the
context of just fixed order perturbation theory, we remind that the
\noun{Nnlops }prediction for $p_{{\scriptscriptstyle \mathrm{T}}}^{{\scriptscriptstyle \mathrm{J}2}}$
is, by contrast, only LO accurate. %
} In the vicinity of the peak, large logarithms enter at every order
in perturbation theory. In this feasibility study we claim to control
these large logarithms nominally at just LL/$\mathrm{NLL}_{\sigma}$
accuracy. The improved $\noun{Hjj}^{\star}$ prediction works so as
to implement unitarity for the 0- and 1-jet inclusive cross sections
by ascribing the mismatch there to missing $\mathrm{NNLL}_{\sigma}$
Sudakov logarithms beyond NLO. The increasing difference of $\noun{Hjj}^{\star}$
with respect to \noun{Hjj-Minlo} in the region $p_{{\scriptscriptstyle \mathrm{T}}}^{{\scriptscriptstyle \mathrm{J}2}}\le20$
GeV, up onto the Sudakov peak, roughly reflects this $\mathrm{NNLL}_{\sigma}$
`profiling' of the $\sim$10-12\% excess in the \noun{Nnlops} total
inclusive cross section over that of \noun{Hjj-Minlo} (see e.g.~figs.~\ref{fig:yH-fully-incl}-\ref{fig:Inclusive-jet-cross-section-akT-R-eq-0.4}). 

\begin{figure}[htbp]
\centering{}\includegraphics[scale=0.9]{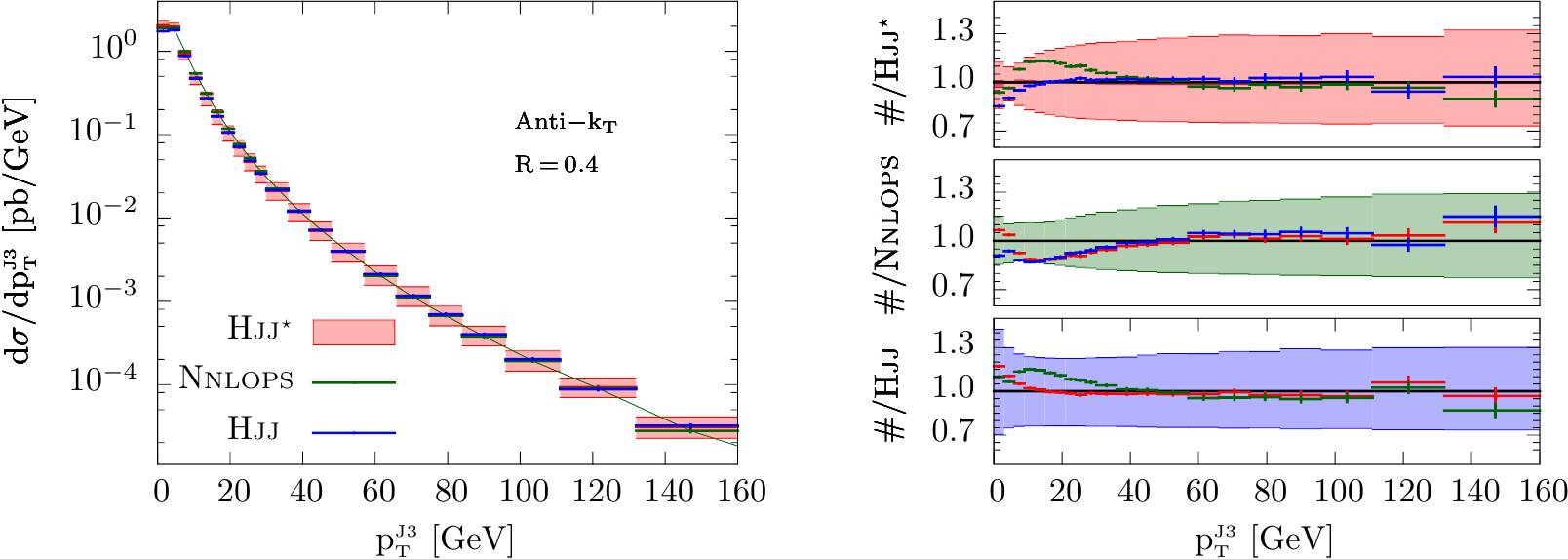}\protect\caption{\label{fig:pTJ3-antikT-R-eq-0.4}Third jet transverse momentum spectrum.}
\end{figure}

In figure \ref{fig:pTJ3-antikT-R-eq-0.4} we plot the transverse momentum
of the third jet. In this case there is, coincidentally, good agreement
of all predictions in the moderate to high $p_{{\scriptscriptstyle \mathrm{T}}}$
domain. This is somewhat fortuitous in the context of the \noun{Nnlops
}simulation, since the third jet in that simulation is generated exclusively
in the parton shower approximation, whereas in $\noun{Hjj}^{\star}$
and \noun{Hjj-Minlo} it has a matched matrix element-parton shower
description. With a view to validating our ideas, what is more relevant
is the observation of the relative behaviour of $\noun{Hjj}^{\star}$
and \noun{Hjj-Minlo}. Here we see, essentially, exactly the same trend
as found in the case of $p_{{\scriptscriptstyle \mathrm{T}}}^{{\scriptscriptstyle \mathrm{J}2}}$,
specifically, identical agreement for $p_{{\scriptscriptstyle \mathrm{T}}}^{{\scriptscriptstyle \mathrm{J}2}}\gtrsim10-15$
GeV, with a steadily increasing excess of $\noun{Hjj}^{\star}$\noun{
}over \noun{Hjj-Minlo} as one looks towards zero transverse momentum.
These aspects are also fully explained and intended, with the same
reasoning as for $p_{{\scriptscriptstyle \mathrm{T}}}^{{\scriptscriptstyle \mathrm{J}2}}$.
The only slight difference here is that the third jet being, by definition,
softer than the second jet, implies that the excess of $\noun{Hjj}^{\star}$\noun{
}over \noun{Hjj-Minlo} is confined to a slightly lower region of the
$p_{{\scriptscriptstyle \mathrm{T}}}^{{\scriptscriptstyle \mathrm{J}3}}$
distribution, than one finds in the $p_{{\scriptscriptstyle \mathrm{T}}}^{{\scriptscriptstyle \mathrm{J}2}}$
case.

Figures \ref{fig:yJ2-antikT-R-eq-0.4} and \ref{fig:yJ3-antikT-R-eq-0.4}
show, respectively, the rapidity spectra of the second and third hardest
$R=0.4$ anti-$k_{t}$ jets, with $p_{{\scriptscriptstyle \mathrm{T}}}\ge25$
GeV. Both figures reveal the $\noun{Hjj}^{\star}$\noun{ }results
agreeing perfectly with those of the `parent' \noun{Hjj-Minlo} simulation.
The \noun{Nnlops }predictions clearly differ in shape and normalization
with respect to the latter but, nevertheless, they remain within the
renormalization and factorization scale uncertainty bands. For what
concerns the normalization of the distributions, the perfect agreement
between $\noun{Hjj}^{\star}$\noun{ }and \noun{Hjj-Minlo }was to be
expected, based on that seen already in the related 2-jet inclusive
cross sections (fig.~\ref{fig:Inclusive-jet-cross-section-akT-R-eq-0.4}).
As with the $p_{{\scriptscriptstyle \mathrm{T}}}^{{\scriptscriptstyle \mathrm{J}2}}$
and $p_{{\scriptscriptstyle \mathrm{T}}}^{{\scriptscriptstyle \mathrm{J}3}}$
spectra, for modest values of the transverse momentum, the tendency
of $\noun{Hjj}^{\star}$ to reproduce \noun{Hjj-Minlo }here is as
intended and desired; the latter being NLO accurate for $\mathrm{y}_{{\scriptscriptstyle \mathrm{J}2}}$
and LO accurate for $\mathrm{y}_{{\scriptscriptstyle \mathrm{J}3}}$,
in contrast to the LO and parton shower accuracy, respectively, afforded
by the\noun{ Nnlops.}

\subsubsection*{Jet rates}

\begin{figure}[htbp]
\begin{centering}
\includegraphics[scale=0.9]{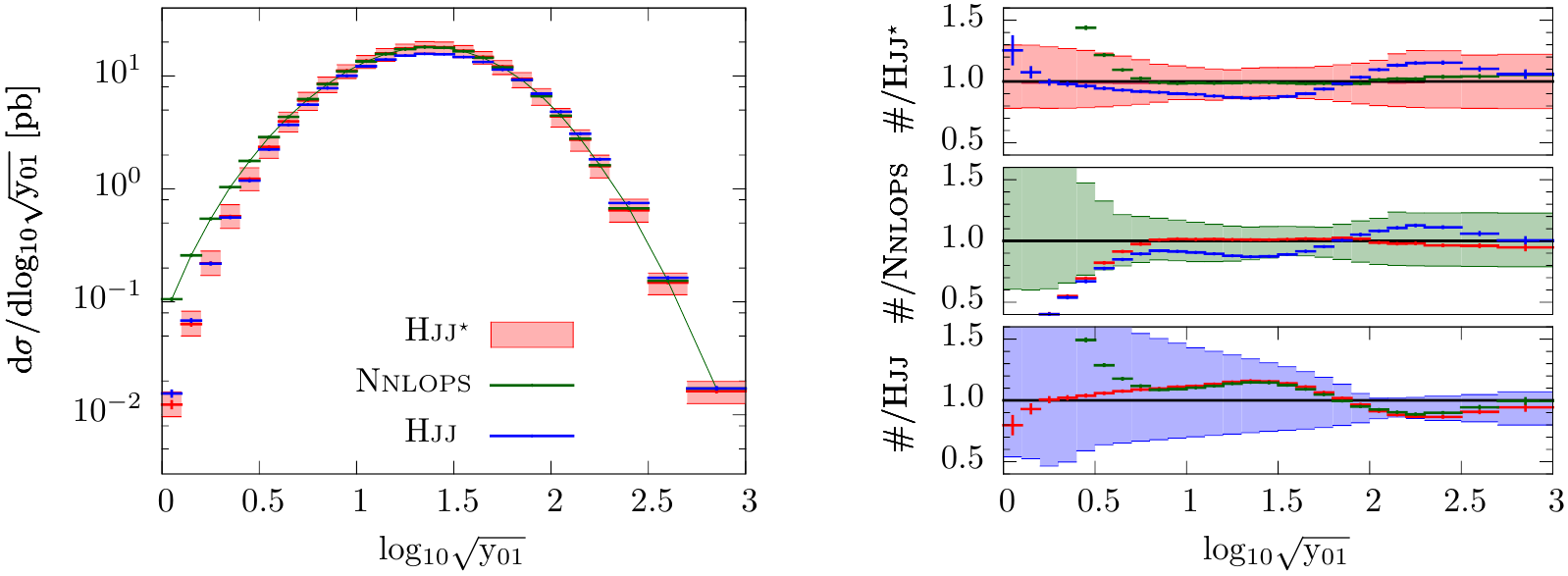} 
\par\end{centering}

\begin{centering}
~ 
\par\end{centering}

\begin{centering}
\includegraphics[scale=0.9]{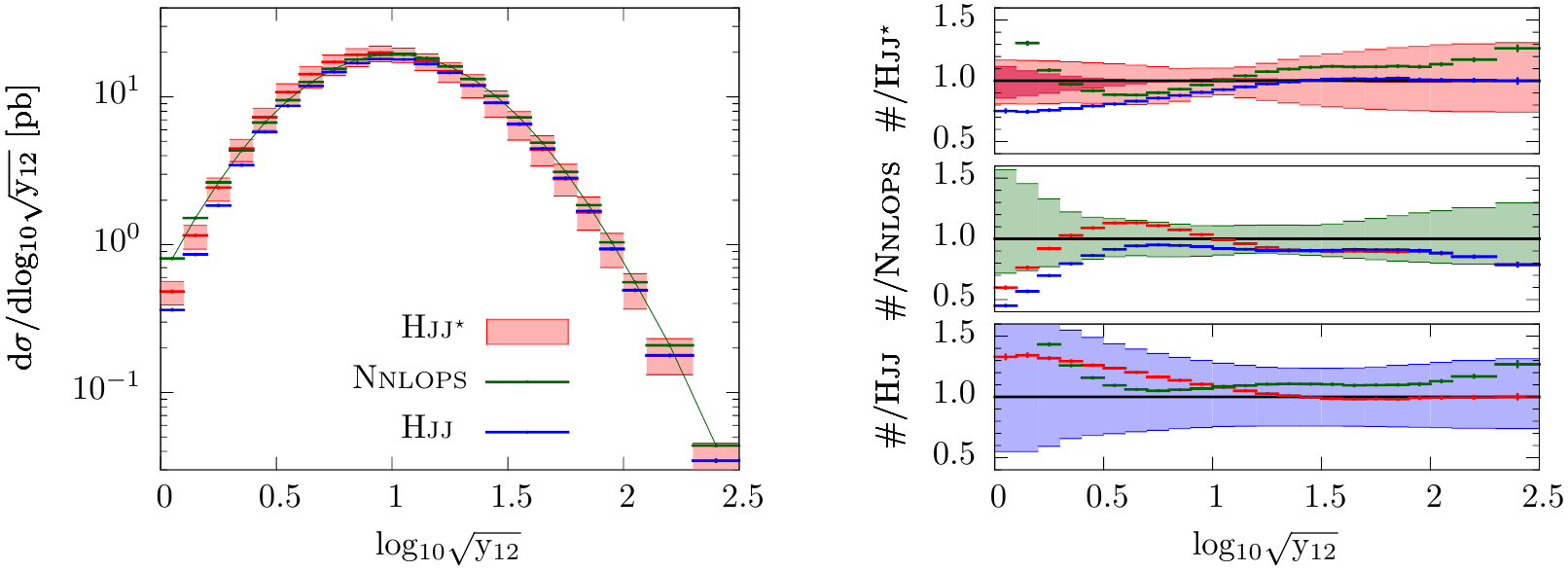}
\par\end{centering}

\protect\caption{\label{fig:log10-sqrt-y01-and-log10-sqrt-y12}In the upper plots we
display the $\log_{10}\sqrt{y_{01}}$ differential jet rate on the
left, while on the right we show the various predictions relative
to the central improved \noun{Hjj-Minlo} ($\noun{Hjj}^{\star}$), $\noun{Nnlops}$ and original \noun{Hjj-Minlo} ($\noun{Hjj}$) ones,
respectively, in the top, middle and bottom panels. In the lower plots
we display the corresponding set of distributions for the $\log_{10}\sqrt{y_{12}}$
differential jet rate. In the making of these plots jets have been
clustered according to the $k_{t}$-jet algorithm, with radius parameter
$R=1$.}
\end{figure}

\begin{figure}[htbp]
\begin{centering}
\includegraphics[scale=0.9]{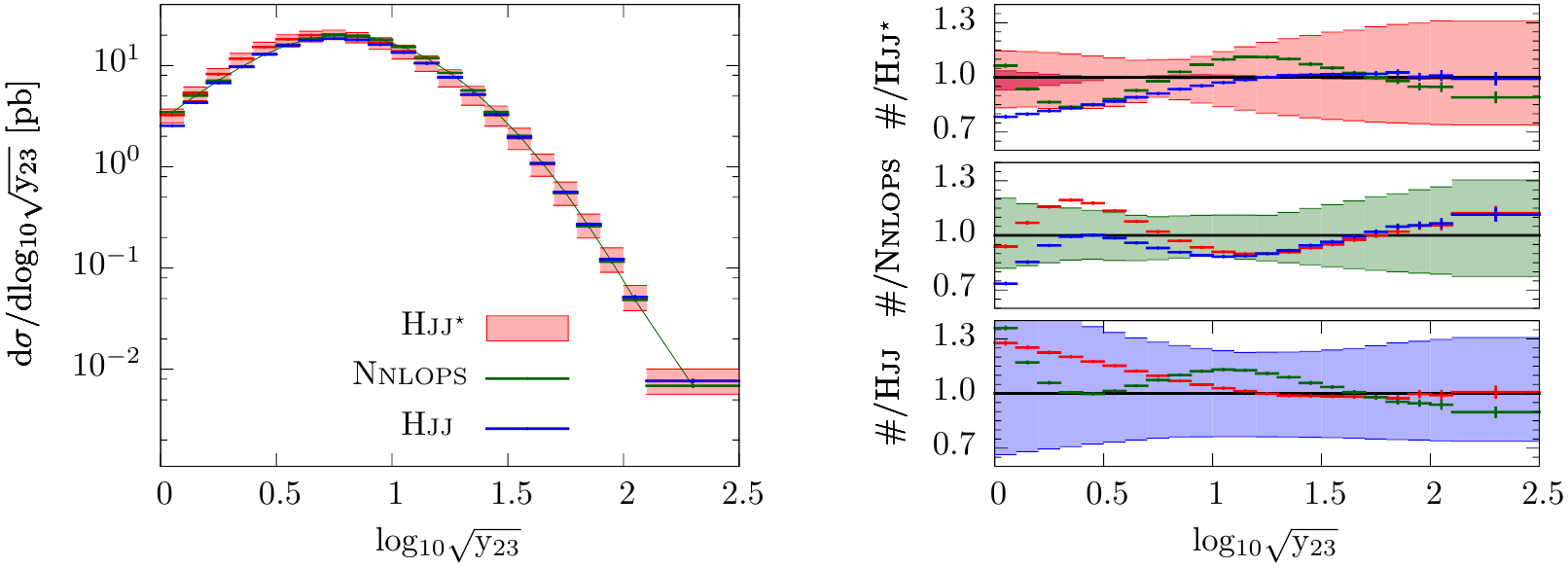} 
\par\end{centering}

\begin{centering}
~ 
\par\end{centering}

\centering{}\includegraphics[scale=0.9]{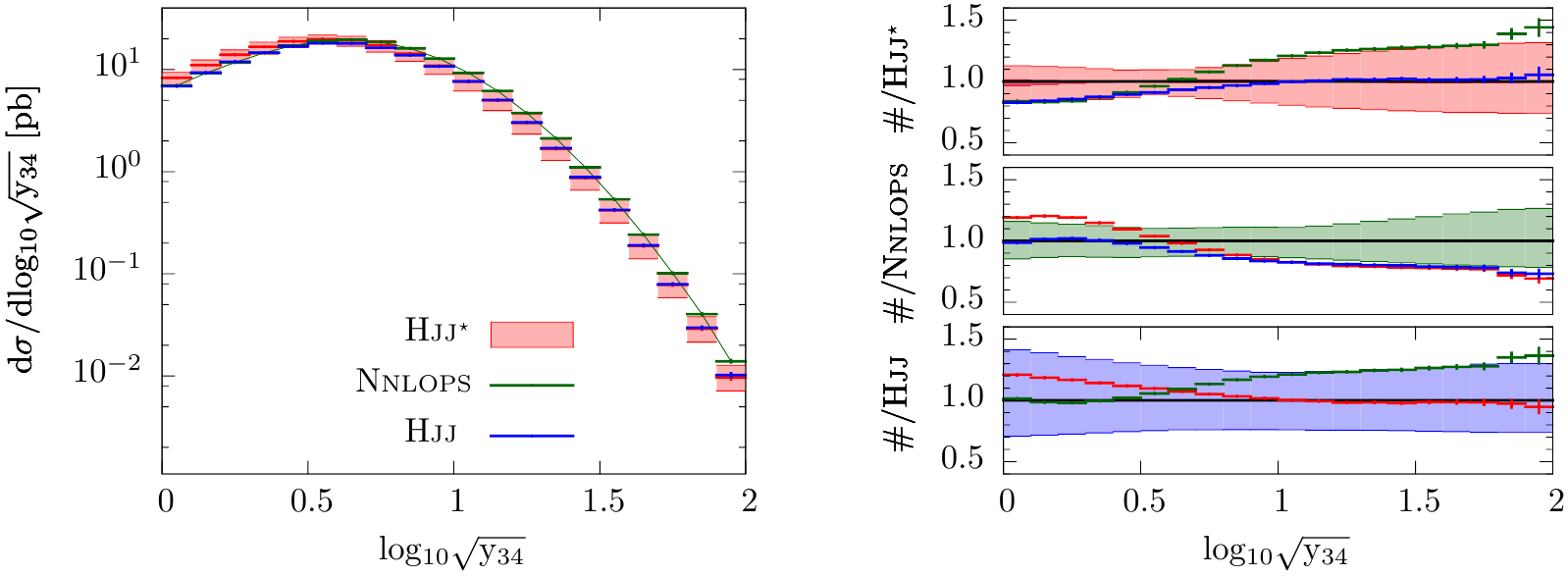}\protect\caption{\label{fig:log10-sqrt-y23-and-log10-sqrt-y34}In the upper-left plot
we show the $\log_{10}\sqrt{y_{23}}$ jet rate, while on the right
we show the various predictions again as ratios with respect to one
another. In the lower plots we display the corresponding set of distributions
for the $\log_{10}\sqrt{y_{34}}$ jet rate. Jets have been constructed
using the $k_{t}$-jet algorithm, with $R=1$.}
\end{figure}

In figures \ref{fig:log10-sqrt-y01-and-log10-sqrt-y12} and \ref{fig:log10-sqrt-y23-and-log10-sqrt-y34}
we present differential jets rates obtained from the exclusive $k_{t}$-jet
clustering algorithm with radius parameter $R=1$. Figure \ref{fig:log10-sqrt-y01-and-log10-sqrt-y12}
shows the $\log_{10}\sqrt{y_{01}}$ and $\log_{10}\sqrt{y_{12}}$
jet rate distributions, while figure \ref{fig:log10-sqrt-y23-and-log10-sqrt-y34}
shows $\log_{10}\sqrt{y_{23}}$ and $\log_{10}\sqrt{y_{34}}$. 

The $\log_{10}\sqrt{y_{01}}$ distribution in figure \ref{fig:log10-sqrt-y01-and-log10-sqrt-y12}
is equivalent to a plot of the transverse momentum of the leading
jet in the event, defined according to the $k_{t}$-jet clustering
algorithm with $R=1$. It is therefore not surprising to find that
the results for this distribution have a markedly similar structure
to those for the leading jet transverse momentum spectrum in fig.~\ref{fig:Leading-jet-transverse-mom};
notwithstanding the fact that in the latter case the jets were defined
according to the anti-$k_{t}$ jet algorithm, with radius parameter
$R=0.4$. We therefore refer the reader back to the discussion surrounding
fig.~\ref{fig:Leading-jet-transverse-mom}, for further explanation
regarding the features of the $\log_{10}\sqrt{y_{01}}$ distribution. 

The $\log_{10}\sqrt{y_{12}}$ in fig.~\ref{fig:log10-sqrt-y01-and-log10-sqrt-y12}
is more interesting, since this distribution is directly affected
by our proposed \noun{Minlo }improvement procedure. One can relatively
quickly gain an appreciation for the pattern of the results here by
noting that there is some reasonable degree of correspondence to be
expected between $\sqrt{y_{12}}$ and $p_{{\scriptscriptstyle \mathrm{T}}}^{{\scriptscriptstyle \mathrm{J}2}}$,
based on how $\sqrt{y_{12}}$ is defined; if all the jet clustering
algorithm did was initial-state clusterings, they would indeed be
exactly the same thing. Despite seeming like an over-simplification,
it is nevertheless the case that the relative behaviours of the three
predictions here are in very good agreement, quantitatively, with
that discussed earlier for the $p_{{\scriptscriptstyle \mathrm{T}}}^{{\scriptscriptstyle \mathrm{J}2}}$
spectrum (fig.~\ref{fig:pTJ2-antikT-R-eq-0.4}). 

As in $p_{{\scriptscriptstyle \mathrm{T}}}^{{\scriptscriptstyle \mathrm{J}2}}$
we see an excess of \noun{Nnlops }with respect to $\noun{Hjj}^{\star}$\noun{
}and \noun{Hjj-Minlo }of $\sim12\%$ in the region $20\,\mathrm{GeV}\lesssim\sqrt{y_{12}}\lesssim100\,\mathrm{GeV}$,
with the latter pair of simulations in perfect agreement. For $\sqrt{y_{12}}\lesssim20\,\mathrm{GeV}$,
as in the corresponding region of the $p_{{\scriptscriptstyle \mathrm{T}}}^{{\scriptscriptstyle \mathrm{J}2}}$
distribution, the $\noun{Hjj}^{\star}$\noun{ }and \noun{Hjj-Minlo
}predictions\noun{ }become increasingly separate, with the former
increasing over the latter, manifesting the restorative effect of
the correction procedure to recover the inclusive 0- and 1-jet \noun{Nnlops
}cross sections. Even the crossing over of the \noun{Nnlops }and $\noun{Hjj}^{\star}$\noun{
}distributions appears to occur at exactly the same place in the $p_{{\scriptscriptstyle \mathrm{T}}}^{{\scriptscriptstyle \mathrm{J}2}}$
and $\log_{10}\sqrt{y_{12}}$ distributions ($\sqrt{y_{12}}\sim9\,\mathrm{GeV}$).
In contrast to the $p_{{\scriptscriptstyle \mathrm{T}}}^{{\scriptscriptstyle \mathrm{J}2}}$
distribution, the $\log_{10}\sqrt{y_{12}}$ plot makes it clearer
when, and to what extent, the correction kicks-in. One can see that
the correction turns on smoothly just before the Sudakov peak, starting
at $\log_{10}\sqrt{y_{12}}\approx1.25$, ($\sqrt{y_{12}}\approx18\,\mathrm{GeV}$),
leading to a 7\% increase in $\noun{Hjj}^{\star}$ over \noun{Hjj-Minlo}
on\noun{ }the\noun{ }Sudakov peak, and ranging up to 25\% at $\sqrt{y_{12}}=1\,\mathrm{GeV}$. 

Lastly, this $\log_{10}\sqrt{y_{12}}$ distribution shows the first
real evidence, so far, of some sensitivity in the $\noun{Hjj}^{\star}$
results to the technical $\rho$ parameter. The conservatively estimated
systematic uncertainty owing to $\rho$ is depicted by the dark-red
band, seen superimposed on the light-red band, in the uppermost ratio
plot. This sensitivity to $\rho$ is, however, rather contained at
the level of $\pm10-15\%$, moreover, it is basically negligible above
$\sqrt{y_{12}}=3\,\mathrm{GeV}$.

Moving on, in the upper half of fig.~\ref{fig:log10-sqrt-y23-and-log10-sqrt-y34}
we have the $\log_{10}\sqrt{y_{23}}$ distribution. The correspondence
of $\sqrt{y_{12}}$ with $p_{{\scriptscriptstyle \mathrm{T}}}^{{\scriptscriptstyle \mathrm{J}2}}$,
which helped to quickly understand the $\log_{10}\sqrt{y_{12}}$ results
above, has an analogon here, namely, that neglecting final-state clusterings
by the jet algorithm, $\sqrt{y_{23}}$ becomes equal to $p_{{\scriptscriptstyle \mathrm{T}}}^{{\scriptscriptstyle \mathrm{J}3}}$.
This analogy continues to appear to hold remarkably well, for describing
the features of $\log_{10}\sqrt{y_{23}}$ in terms of those found
in the $p_{{\scriptscriptstyle \mathrm{T}}}^{{\scriptscriptstyle \mathrm{J}3}}$
distribution of fig.~\ref{fig:pTJ3-antikT-R-eq-0.4}. The arrangement
of the three predictions relative to one another, throughout the $\log_{10}\sqrt{y_{23}}$
distribution, is very much in direct correspondence with what one
can see in the $p_{{\scriptscriptstyle \mathrm{T}}}^{{\scriptscriptstyle \mathrm{J}3}}$
distribution. For example, all three predictions even cross at the
same point in the $\log_{10}\sqrt{y_{23}}$ and $p_{{\scriptscriptstyle \mathrm{T}}}^{{\scriptscriptstyle \mathrm{J}3}}$
distributions: $\sqrt{y_{23}}\approx50\,\mathrm{GeV}$ in fig.~\ref{fig:log10-sqrt-y23-and-log10-sqrt-y34}
and, correspondingly, $p_{{\scriptscriptstyle \mathrm{T}}}^{{\scriptscriptstyle \mathrm{J}3}}\approx50\,\mathrm{GeV}$
in fig.~\ref{fig:pTJ3-antikT-R-eq-0.4}. As was noted in comparing
the $p_{{\scriptscriptstyle \mathrm{T}}}^{{\scriptscriptstyle \mathrm{J}2}}$
and $p_{{\scriptscriptstyle \mathrm{T}}}^{{\scriptscriptstyle \mathrm{J}3}}$
distributions beforehand (figs.~\ref{fig:pTJ2-antikT-R-eq-0.4}-\ref{fig:pTJ3-antikT-R-eq-0.4}),
the effect of our corrective procedure in lifting the $\noun{Hjj}^{\star}$
distribution above that of its `parent' \noun{Hjj-Minlo }simulation,
in the region $\log_{10}\sqrt{y_{12}}<1.25$, directly percolates
into the same lower reaches of $\log_{10}\sqrt{y_{23}}$ (and also
$\log_{10}\sqrt{y_{34}}$). The extent of this lifting in $\log_{10}\sqrt{y_{12}}$
and $\log_{10}\sqrt{y_{23}}$, is quantitatively compatible with that
seen in $p_{{\scriptscriptstyle \mathrm{T}}}^{{\scriptscriptstyle \mathrm{J}2}}$
and $p_{{\scriptscriptstyle \mathrm{T}}}^{{\scriptscriptstyle \mathrm{J}3}}$,
both in terms of its magnitude and the phase space domain over which
it occurs; in particular we note that the separation of the $\noun{Hjj}^{\star}$
and \noun{Hjj-Minlo }distributions starts at a very slightly higher
value of $y_{12}$ than $y_{23}$, the latter being, by definition,
smaller than the former. As with the discussion of the preceding jet
rate variables and transverse momentum spectra, the effect of the
correction procedure is rather modest and it is limited to a region
of phase-space for which all-orders large logarithmic corrections
are significant.

In the lower half of figure \ref{fig:log10-sqrt-y23-and-log10-sqrt-y34}
we show the $\log_{10}\sqrt{y_{34}}$ distribution. In order to have
a non-zero contribution to this observable events must contain at
least four partons. So, in the case of $\noun{Hjj}^{\star}$ and \noun{Hjj-Minlo
}this distribution directly probes, for the first time, radiation
which is exclusively due to the parton shower interfacing. The distribution
is plainly smooth and exhibits no irregularities that might otherwise
signal some problem in that interfacing. The same comments apply here
as above, in regards to the lifting of the $\noun{Hjj}^{\star}$ distribution
with respect to \noun{Hjj-Minlo}, due to the action of our correction
procedure on the $y_{12}$ distribution and the associated feed-down
from that onto the higher multiplicity differential jet rates.

\begin{figure}[htbp]
\begin{centering}
\includegraphics[scale=0.9]{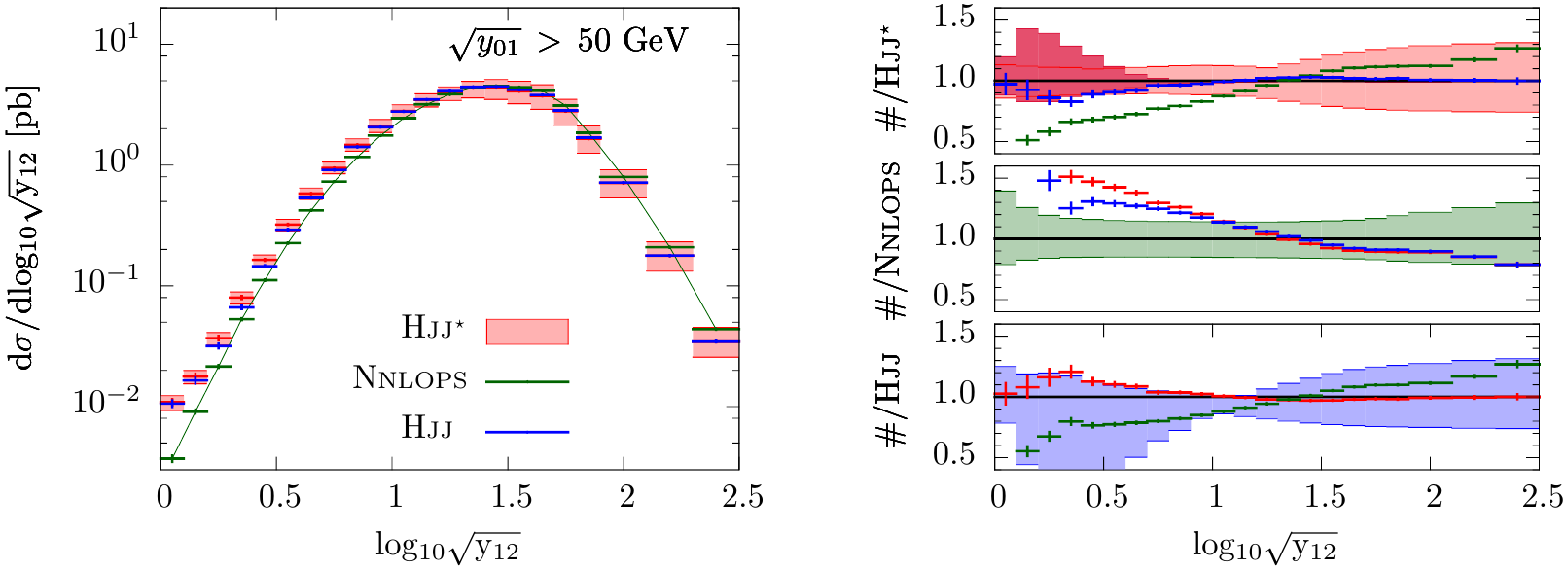} 
\par\end{centering}

\begin{centering}
~ 
\par\end{centering}

\centering{}\includegraphics[scale=0.9]{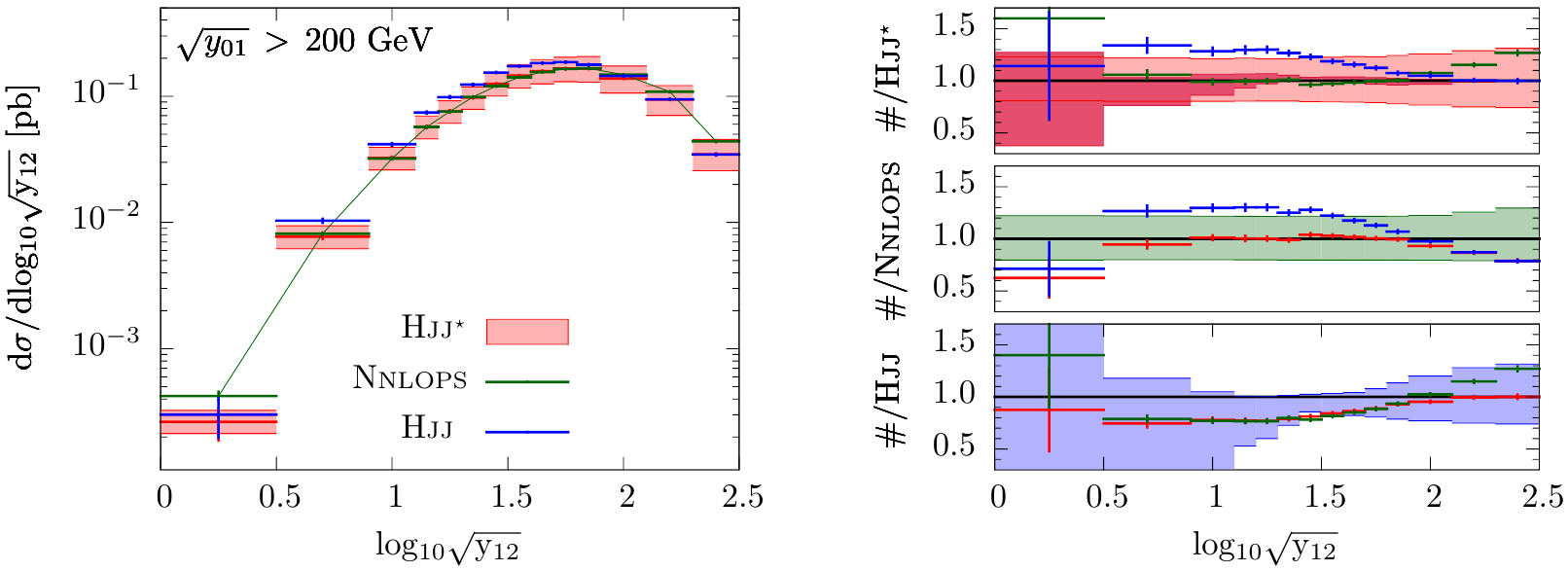}\protect\caption{\label{fig:log10-sqrt-y12-and-y01-gt-some-cuts}The $\log_{10}\sqrt{y_{12}}$
differential jet rates, defined according to the $k_{{\scriptscriptstyle \mathrm{T}}}$-jet
algorithm with jet radius parameter $R=1$, and with cuts of $10$,
$50$ and $200$ GeV imposed on $\sqrt{y_{01}}$.}
\end{figure}

The penultimate set of differential jet rates we wish to present are
given in figure \ref{fig:log10-sqrt-y12-and-y01-gt-some-cuts}.
Here we examine the key jet rate of interest to our studies, given
its role in the proposed correction procedure, $\log_{10}\sqrt{y_{12}}$,
but now subject to additional cuts in the $\sqrt{y_{01}}$ jet rate
variable. These cuts are intended to bring to the fore events for
which there is a hierarchy $y_{12}\ll y_{01}$ and associated large
logarithm $L_{12}$. This aspect is indeed manifested in both $\log_{10}\sqrt{y_{12}}$
distributions in fig.~\ref{fig:log10-sqrt-y12-and-y01-gt-some-cuts}
through the Sudakov peak shifting to higher $y_{12}$ values. The
Sudakov peak in the inclusive distribution of fig.~\ref{fig:log10-sqrt-y01-and-log10-sqrt-y12}
is centred around $\log_{10}\sqrt{y_{12}}=1$ ($\sqrt{y_{12}}=10\textnormal{ }\mathrm{GeV}$),
moving up to $\log_{10}\sqrt{y_{12}}\approx1.5$ ($\sqrt{y_{12}}\approx30\textnormal{ }\mathrm{GeV}$)
on imposing the $\sqrt{y_{01}}>50\mbox{ }\mathrm{GeV}$ cut, as shown
in the uppermost plot in fig.~\ref{fig:log10-sqrt-y12-and-y01-gt-some-cuts},
and further to $\log_{10}\sqrt{y_{12}}\approx1.75$ ($\sqrt{y_{12}}\approx55\textnormal{ }\mathrm{GeV}$)
on imposing the $\sqrt{y_{01}}>200\mbox{ }\mathrm{GeV}$ cut. The
shifting of the peak to higher $y_{12}$ values is a manifestation
of the fact that the cuts imply a proportionate increase in the available
phase space for high $p_{{\scriptscriptstyle \mathrm{T}}}$ emission
of the second pseudoparton.

One of the easiest features to make sense of in fig.~\ref{fig:log10-sqrt-y12-and-y01-gt-some-cuts},
is the excess of the \noun{Nnlops} results over $\noun{Hjj}^{\star}$
and \noun{Hjj-Minlo} predictions in the high $\sqrt{y_{12}}$ region,
with the latter pair of results being indistinguishable there. This
attribute is consistent with the enhancement of the \noun{Nnlops }cross
section over the corresponding \noun{Hjj-Minlo }and $\noun{Hjj}^{\star}$
results, in both the inclusive 2-jet cross section, with high jet
$p_{{\scriptscriptstyle \mathrm{T}}}$ thresholds (fig.~\ref{fig:Inclusive-jet-cross-section-akT-R-eq-0.4}),
and the transverse momentum spectrum of the second hardest jet (fig.~\ref{fig:pTJ2-antikT-R-eq-0.4}).
In the latter distribution the discrepancy increases with radiation
hardness, as it does in fig.~\ref{fig:log10-sqrt-y12-and-y01-gt-some-cuts}.
Technically, the agreement of $\noun{Hjj}^{\star}$ and \noun{Hjj-Minlo}
in this limit is also easy to understand, since in these regions $L_{12}$
is not large and the \noun{Minlo} correction procedure `switches off',
with the \noun{Nnlops }prediction being categorically inferior to
\noun{Hjj-Minlo }there. Specifically, the $h(L_{12})$ function (eq.~\ref{eq:sect26-freezing-out-the-weight-factor-at-aSL2-gtrsim-1})
tends to zero. 

Looking towards the Sudakov peak regions in fig.~\ref{fig:log10-sqrt-y12-and-y01-gt-some-cuts},
where the great bulk of the cross section is centred, one expects,
by virtue of the fact that our method is to return inclusive 0- and
1-jet \noun{Nnlops }predictions, the \noun{Nnlops }and $\noun{Hjj}^{\star}$
predictions to agree well there. The results in figure \ref{fig:log10-sqrt-y12-and-y01-gt-some-cuts}
support this simple reasoning quite well. 

Turning back to the 1-jet inclusive cross section predictions in fig.~\ref{fig:Inclusive-jet-cross-section-akT-R-eq-0.4},
with a 50 GeV jet $p_{{\scriptscriptstyle \mathrm{T}}}$ threshold,
there is a relatively small difference between\noun{ Hjj-Minlo }and
\noun{Nnlops} predictions, this implies that, on average, the $\delta(\Phi_{{\scriptscriptstyle \mathcal{B}\mathrm{J}}})$
term in eq.~\ref{eq:sect26-first-delta-defn} is very small in the
context of that observable, which is in some reasonable degree of
correspondence with the cumulant of the first distribution in figure
\ref{fig:log10-sqrt-y12-and-y01-gt-some-cuts} ($\sqrt{y_{01}}>50\mbox{ }\mathrm{GeV}$).
Granted this point, it is then no surprise to observe that the $\noun{Hjj}^{\star}$
and \noun{Hjj-Minlo }predictions are essentially in perfect agreement
all the way down to $\sqrt{y_{12}}\sim3\mbox{ GeV}$, exhibiting only
small differences beyond that point. 

The difference in normalization of the \noun{Hjj-Minlo }and \noun{Nnlops
}predictions in the case of the $\sqrt{y_{01}}>200\mbox{ }\mathrm{GeV}$
cut, can be anticipated by looking at the difference in the respective
leading jet $p_{{\scriptscriptstyle \mathrm{T}}}$ spectra for $p_{{\scriptscriptstyle \mathrm{T}}}>200\mbox{ }\mathrm{GeV}$
(fig.~\ref{fig:Leading-jet-transverse-mom}), revealing a fairly
flat 15\% surplus of \noun{Hjj-Minlo }over \noun{Nnlops}. Indeed a
15\% excess of\noun{ Hjj-Minlo }over \noun{Nnlops }is what we also
see here in the vicinity of the Sudakov peak, in the lower plot of
fig.~\ref{fig:log10-sqrt-y12-and-y01-gt-some-cuts}. In this
region and that below, both dominated by large logarithmic corrections,
one sees the improved $\noun{Hjj}^{\star}$ result nicely following
the \noun{Nnlops} results.

Before leaving the discussion of figure \ref{fig:log10-sqrt-y12-and-y01-gt-some-cuts}
we must remark on the systematic uncertainty coming from the $\rho$
parameter (dark-red band). Indeed these observables have been mainly
studied to try to expose and stress-test this aspect. The predictions
of fig \ref{fig:log10-sqrt-y12-and-y01-gt-some-cuts} show the
biggest $\rho$ dependence of any in this paper. Sure enough, demanding
that $\sqrt{y_{01}}$ be 50 or 200 GeV and then looking down at the
$1\mbox{ GeV}\le\sqrt{y_{12}}\le3\mbox{ GeV}$ (i.e.~$0\le\log_{10}\sqrt{y_{12}}\le0.5$)
we see what looks like a sizable $\rho$ uncertainty. In the $\sqrt{y_{01}}>50\mbox{ }\mathrm{GeV}$
case we see this uncertainty rises up to $_{-20\%}^{+50\%}$ at $\sqrt{y_{12}}\approx1.5\mbox{ GeV}$.
In the same region of the $\sqrt{y_{01}}>200\mbox{ }\mathrm{GeV}$
distribution we see an uncertainty similar in magnitude, however,
for this plot the conclusion is less precise, due to the appearance
of large statistical uncertainties. Taking together the following
points, we believe we can now conclude that the uncertainty due to
our $\rho$ parameter is in general negligible: i.) we took a rather
conservative approach to assessing the uncertainty due to $\rho$,
varying it from 1 to $27$, ii.) we constructed observables to isolate
and expose potential problems owing to $\rho$, we found no pathologies,
and the latter uncertainty only showed up in the very deep Sudakov
region, where theoretical control is very limited.

\begin{figure}[htbp]
\begin{centering}
\includegraphics[scale=0.9]{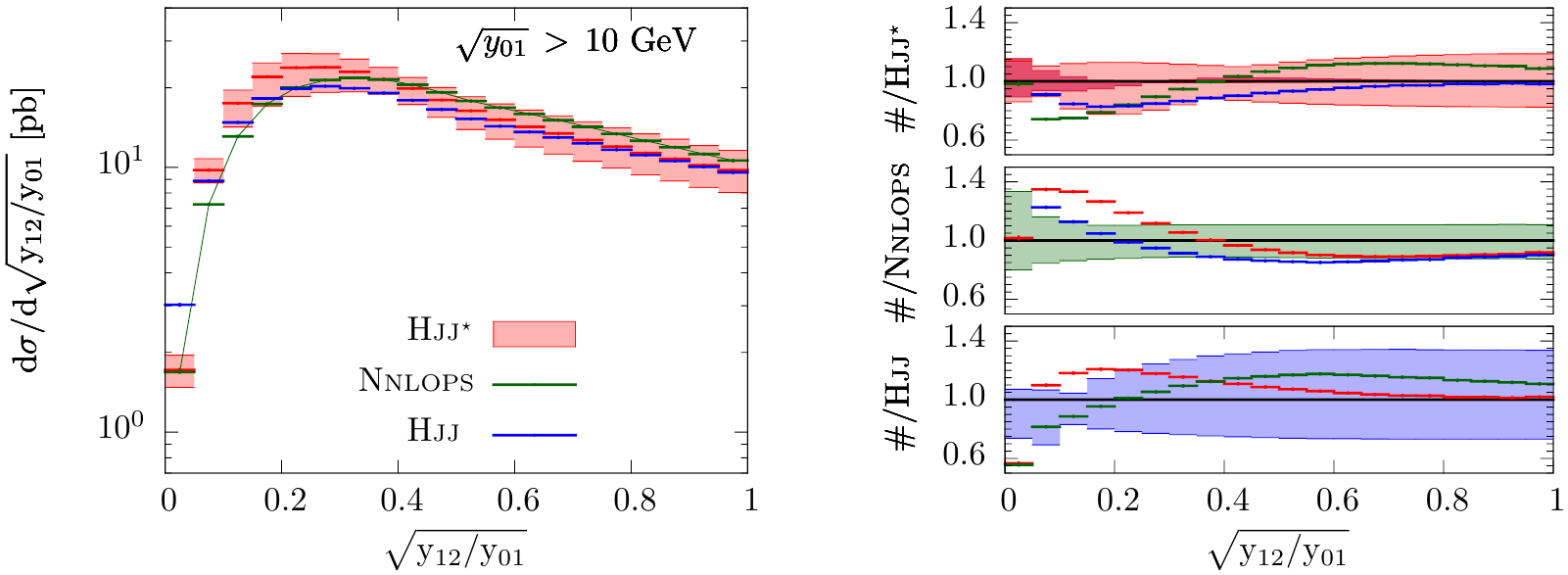} 
\par\end{centering}

\begin{centering}
~ 
\par\end{centering}

\begin{centering}
~ 
\par\end{centering}

\begin{centering}
\includegraphics[scale=0.9]{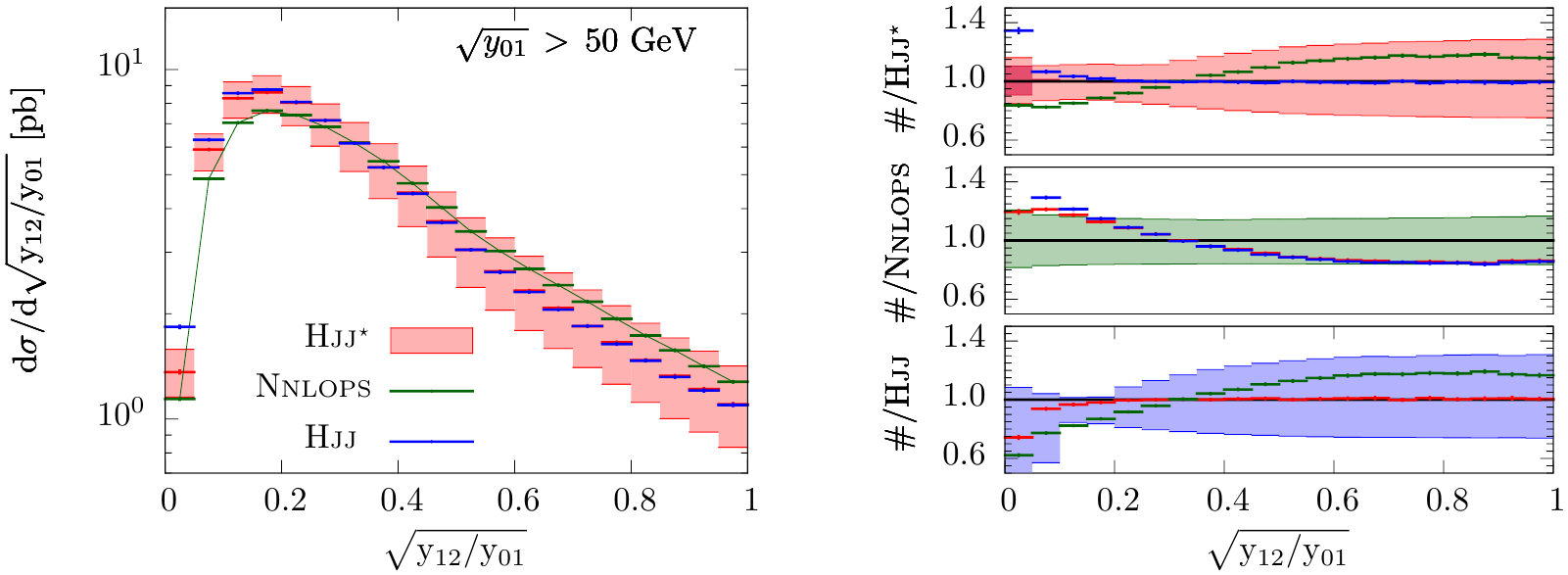} 
\par\end{centering}

\begin{centering}
~ 
\par\end{centering}

\begin{centering}
~ 
\par\end{centering}

\centering{}\includegraphics[scale=0.9]{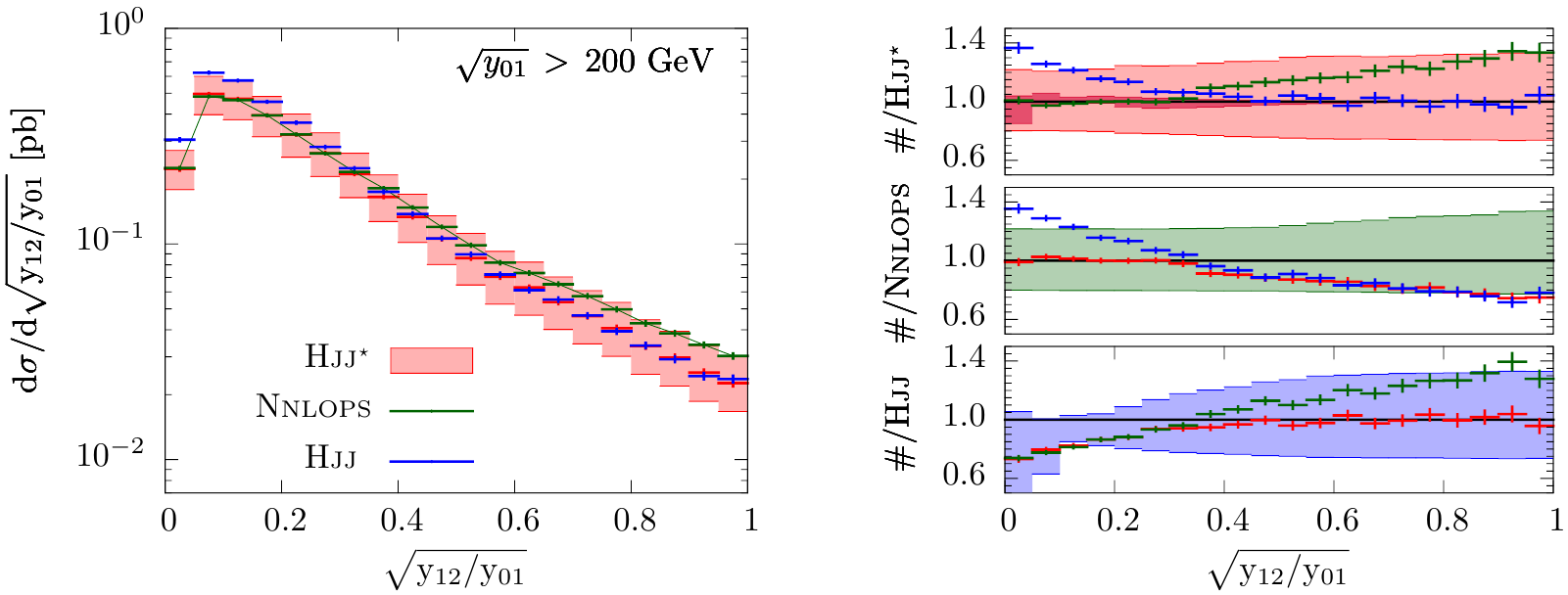}\protect\caption{\label{fig:sqrt-y12-over-y01-with-cuts-on-sqrt-y01}Ratios of differential
jet rates, $\sqrt{y_{12}/y_{01}}$, defined according to the $k_{{\scriptscriptstyle \mathrm{T}}}$-jet
algorithm with jet radius parameter $R=1$, and with cuts of $10$,
$50$ and $200$ GeV imposed on $\sqrt{y_{01}}$.}
\end{figure}

We conclude the presentation of results on jet rates with the $\sqrt{y_{12}/y_{01}}$
distributions in fig.~\ref{fig:sqrt-y12-over-y01-with-cuts-on-sqrt-y01}.
The latter quantity is precisely that which our \noun{Minlo }improvement
procedure directly modifies, in order to achieve agreement with the
inclusive $\Phi_{{\scriptscriptstyle \mathcal{B}\mathrm{J}}}$ distribution
of the \noun{Nnlops (}see again sects\noun{.~\ref{sub:Variation-on-MinloPrime}
}and\noun{ \ref{sec:Merging-without-a-merging-scale-three-units}).}
The three $\sqrt{y_{12}/y_{01}}$ plots in fig.~\ref{fig:sqrt-y12-over-y01-with-cuts-on-sqrt-y01}
are also in rough correspondence with those for $\log_{10}\sqrt{y_{12}}$
in figs.~\ref{fig:log10-sqrt-y01-and-log10-sqrt-y12} and \ref{fig:log10-sqrt-y12-and-y01-gt-some-cuts}.
Indeed, the arrangement of the three predictions relative to one another
in fig.~\ref{fig:log10-sqrt-y12-and-y01-gt-some-cuts}, for the $\sqrt{y_{01}}>50\mbox{ GeV}$
and $\sqrt{y_{01}}>200\mbox{ GeV}$ cuts, is essentially the same
as that which one finds for the same $\sqrt{y_{01}}$ cuts applied
to the $\sqrt{y_{12}/y_{01}}$ distribution in fig.~\ref{fig:sqrt-y12-over-y01-with-cuts-on-sqrt-y01}.
This correspondence is to expected, if one assumes that the bulk of
events making the distributions in both cases ($\log_{10}\sqrt{y_{12}}$
and $\sqrt{y_{12}/y_{01}}$) is dominated by those having $\sqrt{y_{01}}$
close to the cut; this assumption is reasonably fair, given that $\sqrt{y_{01}}$
falls off quickly, like the leading jet $p_{{\scriptscriptstyle \mathrm{T}}}$
spectrum. Crudely speaking, this makes the denominator of $\sqrt{y_{12}/y_{01}}$
a constant and, by implication, $\sqrt{y_{12}/y_{01}}$ tends to look
the same as scaled a plot of $\sqrt{y_{12}}$. We also consider that
the $\sqrt{y_{12}/y_{01}}$ distribution for $\sqrt{y_{01}}>10\mbox{ GeV}$
bears a considerable resemblance to that of the inclusive $\log_{10}\sqrt{y_{12}}$
one in fig.~\ref{fig:log10-sqrt-y01-and-log10-sqrt-y12}. This makes
sense on the basis that the $\sqrt{y_{01}}$ cut on the former distribution
is loose to the point of being no cut at all. This being the case,
we refer back to our discussion on the features of the aforementioned
$\log_{10}\sqrt{y_{12}}$ plots, for explanation of the structures
in the $\sqrt{y_{12}/y_{01}}$ ones. 

\subsubsection*{Higgs-jet and dijet correlations}

In this subsection we move to check observables more sensitive directional
correlations between the Higgs boson and jets in the event. Such variables
are routinely encountered in experimental analysis relating to Higgs
production via vector boson fusion. In leaving the jet rate variables
behind, we return again to define all jets according to the anti-$k_{t}$
clustering algorithm with radius parameter $R=0.4$, for all of the
remaining numerical results in this work.

\begin{figure}[htbp]
\begin{centering}
\includegraphics[scale=0.9]{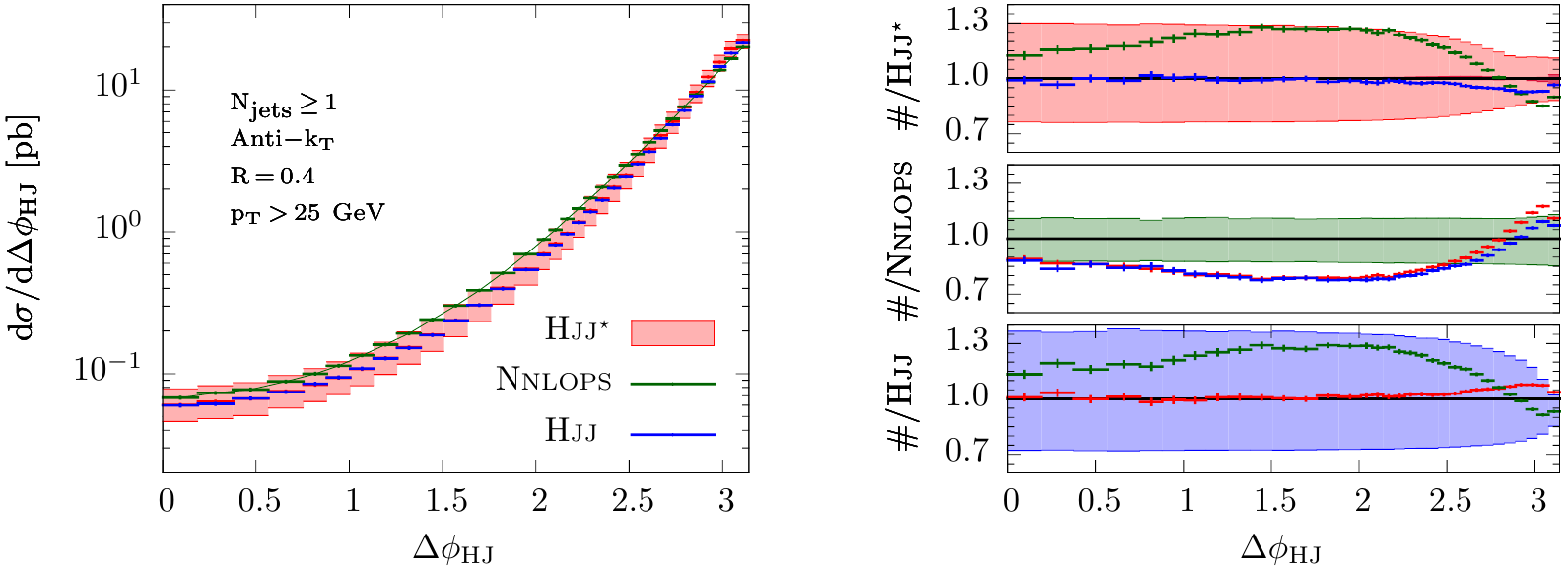}
\par\end{centering}

\protect\caption{\label{fig:Delta-Phi-HJ}Azimuthal angle between the Higgs boson and
leading jet, with jets defined according to the anti-$k_{t}$ algorithm
with a $25$ GeV transverse momentum threshold and radius parameter
$R=0.4$. }
\end{figure}

We start with the azimuthal separation of the Higgs boson and the
leading jet, $\Delta\phi_{{\scriptscriptstyle \mathrm{HJ}}}$, for
events containing at least one jet, in fig.~\ref{fig:Delta-Phi-HJ}.
The region $\Delta\phi_{{\scriptscriptstyle \mathrm{HJ}}}\approx\pi$
is dominated by configurations consisting of a hard underlying Higgs-plus-one
jet configuration, accompanied by additional soft radiations. Decreasing
$\Delta\phi_{{\scriptscriptstyle \mathrm{HJ}}}$ implies an increased
amount of radiation beyond that in the hard underlying Higgs-plus-one
jet configuration (to balance momentum in the transverse plane). Indeed,
if we assume that this extra radiation is collimated into a single
would-be jet, then already in the vicinity of $\Delta\phi_{{\scriptscriptstyle \mathrm{HJ}}}\approx2.1$
the distribution is becoming dominated by Mercedes-star configurations
of the Higgs, jet and the would-be jet, as well as others involving
yet greater angular separation of the leading jet and would-be jet.
Bearing in mind the latter point, the near perfect agreement of the
\noun{Hjj-Minlo }and $\noun{Hjj}^{\star}$ predictions for $\Delta\phi_{{\scriptscriptstyle \mathrm{HJ}}}\lesssim2.1$
is expected and desired; both being NLO accurate in the description
of 2-jet observables. In the region $\Delta\phi_{{\scriptscriptstyle \mathrm{HJ}}}>2.1$
we see the $\noun{Hjj}^{\star}$ result gently lifts off the \noun{Hjj-Minlo
}one. This lift-off is qualitatively expected, on the basis that the
integral of this distribution must equal the inclusive 1-jet cross
section, with a 25 GeV jet transverse momentum cut, and we know that
the inclusive 1-jet cross section from $\noun{Hjj}^{\star}$ (and
to a lesser extent \noun{Nnlops) }exhibits a 5\% enhancement over
that of \noun{Hjj-Minlo} (see again fig.~\ref{fig:Inclusive-jet-cross-section-akT-R-eq-0.4}).

\begin{figure}[htbp]
\begin{centering}
\includegraphics[scale=0.9]{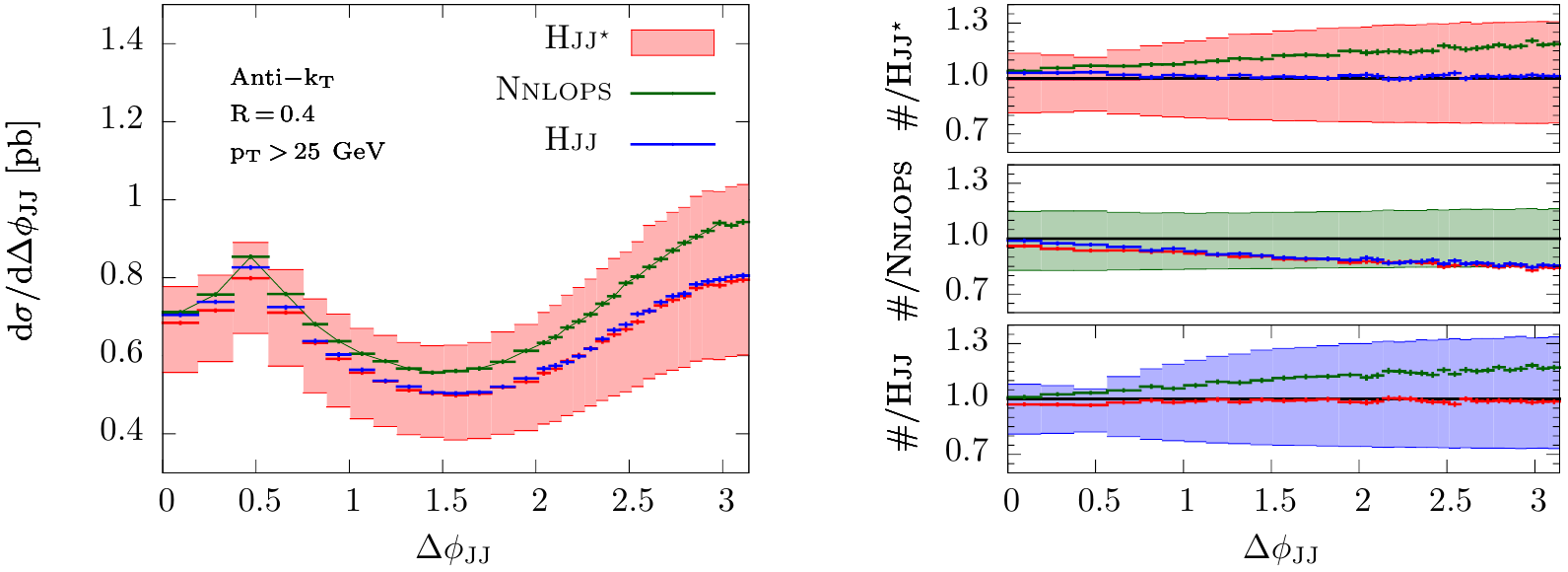} 
\par\end{centering}

\begin{centering}
~ 
\par\end{centering}

\centering{}\includegraphics[scale=0.9]{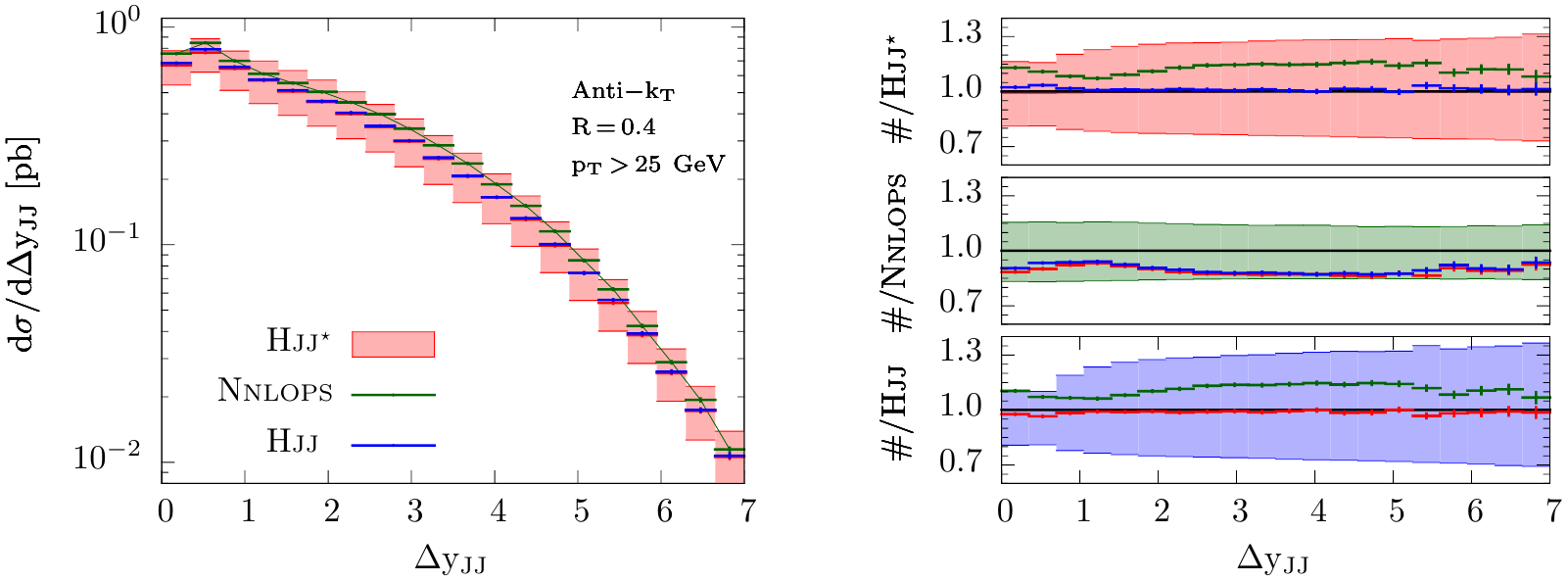}\protect\caption{\label{fig:DeltaPhiJJ-DeltaYJJ}Azimuthal (top) and rapidity (bottom)
separation between the two leading jets, with jets defined according
to the anti-$k_{t}$ algorithm with a $25$ GeV transverse momentum
threshold and radius parameter $R=0.4$. }
\end{figure}

\begin{figure}[htbp]
\centering{}\includegraphics[scale=0.9]{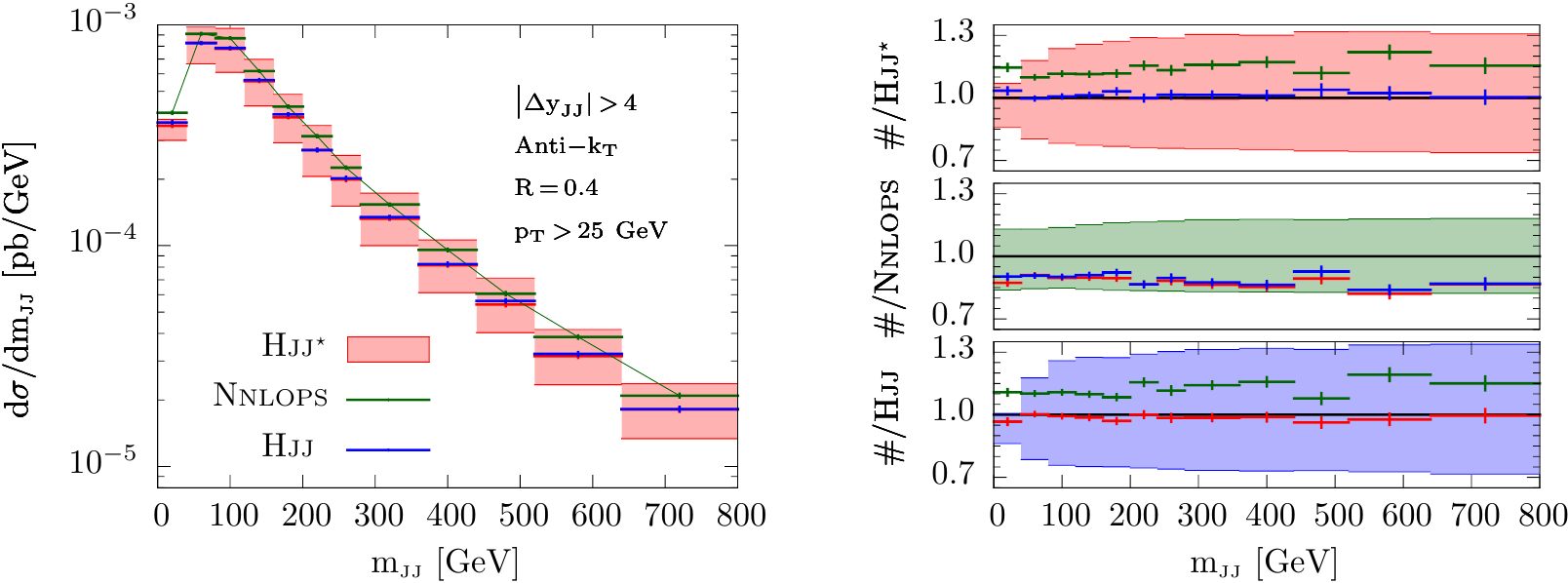}\protect\caption{\label{fig:mJJ-and-mJJ-with-DeltaYJJ-gt-4}In this plot we show the
invariant mass of the two leading jets, for events in which they have
a rapidity separation greater than four. The jets have been defined
according to the anti-$k_{t}$ clustering algorithm, with radius parameter
$R=0.4$, and with a transverse momentum threshold of 25 GeV.}
\end{figure}

In figure \ref{fig:DeltaPhiJJ-DeltaYJJ} we display the azimuthal
separation of the two leading jets in the uppermost plot and their
rapidity separation in the lower one. In figure \ref{fig:mJJ-and-mJJ-with-DeltaYJJ-gt-4}
we have further plotted the invariant mass of the two leading jets,
for events in which they are separated by at least four units of rapidity.
All of these distributions demand the presence of at least two jets
in the final state. From the analysis of our foregoing results, we
understand that for a global jet $p_{{\scriptscriptstyle \mathrm{T}}}$
threshold of 25 GeV, we can expect that 2-jet inclusive observables,
such as these, are dominated by events with no strong hierarchy of
scales $y_{12}\ll y_{01}$. Consequently, we expect, and we find,
that our corrective procedure has no effect, with the $\noun{Hjj}^{\star}$
and \noun{Hjj-Minlo }results being indistinguishable from one another
throughout. This is again our desired behaviour given that the \noun{Hjj-Minlo
}prediction is nominally NLO accurate for these observables, while
the \noun{Nnlops }is similarly just LO. Lastly, we add that the same
conclusions hold for the $m_{{\scriptscriptstyle \mathrm{JJ}}}$ distribution
when the $\left|\Delta\mathrm{y}_{{\scriptscriptstyle \mathrm{JJ}}}\right|>4$
rapidity separation cut is not imposed, in particular, the $\noun{Hjj}^{\star}$
and \noun{Hjj-Minlo }results remain indistinguishable, with the \noun{Nnlops
}continuing to exhibit the same relative discrepancy (albeit within
a smaller scale uncertainty band).

\subsubsection*{Jet binned Higgs boson transverse momentum distribution}

\begin{figure}[htbp]
\begin{centering}
\includegraphics[scale=0.9]{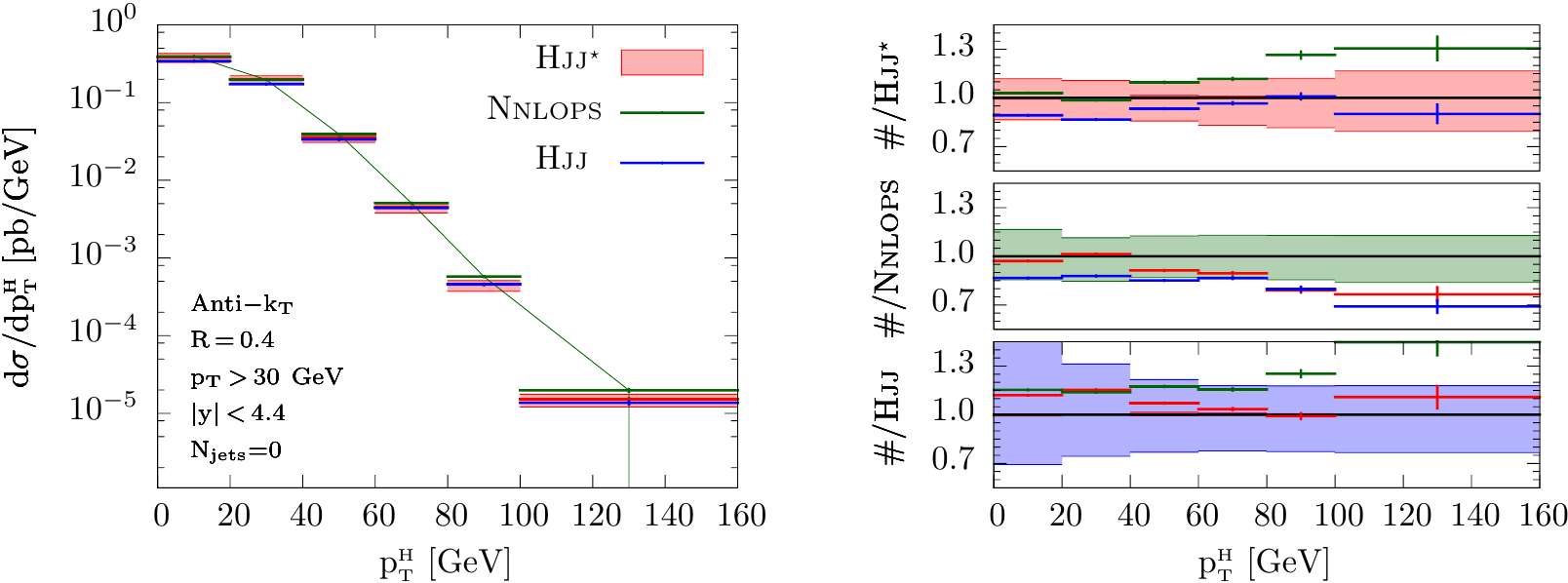}
\par\end{centering}

\begin{centering}
~
\par\end{centering}

\begin{centering}
\includegraphics[scale=0.9]{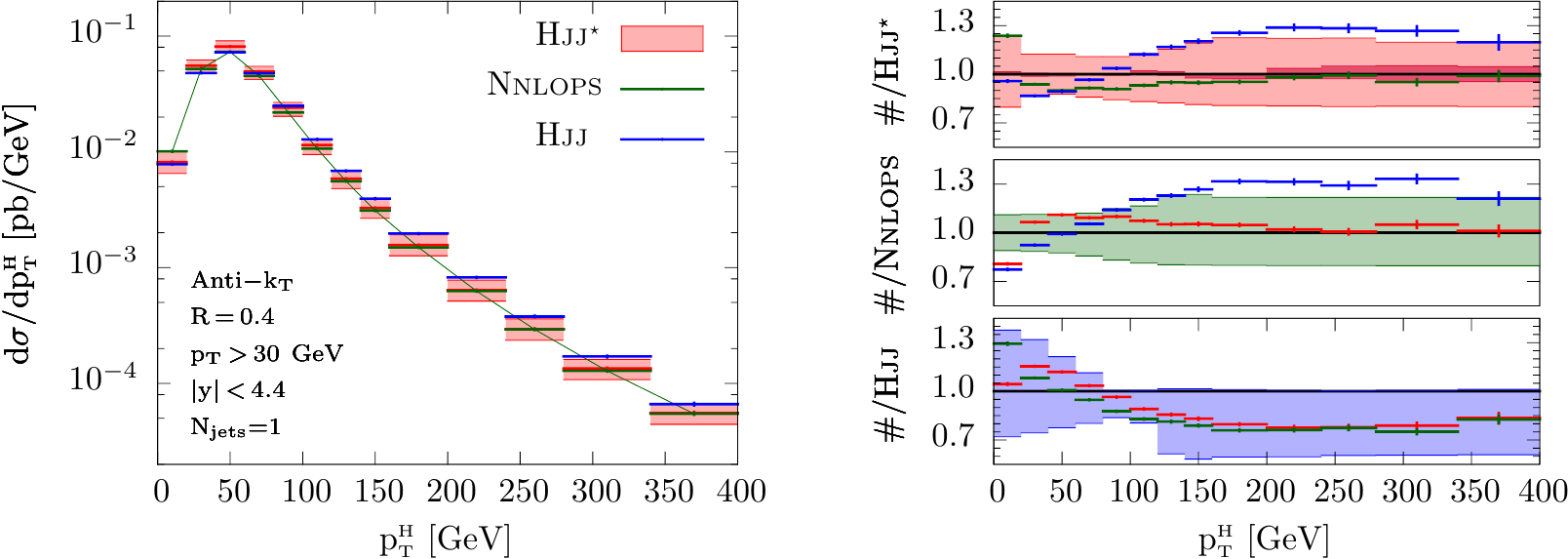}
\par\end{centering}

\protect\caption{\label{fig:ATLAS-Higgs-pT-in-0j-and-1j-bins}In the upper plot we
show the transverse momentum distribution of the Higgs boson in 0-jet
events. Jets are here constructed according to the anti-$k_{t}$ clustering
algorithm, for a radius parameter $R=0.4$. Jets are required to have
transverse momentum $p_{{\scriptscriptstyle \mathrm{T}}}\ge30$ GeV
and rapidity $\left|\mathrm{y}\right|\le4.4$. The corresponding distribution
in the case of 1-jet events is shown below.}
\end{figure}

\begin{figure}[htbp]
\begin{centering}
\includegraphics[scale=0.9]{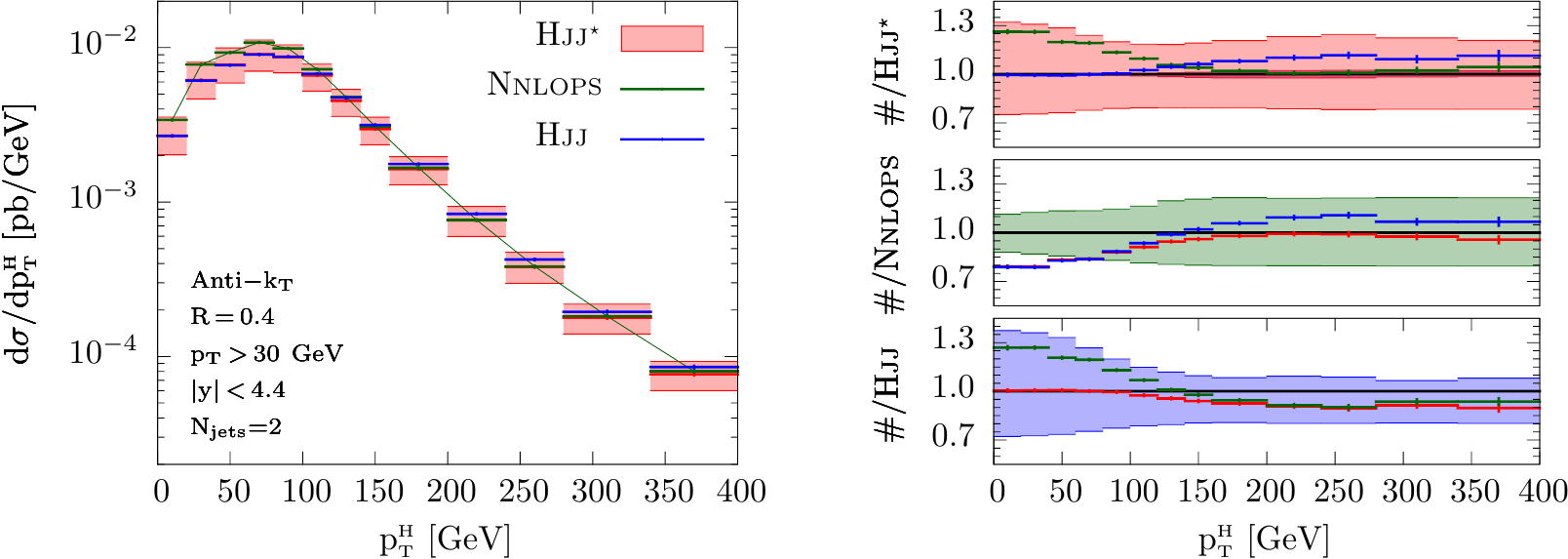}
\par\end{centering}

\begin{centering}
~
\par\end{centering}

\centering{}\includegraphics[scale=0.9]{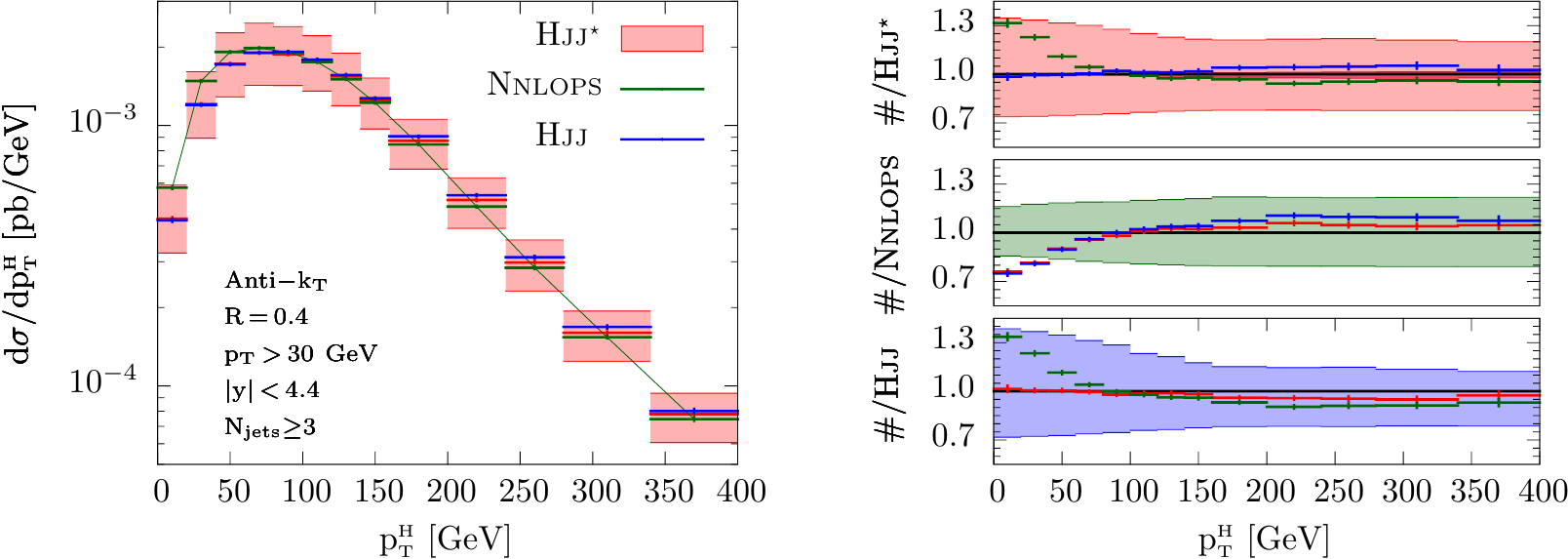}\protect\caption{\label{fig:ATLAS-Higgs-pT-in-2j-and-ge-3j-bins}In the upper plot
we show the transverse momentum distribution of the Higgs boson in
2-jet events. Jets are here constructed according to the anti-$k_{t}$
clustering algorithm, for a radius parameter $R=0.4$. Jets are required
to have transverse momentum $p_{{\scriptscriptstyle \mathrm{T}}}\ge30$
GeV and rapidity $\left|\mathrm{y}\right|\le4.4$. The corresponding
distribution in the case of $\ge3$-jet events is shown underneath.}
\end{figure}

The final results we present, in figs.~\ref{fig:ATLAS-Higgs-pT-in-0j-and-1j-bins}-\ref{fig:ATLAS-Higgs-pT-in-2j-and-ge-3j-bins},
are of the Higgs boson transverse momentum spectrum, in events with
a given exclusive jet multiplicity. The jets in question were defined
according to the anti-$k_{t}$ jet algorithm with $R=0.4$ and a jet
transverse momentum threshold of 30 GeV.

Figure~\ref{fig:ATLAS-Higgs-pT-in-0j-and-1j-bins} shows the Higgs
boson's transverse momentum in 0-jet events in the upper plot and
in 1-jet events in the lower one. In the case of the 0-jet events
we see the $\noun{Hjj}^{\star}$ prediction aligns itself with the
superior (NNLO) \noun{Nnlops }result in the low transverse momentum
domain. On the other hand, as soon as the Higgs boson has reaches
a transverse momentum in excess of that of the jet defining $p_{{\scriptscriptstyle \mathrm{T}}}$
threshold, we see that $\noun{Hjj}^{\star}$ quickly comes into agreement
with \noun{Hjj-Minlo}. This behaviour is also as intended, since in
the latter region, momentum conservation combined with the requirement
that there be no resolved jets, dictates that the Higgs boson must
be considered as recoiling against multiple hard radiations which
are widely separated in angle from one another, all with $p_{{\scriptscriptstyle \mathrm{T}}}<30\mbox{ \textrm{GeV}}$.
The latter class of `hedgehog' configurations is described more accurately
by the higher multiplicity \noun{Hjj-Minlo }simulation.

Turning to the Higgs transverse momentum in the 1-jet events, we see
the results we naively expect in the region $p_{{\scriptscriptstyle \mathrm{T}}}^{{\scriptscriptstyle \mathrm{H}}}>100\text{ GeV}$,
with \noun{Nnlops }and $\noun{Hjj}^{\star}$ in very good agreement.
In the region surrounding the peak of the distribution at $p_{{\scriptscriptstyle \mathrm{T}}}^{{\scriptscriptstyle \mathrm{H}}}\sim50\text{ GeV}$,
$\noun{Hjj}^{\star}$ continues\noun{ }to agree well with \noun{Hjj-Minlo},
but not quite as nicely as before. The slight excess of the\noun{
}$\noun{Hjj}^{\star}$ prediction over the \noun{Nnlops }around this
peak follows the same explanation as for the similarly sized enhancement
of the exclusive 1-jet cross section of the former over the latter,
in the discussion surrounding fig.~\ref{fig:Exclusive-jet-cross-section-akT-R-eq-0.4}.
There we explained that our correction procedure led to an enhanced
1-jet exclusive cross section, by acting to recover the inclusive
1-jet cross section of the \noun{Nnlops, }while maintaining the 2-jet
inclusive cross section of \noun{Hjj-Minlo}; since the 2-jet inclusive
cross section of \noun{Hjj-Minlo} was low with respect to that of
the \noun{Nnlops}, the $\noun{Hjj}^{\star}$ 1-jet exclusive cross
section therefore had to be high. Remarkably, on the other hand, we
note that for the lowest bin in the $\mathrm{N_{jets}}=1$ $p_{{\scriptscriptstyle \mathrm{T}}}^{{\scriptscriptstyle \mathrm{H}}}$
plot, it is in fact natural and correct that the $\noun{Hjj}^{\star}$
distribution is found to be in complete agreement with \noun{Hjj-Minlo},
for in that region the recoil of the leading jet can no longer be
balanced by the Higgs boson, and instead extra radiation must be present
to this end. 

Lastly, we look to the Higgs boson transverse momentum distributions
in the exclusive 2-jet events and inclusive 3-jet events, in the upper
and lower plots of fig.~\ref{fig:ATLAS-Higgs-pT-in-2j-and-ge-3j-bins}.
For both the exclusive 2-jet and inclusive 3-jet $p_{{\scriptscriptstyle \mathrm{T}}}^{{\scriptscriptstyle \mathrm{H}}}$
spectra, we see that $\noun{Hjj}^{\star}$ agrees perfectly with the \noun{Hjj-Minlo
}generator in the low transverse momentum domain. In the high transverse
momentum regions we find that all three predictions agree rather well
with one another. In the exclusive 2-jet case at high $p_{{\scriptscriptstyle \mathrm{T}}}^{{\scriptscriptstyle \mathrm{H}}}$,
we can, however, clearly see that the correction procedure has driven
$\noun{Hjj}^{\star}$ to reproduce \noun{Nnlops }rather than \noun{Hjj-Minlo}.
We believe that this too is again the desired result and that the
\noun{Nnlops }prediction is superior to that of \noun{Hjj-Minlo} in
this particular kinematic domain. This assumption is based on the
fact that in high $p_{{\scriptscriptstyle \mathrm{T}}}^{{\scriptscriptstyle \mathrm{H}}}$
$\mathrm{N}_{\mathrm{jets}}=2$ events, the leading jet has a transverse
momentum which is bounded from below by approximately half that of
the Higgs boson, moreover, it will tend to have a transverse momentum
close to that of the Higgs. Thus, the $p_{{\scriptscriptstyle \mathrm{T}}}^{{\scriptscriptstyle \mathrm{H}}}\sim\mathrm{200}\text{ GeV}$
region of the $\mathrm{N}_{\mathrm{jets}}=2$ events $p_{{\scriptscriptstyle \mathrm{T}}}^{{\scriptscriptstyle \mathrm{H}}}$
spectrum, will be dominated by events with $\sqrt{y_{01}}\sim200\mbox{ GeV}$.
Referring back to the $\log_{10}\sqrt{y_{12}}$ plot of fig.~\ref{fig:log10-sqrt-y12-and-y01-gt-some-cuts},
with the $\sqrt{y_{01}}>200\text{ GeV}$ cut imposed, we can then
understand that nearly all such events will come with a second $p_{{\scriptscriptstyle \mathrm{T}}}>30\mbox{ GeV}$
jet `for free', i.e.~the $p_{{\scriptscriptstyle \mathrm{T}}}^{{\scriptscriptstyle \mathrm{H}}}$
spectrum with $\mathrm{N}_{\mathrm{jets}}=2$,
for $p_{{\scriptscriptstyle \mathrm{T}}}^{{\scriptscriptstyle \mathrm{H}}}\gtrsim\mathrm{200}\text{ GeV}$,
becomes essentially the $\mathrm{N}_{\mathrm{jets}}\ge1$ $p_{{\scriptscriptstyle \mathrm{T}}}^{{\scriptscriptstyle \mathrm{H}}}$
distribution. Hence, we believe the \noun{Nnlops}/$\noun{Hjj}^{\star}$
prediction to be more accurate than \noun{Hjj-Minlo} in this case.

\section{Conclusion\label{sec:Conclusion}}

In this work we have revisited the \noun{Minlo} and \noun{Minlo}$^{\prime}$
frameworks. Our main aim has been to address the issue of how to extend
the accuracy of existing \noun{Minlo} simulations up to that of \noun{Minlo$^{\prime}$}.
We focused on \noun{Minlo} simulations of $\mathcal{B}$+2-jet production
($\mathcal{B}\noun{jj}$), with $\mathcal{B}$ a colourless system,
as prototypical `complex processes', however, our ideas are more widely
applicable. For the latter generators, which are NLO accurate in the
description of $\mathcal{B}$+2-jet ($\mathcal{B}\noun{jj}$) inclusive
observables, promotion to \noun{Minlo$^{\prime}$} accuracy amounts
to the requirement that $\mathcal{B}$+1-jet ($\mathcal{B}\noun{j}$)
inclusive quantities also be recovered at NLO. We have also considered
how to go further in this framework, and obtain (N)NLO accuracy for
inclusive $\mathcal{B}$-production observables from $\mathcal{B}\noun{jj}$-\noun{Minlo}.

In existing \noun{Minlo$^{\prime}$ }simulations the two-fold
NLO accuracy is obtained by constructing a Sudakov form factor which
returns the relevant inclusive NLO $\mathcal{B}m\noun{j}$ cross sections, differential in the
underlying Born phase space, starting from NLO $\mathcal{B}n\noun{j}$ cross sections, with $m = n-1$.
While the form factors are
explicitly constructed from high-order resummation ingredients, the
accuracy of the resummation in the resulting \noun{Minlo$^{\prime}$}
simulation is the same as in the initial \noun{Minlo} one. The net
effect of the modifications is to carefully unitarize the inclusive
cross section, differential in the Born kinematics. This is very similar to
the working of \noun{Powheg} Sudakov form factors in \noun{Nlops
}matching.\footnote{Indeed, in formulating the original \noun{Minlo$^{\prime}$}
method, the \noun{Minlo} cross section was initially cast in the form
of the \noun{Powheg} hardest emission cross section.} The latter contain NLL and even power suppressed terms
in the exponent in order to recover NLO accuracy, despite being,
in general, just LL accurate. 

We started in this work by trying to clarify to what extent the $\mathcal{B}\noun{jj}$-\noun{Minlo}
simulations already achieve the aforementioned \noun{Minlo$^{\prime}$}
accuracy, and to see how to improve them in this direction by better
understanding the relevant resummation.
We used the \noun{Caesar }formalism to derive a \noun{$\mathrm{NNLL}_{\sigma}$
}resummation formula for the $0\rightarrow1$-$k_{t}$-jet rate $y_{01}$
and, separately, the $1\rightarrow2$ jet rate $y_{12}$; including
leading multiple emission corrections in the exclusive $k_{t}$-algorithm.
The \noun{$\mathrm{NNLL}_{\sigma}$}
formula reveals existing $\mathcal{B}\noun{jj}$-\noun{Minlo }simulations
miss $\mathrm{NLL}_{\sigma}$ terms in their $y_{12}$ Sudakov form
factor exponents, associated to soft-wide angle gluon emission from
the underlying $\mathcal{B}\noun{j}$ state
in the kinematic domain where \noun{Caesar} is valid, $y_{01}\gtrsim m_{{\scriptscriptstyle \mathrm{H}}}^{2}$.
The Sudakov form factors in existing $\mathcal{B}\noun{jj}$-\noun{Minlo}
codes also miss the \noun{$\mathrm{NNLL}_{\sigma}$} multiple emission
corrections in the resummation formula. With these clarifications
one could formally improve $\mathcal{B}\noun{jj}$-\noun{Minlo} codes
towards $\mathcal{B}\noun{jj}$-\noun{Minlo}$^{\prime}$, implementing
improved Sudakov form factors to that end.

We derived the
fixed order expansion of the $\mathrm{NNLL}_{\sigma}$ \noun{Caesar}
formula and from this we showed how our \noun{Minlo }procedure applied
to the $\mathcal{B}$\noun{j(j) }NLO computations returns a matched,
resummed, NLO accurate jet resolution spectrum. In doing so we also
assumed the presence of unknown $\mathrm{N^{3}LL}_{\sigma}$ and $\mathrm{NNLL}_{\sigma}$
terms in our initial fixed order expansion formula; the former owe
to the limitations of our initial resummation formula truncated at
NLO, while we allowed for the presence of the latter in anticipation
of a breakdown of the \noun{Caesar} framework in considering the region
$y_{01}\ll m_{{\scriptscriptstyle \mathrm{H}}}^{2}$ later on. Upon integration
over the \noun{Minlo} jet resolution spectrum, we determine how the
distribution of the Born kinematics differs from that of conventional
NLO on account of those unknown terms, which were tracked and contained.
We demonstrate how such unwanted terms are removed in the original
\noun{Minlo$^{\prime}$} approach and, based on that, we introduced an
approximately equivalent procedure, which promotes the  \noun{Minlo}
simulations to \noun{Minlo$^{\prime}$} accuracy without the need of
analytic expressions for higher order terms in the Sudakov form factor
--- terms which are in general unknown.
To this end we solve the condition that the missing higher order Sudakov
contribution must be such that \noun{$\mathcal{B}m$j-Minlo} recover the
(N)NLO results for \noun{$\mathcal{B}n$j} Born kinematics, when suitably
integrated, to determine a numerical approximation to the former. The
Sudakov correction so-derived renders \noun{$\mathcal{B}m$j-Minlo}
(N)NLO accurate for \noun{$\mathcal{B}n$j} inclusive observables, while
maintaining NLO accuracy for \noun{$\mathcal{B}m$j} ones.

This procedure is a useful extension of the \noun{Minlo }framework.
Despite the fact that \noun{Minlo}$^{\prime}$ is proven to recover
conventional NLO results for inclusive quantities up to NNLO ambiguities,
in the original \noun{Minlo}$^{\prime}$ paper modest numerical disagreements
were found, between the predictions of $\mathcal{B}\noun{j}$\noun{-Minlo}$^{\prime}$
and conventional NLO, for W and Z production. With the suggested extension
of the \noun{Minlo }method in this paper, by construction, the predictions
for fully inclusive observables become essentially identical to the
standard NLO predictions.

A more important benefit of the proposed extension is that one can
begin to make \noun{Minlo}$^{\prime}$ simulations, merging two units of
multiplicity at NLO, or one NLO and one NNLO, without a merging scale,
for complex processes, provided the resummation in the \noun{Minlo}
program entering the correction procedure is $\mathrm{NLL}_{\sigma}$
accurate. The main limitation to be faced in practice will occur for
processes in which the dimensionality of the underlying Born phase space
is high; in this case the numerical determination of the
$\delta\left(\Phi\right)$ term (eq.~\ref{eq:sect26-first-delta-defn})
which corrects the Sudakov form factor will become challenging. We
should be clear though that, broadly speaking, the main objective in
this work has been to demonstrate that one can use precision inclusive
cross sections to determine approximate corrections
to the Sudakov form factor of \noun{Minlo} simulations, such that they
return the correct (N)NLO Born kinematics; this in turn leads to (N)NLO
accuracy for arbitrary infrared-safe observables which nominally
receive their leading contributions from parton multiplicities lower
than that included in the initial \noun{Minlo} simulation. The precise
way we have done this, discussed in the second part of section
\ref{sub:Removal-of-spurious-terms}, and our implementation of it, is
undoubtedly just one option out of many, and can be simply considered
as a practical, working, proof-of-concept at this point. Even in the
worst case, should the dimensionality of the Born kinematics become too
much, the method here still has the potential to greatly improve results,
in approximating the full $\delta\left(\Phi\right)$ term of
eq.~\ref{eq:sect26-first-delta-defn} by a carefully dimensionally reduced
version of it.

The loose requirement on the accuracy of the \noun{Minlo} resummation
has an additional useful property: the method can be applied in regions
of phase-space where the underlying Born itself has disparate kinematic
scales associated with it; in such regions achieving high accuracy
resummation is currently a formidable challenge. In particular, our
improvement procedure remains valid for $\mathcal{B}\noun{jj}$-\noun{Minlo}
in regions of phase-space where $y_{01}$ is much smaller than $m_{{\scriptscriptstyle \mathcal{B}}}^{2}$,
where both large logarithms of $m_{{\scriptscriptstyle \mathcal{B}}}^{2}/y_{01}$
and $y_{01}/y_{12}$ require resummation.
To this end we first argued that the \noun{Caesar }$y_{01}/y_{12}$
resummation remains \noun{$\mathrm{NLL}_{\sigma}$ }accurate in the
region $y_{01}\ll m_{{\scriptscriptstyle \mathcal{B}}}^{2}$. Our
argument is largely based on the finding that the \noun{Caesar }Sudakov
form factor for this variable is the same as that prescribed by the
coherent branching formalism at $\mathrm{NLL}_{\sigma}$ (except for
a subset of the soft-wide-angle radiation, which is beyond the accuracy
of the latter). We also note that the leading $\mathrm{NLL}_{\sigma}$
terms in the fixed order expansion of the \noun{Caesar} resummation
formula are obtained by integrating a \emph{single} soft/collinear
emission over the region $y_{12}<y_{01}$, i.e.~they are the same
whether $y_{01}\ll m_{{\scriptscriptstyle \mathcal{B}}}^{2}$, or
not. Taking the latter two points together, it follows that the \noun{Caesar}
$y_{01}/y_{12}$ resummation must hold at the $\mathrm{NLL}_{\sigma}$
level, even in the difficult regions.

The fact that $y_{12}$ resummation
works in all regions of phase-space implies one can do a `nested'
\noun{Minlo}$^{\prime}$ simulation with the help of our proposed
extension: instead of using unitarity such that partially integrated
$\mathcal{B}\noun{jj}$ distributions become equal to the NLO $\mathcal{B}\noun{j}$
distributions, we train them on \noun{Minlo}$^{\prime}$ $\mathcal{B}\noun{j}$
distributions. This makes them also NLO correct in inclusive $\mathcal{B}$
distributions, since the latter are obtained on suitably integrating
over radiation in $\mathcal{B}\noun{j}$-\noun{Minlo}$^{\prime}$.
In this way the extension of the \noun{Minlo}$^{\prime}$ method to
the merging of more than two multiplicities is realised.

As a feasibility study for the latter, we have applied our correction
procedure to \noun{Hjj-Minlo}. We start from a LHE event file for
the \noun{Hjj-Minlo} simulation and reweight the events such that
distributions differential in $\Phi_{{\scriptscriptstyle \mathrm{HJ}}}$
become equal to the existing NNLO-improved \noun{Hj-Minlo$^{\prime}$}
calculation, without hampering the formal accuracy of the \noun{Hjj-Minlo}
simulation. We therefore made predictions that are NNLO accurate for
inclusive Higgs boson production, and NLO accurate for \noun{Hj} and
\noun{Hjj} observables. Since the LHE events are ultimately generated
according to the \noun{Powheg Nlops }matching procedure they may,
of course, be showered in the usual way. Our numerical results are
very encouraging: for inclusive observables in \noun{H} and \noun{Hj}
production, we recover the results of the \noun{Hnnlops} simulation,
while for observables in which $y_{12}\sim y_{01}$
we recover the \noun{Hjj-Minlo} predictions, with smooth interpolation
between them.

There is ample freedom in the functional form of the reweighting factor
which is formally beyond the accuracy of the method. We have explored
(some of) its dependence and seen essentially no visible effects of it
in the many distributions we have examined. The distribution which
displayed most sensitivity to this ambiguity was, unsurprisingly,
$\log_{10}\sqrt{y_{12}}$. Even for this variable, the sensitivity is
located in the deep Sudakov region, mostly well-below the Sudakov peak,
a region which is anyway very sensitive to higher order resummation and
non-perturbative effects.

There are a number of aspects of this work which can be explored further
and refined. It is clear, for example, that it is interesting to consider
our approach in application to other processes. We have shown the method
can work well for a process with 3 final-state particles (\noun{Hjj}),
thus it seems reasonable to expect similar quality results in application
to processes with equal multiplicity, e.g. trijet, and jet-associated
single-top/top-pair production (with some approximation in the handling
of top decays). In fact these processes are in one sense less demanding
than that which we demonstrated, in so far as we dealt with a process
for which two jets could become unresolved, moreover, this was handled
while mapping onto an NNLO calculation of Higgs production. On the other
hand, for high multiplicity processes like VBF Higgs-plus-3-jet production,
the dimensionality of the phase space  combined with the problems to be
anticipated in obtaining high statistics for determining
$\delta\left(\Phi\right)$, would likely prove too cumbersome in practice,
at least for our proof-of-concept implementation. Nevertheless, in the
absence of a better alternative, we would still advocate trying the latter
method in some approximate form, e.g. applying it on only a carefully
selected subset of the variables which parametrize the underlying Born
kinematics. Depending on the initial circumstances this may lead to very
desirable improvements.

Relatedly, on a technical level
it is worth considering a more sophisticated approach to our implementation,
e.g.~using an adaptive, optimised, grid parametrization procedure
for the underlying Born phase space, together with advanced interpolation
procedures for computing $\delta\left(\Phi\right)$. Having said that,
it is perhaps a good indicator of the potential of this approach,
that it appears to have worked remarkably well even with just a basic
implementation using hand-made rigid grids. From a theoretical perspective,
one may wish to consider improvements to the $\mathcal{B}\noun{jj}$-\noun{Minlo}$^{\prime}$
codes based on our comparison of their inherent resummation and that
of the \noun{Caesar }framework, such as inclusion of soft-wide-angle
($\Delta S_{1}$) and multiple emission ($\mathcal{F}_{2}$) terms
in the Sudakov form factors. Our numerical studies in this paper suggest
that these inclusions would be of really quite limited interest though.

A further investigation would be to consider
the effect of breaking the reweighting procedure for $\mathcal{B}\noun{jj}$-\noun{Minlo}
into two phases: in the first stage just the inclusive $\Phi_{{\scriptscriptstyle \mathcal{B}}}$
distribution of $\mathcal{B}\noun{jj}$-\noun{Minlo} is corrected
to that of $\mathcal{B}\noun{j}$-\noun{Minlo}$^{\prime}$, by adjusting
the $y_{01}$ distribution; in the second stage the procedure is applied
to the $\mathcal{B}\noun{jj}$-\noun{Minlo} output from the first
stage, in exactly the same way as set out in sects.~\ref{sec:Merging-without-a-merging-scale-three-units}-\ref{sec:Feasibility-study}.
At the $\mathrm{NLL}_{\sigma}$ level, there is no distinguishing
between the latter approach and that which we carried out. On the
other hand, it is clear that, at some level, the effective Sudakov
form factor correction that we derive for the $y_{01}/y_{12}$ resummation
will make up for what might better be considered as deficiencies in
the $y_{01}$ Sudakov form factor. Nevertheless, from our numerical
studies here, we expect that this change would only register much
like the $\rho$-parameter variations that we assessed, i.e.~we believe
it will only become visible in $y_{12}$ regions which are under poor
theoretical control (the deep Sudakov region). It is also important
not to over emphasise this point in view of the fact that the correction
procedure obtains the $\Phi_{{\scriptscriptstyle \mathcal{B}}}$,
$y_{01}$ and $\mathrm{y}_{{\scriptscriptstyle \mathrm{J}}}$ distributions
of $\mathcal{B}\noun{j}$-\noun{Minlo}$^{\prime}$, by construction,
in any case. Nevertheless, this alternative may prove advantageous in
other applications of the method.

\section*{Acknowledgements}

It's a pleasure to thank Andrea Banfi, Giulia
Zanderighi and Emanuele Re for helpful discussions. We also thank
Andrea, Giulia and Gavin Salam for a copy of their \noun{Caesar }program. 

\smallskip{}

\noindent KH wishes to thank CERN-TH for its kind hospitality during
the course of this work. KH also gratefully acknowledges the Mainz
Institute for Theoretical Physics (MITP) for its hospitality and support
while part of this work was carried out.

\smallskip{}

\noindent RF is supported by the Alexander von Humboldt Foundation, in
the framework of the Sofja Kovaleskaja Award Project ``Event
Simulation for the Large Hadron Collider at High Precision'', endowed
by the German Federal Ministry of Education and Research.

\smallskip{}

\noindent The authors acknowledge the use of the UCL Legion High Performance
Computing Facility (Legion@UCL), and associated support services,
in the completion of this work.

\smallskip{}

\noindent This work used the DiRAC Complexity system, operated by
the University of Leicester IT Services, which forms part of the STFC
DiRAC HPC Facility (www.dirac.ac.uk). This equipment is funded by
BIS National E-Infrastructure capital grant ST/K000373/1 and STFC
DiRAC Operations grant ST/K0003259/1. DiRAC is part of the National
E-Infrastructure.

\appendix

\section{Appendix}

\subsection{Derivation of \texorpdfstring{$\mathrm{LL}$/$\mathrm{NNLL}_{\sigma}$}{LL/NNLLo} jet resolution
spectra\label{sub:Derivation-of-NNLLsigma-jet-resolution-spectra}}

In this appendix we give details on how our general $\mathrm{NNLL}_{\sigma}$
jet resolution spectrum is arrived at from the results of refs.~\cite{Banfi:2004yd}
and \cite{Banfi:2010xy}.

In eqs.~3.14-3.17 of ref.~\cite{Banfi:2010xy} the $\mathrm{NLL}$
resummed cumulant cross section, matched to NLO, to yield also $\mathrm{NNLL}_{\sigma}$
accuracy is given as 
\begin{equation}
\frac{d\Sigma_{{\scriptscriptstyle \mathcal{R}}}\left(L\right)}{d\Phi}=\frac{d\sigma_{0}}{d\Phi}\,\left(1+\alpha_{{\scriptscriptstyle \mathrm{S}}}C_{1}\right)\, f\left(v\right)\,,\label{eq:app1-CAESAR-dSigmaR-NNLLsigma}
\end{equation}
with $d\sigma_{0}/d\Phi$ the leading order cross section fully differential
in the Born kinematics (denoted $\mathcal{B}$ in \cite{Banfi:2004yd,Banfi:2010xy}),
and with $f\left(v\right)$ encoding the resummation. The term $\alpha_{{\scriptscriptstyle \mathrm{S}}}C_{1}$
is the matching coefficient defined by (eq.~3.16 of ref.~\cite{Banfi:2010xy})
\begin{equation}
\alpha_{{\scriptscriptstyle \mathrm{S}}}C_{1}=\lim_{L\rightarrow\infty}\,\left(\frac{d\Sigma_{{\scriptscriptstyle \mathrm{NLO}}}\left(L\right)}{d\Phi}-\left.\frac{d\Sigma_{{\scriptscriptstyle \mathcal{R},1}}\left(L\right)}{d\Phi}\right|_{{\scriptscriptstyle \bar{\chi}_{1}\rightarrow0}}\right)\,/\,\frac{d\sigma_{0}}{d\Phi}\,,\label{eq:app1-CAESAR-aS-C1-matching-coefficient}
\end{equation}
where the $d\Sigma_{{\scriptscriptstyle \mathrm{NLO}}}\left(L\right)$
in our notation corresponds to $d\Sigma_{1}\left(v\right)$ of \cite{Banfi:2010xy},
and with our $d\Sigma_{{\scriptscriptstyle \mathcal{R},1}}\left(L\right)$
corresponding to $d\Sigma_{r,1}\left(v\right)$ of \cite{Banfi:2010xy}.
Thus the matching coefficient of \cite{Banfi:2010xy} is in our notation
$\alpha_{{\scriptscriptstyle \mathrm{S}}}C_{1}=\bar{\alpha}_{{\scriptscriptstyle \mathrm{S}}}\bar{\chi}_{1}\left(\Phi\right)$.
The function $f\left(v\right)$ is the main result of ref.~\cite{Banfi:2004yd}
(see eq.~3.6 therein) and is comprised as follows (taking $b_{\ell}=0$,
$a_{\ell}=2$, $d_{\ell}=g_{\ell}=1$, and hence $\bar{d}_{\ell}=1$,
as appropriate for the $k_{t}$-jet resolution variables considered in our work, namely, $V\left(\left\{ \tilde{p}\right\} ,k\right)=(k_{t}^{\left(\ell\right)}/Q)^{2}$):
\begin{eqnarray}
f\left(v\right) & = & \mathcal{F}\left(R^{\prime}\right)\, S\left(T\left(\frac{L}{2}\right)\right)\,\exp\left[-\sum_{\ell=1}^{n}\,\left[C_{\ell}r_{\ell}\left(L\right)+B_{\ell}C_{\ell}T\left(\frac{L}{2}\right)\right]\right]\,\prod_{\ell=1}^{n_{i}}\,\frac{q^{\left(\ell\right)}\left(x_{\ell},\mu_{{\scriptscriptstyle F}}^{2}v\right)}{q^{\left(\ell\right)}\left(x_{\ell},\mu_{{\scriptscriptstyle F}}^{2}\right)}.\label{eq:app1-CAESAR-fv-formula-i}
\end{eqnarray}
In ref.~\cite{Banfi:2004yd}, between eqs.~A~1.18 and A~1.19,
it is stated that for processes with less than four colour-charged
legs in the hard underlying Born kinematics, $S\left(t\right)=\exp\left[S_{1}t\right]$,
with $S_{1}$ as given in our eqs.~\ref{eq:sect22-n-eq-2-S1-defn},
\ref{eq:sect22-n-eq-3-S1-defn}.

From ref.~\cite{Banfi:2004yd} eqs.~2.21-2.22, (taking $b_{\ell}=0$,
$a_{\ell}=2$, $d_{\ell}=g_{\ell}=1$, and hence $\bar{d}_{\ell}=1$),
we also have 
\begin{equation}
r_{\ell}\left(L\right)=\int_{y}^{Q^{2}}\,\frac{dk_{t}^{2}}{k_{t}^{2}}\,\bar{\alpha}_{{\scriptscriptstyle \mathrm{S,CMW}}}\left(k_{t}^{2}\right)\,\ln\frac{Q^{2}}{k_{t}^{2}}\,,\qquad\mbox{and}\qquad T\left(\frac{L}{2}\right)=\int_{y}^{Q^{2}}\,\frac{dk_{t}^{2}}{k_{t}^{2}}\,2\bar{\alpha}_{{\scriptscriptstyle \mathrm{S}}}\left(k_{t}^{2}\right)\,,\label{eq:app1-CAESAR-rell-and-T-integral-forms}
\end{equation}
where $\bar{\alpha}_{{\scriptscriptstyle \mathrm{S,CMW}}}$ is the
so-called Bremsstrahlung (CMW) scheme for the strong coupling constant
(as written on pg.~17 in \cite{Banfi:2004yd}) 
\begin{equation}
\bar{\alpha}_{{\scriptscriptstyle \mathrm{S,CMW}}}=\bar{\alpha}_{{\scriptscriptstyle \mathrm{S,\overline{MS}}}}+K\bar{\alpha}_{{\scriptscriptstyle \mathrm{S}}}^{2}\,.\label{eq:app1-CAESAR-aS-CMW-defn}
\end{equation}
Inserting the expressions for $\bar{\alpha}_{{\scriptscriptstyle \mathrm{S,CMW}}}$,
$r_{\ell}\left(L\right)$, $T\left(\frac{L}{2}\right)$, $S\left(T\left(\frac{L}{2}\right)\right)$
into that for $f\left(v\right)$ gives, with no approximations, 
\begin{eqnarray}
f\left(v\right) & = & \mathcal{F}\left(R^{\prime}\right)\,\exp\left[-\int_{0}^{L}\, dL^{\prime}\,\bar{\alpha}_{{\scriptscriptstyle \mathrm{S}}}^{2}\left(y^{\prime}\right)\,2G_{12}\,\left[\,4\mathcal{F}_{2}G_{12}\,\right]L^{\prime}\right]\,\left[\exp\left[-R\left(v\right)\right]\,\prod_{\ell=1}^{n_{i}}\,\frac{q^{\left(\ell\right)}\left(x_{\ell},\mu_{{\scriptscriptstyle F}}^{2}v\right)}{q^{\left(\ell\right)}\left(x_{\ell},\mu_{{\scriptscriptstyle F}}^{2}\right)}\right]\,.\label{eq:app1-CAESAR-fv-formula-ii}
\end{eqnarray}
with $-R\left(v\right)$ here as given in our eq.~\ref{eq:sect22-Big-Sudakov-minus-Ry}.

Now we start to make approximations and deviate from ref.~\cite{Banfi:2004yd},
breaking the NLL resummation by $\mathrm{N^{3}LL_{\sigma}}$ terms. Consider that $\mathcal{F}\left(R^{\prime}\right)$
resums single log terms as 
\begin{equation}
\mathcal{F}\left(R^{\prime}\right)=1+\mathcal{F}_{2}R^{\prime2}+...+\mathcal{O}\left(\mathcal{F}_{n}R^{\prime n}\right)+...\,,\label{eq:app1-CAESAR-FR-expansion}
\end{equation}
$R^{\prime}=\partial_{L}R\left(v\right)$ and so $R^{\prime n}$ is
$\mathcal{O}\left(\bar{\alpha}_{{\scriptscriptstyle \mathrm{S}}}^{n}L^{n}\right)$.
Neglecting terms of $\mathrm{N^{3}LL_{\sigma}}$ accuracy we can simply
replace $R^{\prime}=\bar{\alpha}_{{\scriptscriptstyle \mathrm{S}}}\left(y\right)\,2G_{12}\, L$
\begin{eqnarray}
\mathcal{F}\left(R^{\prime}\right) & = & \exp\left[\,\int_{0}^{L}dL^{\prime}\,\bar{\alpha}_{{\scriptscriptstyle \mathrm{S}}}^{2}\left(y\right)\,2G_{12}\,\left[\,4\mathcal{F}_{2}G_{12}\,\right]\, L^{\prime}\right]\,,\label{eq:app1-CAESAR-FR-naive-exponentation}
\end{eqnarray}
and hence 
\begin{eqnarray}
f\left(v\right) & = & \exp\left[-R\left(v\right)\right]\,\prod_{\ell=1}^{n_{i}}\,\frac{q^{\left(\ell\right)}\left(x_{\ell},\mu_{{\scriptscriptstyle F}}^{2}v\right)}{q^{\left(\ell\right)}\left(x_{\ell},\mu_{{\scriptscriptstyle F}}^{2}\right)}\,+\,\mathcal{O}\left(\mathrm{N^{3}LL}_{\sigma}\right).\label{eq:app1-fv-LL-NNLLsigma}
\end{eqnarray}
We then have the $\mathrm{LL}$/$\mathrm{NNLL}_{\sigma}$ resummed
cumulant expression 
\begin{equation}
\frac{d\Sigma_{{\scriptscriptstyle \mathcal{R}}}\left(L\right)}{d\Phi}=\frac{d\sigma_{0}}{d\Phi}\,\left(1+\bar{\alpha}_{{\scriptscriptstyle \mathrm{S}}}\left(\mu_{{\scriptscriptstyle R}}^{2}\right)\bar{\chi}_{1}\left(\Phi\right)\right)\,\left[\exp\left[-R\left(v\right)\right]\,\prod_{\ell=1}^{n_{i}}\,\frac{q^{\left(\ell\right)}\left(x_{\ell},\mu_{{\scriptscriptstyle F}}^{2}v\right)}{q^{\left(\ell\right)}\left(x_{\ell},\mu_{{\scriptscriptstyle F}}^{2}\right)}\right]\,.\label{eq:app1-starting-point-cumulant-LL-NNLLsigma}
\end{equation}

Recall that $\bar{\chi}_{1}$ encodes hard-virtual and hard-collinear
splitting corrections, and that these contributions contain terms
which cancel the $\mu_{{\scriptscriptstyle R}}$ and $\mu_{{\scriptscriptstyle F}}$
dependence of $d\sigma_{0}/d\Phi_{{\scriptscriptstyle \mathcal{B}}}$.
We may separate these parts as follows: 
\begin{eqnarray}
\bar{\chi}_{1}\left(\Phi\right) & = & \mathcal{H}_{1}\left(\Phi,\mu_{{\scriptscriptstyle R}}^{2},Q^{2}\right)+\sum_{\ell=1}^{n_{i}}\,\frac{\left[\boldsymbol{\mathcal{C}}_{1}\otimes\boldsymbol{\mathrm{q}}^{\left(\ell\right)}\right]_{i}\left(x_{\ell},\mu_{{\scriptscriptstyle F}}^{2}\right)}{q^{\left(\ell\right)}\left(x_{\ell},\mu_{{\scriptscriptstyle F}}^{2}\right)}\,,\label{eq:app1-chibar-decomposed-to-H-and-Cs}
\end{eqnarray}
where 
\begin{eqnarray}
\mathcal{H}_{1}\left(\Phi,\mu_{{\scriptscriptstyle R}}^{2},Q^{2}\right) & = & H_{1}\left(\Phi\right)+2\mathrm{q}\bar{\beta}_{0}\ln\frac{\mu_{{\scriptscriptstyle R}}}{Q}+\left[G_{11}+2S_{1}-2G_{12}\ln\frac{Q_{qq^{\prime}}}{Q}-\mathrm{q}\bar{\beta}_{0}\right]\,2\ln\frac{Q_{qq^{\prime}}}{Q}\,,\label{eq:app1-curly-H-decomposition}
\end{eqnarray}
and 
\begin{equation}
\mathcal{C}_{1,ij}\left(z,\mu_{{\scriptscriptstyle F}}^{2},Q^{2}\right)=C_{1,ij}\left(z\right)-2\ln\frac{\mu_{{\scriptscriptstyle F}}}{Q}\, P_{ij}\left(z\right)\,,\quad\quad C_{1,ij}\left(z\right)=-P_{ij}^{\epsilon}\left(z\right)-\delta_{ij}\delta\left(1-z\right)A_{ij}\frac{\pi^{2}}{12}\,.\label{eq:app1-curly-C-decomposition}
\end{equation}
We underline that in the relation between $\bar{\chi}$, $\mathcal{H}$
and $\mathcal{C}_{1}$, eq.~\ref{eq:app1-chibar-decomposed-to-H-and-Cs},
the renormalization scale in $\mathcal{H}_{1}$ is set to $\mu_{{\scriptscriptstyle R}}$
and in $\mathcal{C}_{1}$, which is convoluted with a PDF, the explicit
factorization scale therein is $\mu_{{\scriptscriptstyle F}}$, i.e.~$\mathcal{C}_{1}$
in eq.~\ref{eq:app1-chibar-decomposed-to-H-and-Cs}
is precisely as it is written in eq.~\ref{eq:app1-curly-C-decomposition}.
Equation~\ref{eq:app1-curly-H-decomposition} basically defines $H_{1}$
as what remains of $\mathcal{H}_{1}$ when its $Q$ and
$\mu_{{\scriptscriptstyle R}}$ dependence is removed, $\mathrm{q}$ being the
number of powers of $\bar{\alpha}_{{\scriptscriptstyle \mathrm{S}}}$ associated
to the Born process.
The $P_{ij}^{\epsilon}\left(z\right)$ terms are the $\mathcal{O}\left(\epsilon\right)$
parts of the LO splitting function $P_{ij}\left(z\right):$ 
\begin{equation}
\begin{array}{ll}
P_{qq}^{\epsilon}\left(z\right)=-C_{F}\left(1-z\right)\,,\quad\quad & A_{qq}=C_{F}\,,\\
P_{gq}^{\epsilon}\left(z\right)=-C_{F}z\,, & A_{gq}=0\,,\\
P_{qg}^{\epsilon}\left(z\right)=-z\left(1-z\right)\,, & A_{qg}=0\,,\\
P_{gg}^{\epsilon}\left(z\right)=0\,, & A_{gg}=C_{A}\,.
\end{array}\label{eq:app1-pij-epsilons-and-Aijs}
\end{equation}
Equations \ref{eq:app1-chibar-decomposed-to-H-and-Cs} and \ref{eq:app1-curly-H-decomposition}
serve to define $\mathcal{H}_{1}$. Whereas $\mathcal{H}_{1}$ is
dependent on the virtual corrections to the underlying hard scattering
process, the $\mathcal{C}_{1}$ pieces are due to collinear splitting
and only depend on the flavour of the (incoming) legs in the Born
configuration. While we strictly only aim for $\mathrm{NNLL}_{\sigma}$
accuracy in our initial resummation formula, to better enable comparison/extension
with existing $\mathrm{NNLL}$ work, without affecting any $\mathrm{NNLL}_{\sigma}$
terms, we opted to replace in our resummation formula (eq.~\ref{eq:app1-starting-point-cumulant-LL-NNLLsigma})
\begin{eqnarray}
1+\bar{\alpha}_{{\scriptscriptstyle \mathrm{S}}}\left(\mu_{{\scriptscriptstyle R}}^{2}\right)\bar{\chi}_{1}\left(\Phi\right) & \rightarrow & \left[1+\bar{\alpha}_{{\scriptscriptstyle \mathrm{S}}}\left(\mu_{{\scriptscriptstyle R}}^{2}\right)\,\mathcal{H}_{1}\left(\mu_{{\scriptscriptstyle R}}^{2}\right)\right]\left[1+\sum_{\ell=1}^{n_{i}}\,\bar{\alpha}_{{\scriptscriptstyle \mathrm{S}}}\left(\mu_{{\scriptscriptstyle R}}^{2}v\right)\,\frac{\left[\boldsymbol{\mathcal{C}}_{1}\otimes\boldsymbol{\mathrm{q}}^{\left(\ell\right)}\right]_{i}\left(x_{\ell},\mu_{{\scriptscriptstyle F}}^{2}v\right)}{q_{i}^{\left(\ell\right)}\left(x_{\ell},\mu_{{\scriptscriptstyle F}}^{2}v\right)}\right]\,.\quad\label{eq:app1-replacement-for-chibar}
\end{eqnarray}
Note in particular the introduction of the $v$ dependence in the renormalisation and factorisation scales in the final term.
From here eq.~\ref{eq:sect22-NNLL_sigma-differential-xsecn-i} follows
immediately on differentiation with respect to $L$.

\subsection{Fixed order expansion of resummation and \noun{Minlo} formulae\label{sub:Fixed-order-expansion}}

Here in eqs.~\ref{eq:app3-dsigmaS-FO-expansion-Hprime-11-10}-\ref{eq:app3-dsigmaS-FO-expansion-Hprime-20},
we record the $H_{nm}$ coefficients of the $\bar{\alpha}_{{\scriptscriptstyle \mathrm{S}}}^{n}L^{m}$,
$v\rightarrow0$ singular, terms in the NLO cross section, eqs.~\ref{eq:sect24-NLO-decomposed-as-S-plus-F},
\ref{eq:sect24-dsigmaS-FO-expansion-master-formula}, obtained by
expanding the $\mathrm{NNLL}_{\sigma}$ resummation formula eq.~\ref{eq:sect22-NNLL_sigma-differential-xsecn-i}:
\begin{eqnarray}
H_{11} & = & 2G_{12}\,,\quad\quad H_{10}=G_{11}+2S_{1}-\sum_{\ell=1}^{n_{i}}\,\frac{\left[\mathrm{\boldsymbol{\mathrm{P}}}\otimes\boldsymbol{\mathrm{q}}^{\left(\ell\right)}\right]_{i}}{q^{\left(\ell\right)}}\,,\label{eq:app3-dsigmaS-FO-expansion-Hprime-11-10}\\
\nonumber \\
H_{23} & = & 2G_{12}^{2}\,,\quad\quad H_{22}=\bar{\beta}_{0}H_{11}+3G_{12}\, H_{10}\,,\label{eq:app3-dsigmaS-FO-expansion-Hprime-23-22}\\
\nonumber \\
H_{21} & = & \left[K+4\mathcal{F}_{2}G_{12}+2\bar{\beta}_{0}\ln\frac{\mu_{{\scriptscriptstyle R}}}{Q}\right]H_{11}+\left[G_{11}+2S_{1}\right]^{2}+\bar{\chi}_{1}\, H_{11}+\bar{\beta}_{0}H_{10}\label{eq:app3-dsigmaS-FO-expansion-Hprime-21}\\
 & - & 2\left(G_{11}+2S_{1}\right)\sum_{\ell=1}^{n_{i}}\,\frac{\left[\mathrm{\boldsymbol{\mathrm{P}}}\otimes\boldsymbol{\mathrm{q}}^{\left(\ell\right)}\right]_{i}}{q^{\left(\ell\right)}}+2\sum_{\ell_{1}}^{n_{i}}\sum_{\ell_{2}<\ell_{1}}^{n_{i}}\,\frac{\left[\boldsymbol{\mathrm{P}}\otimes\boldsymbol{\mathrm{q}}^{\left(\ell_{1}\right)}\right]_{i}}{q^{\left(\ell_{1}\right)}}\,\frac{\left[\boldsymbol{\mathrm{P}}\otimes\boldsymbol{\mathrm{q}}^{\left(\ell_{2}\right)}\right]_{j}}{q^{\left(\ell_{2}\right)}}\nonumber \\
 & + & \sum_{\ell=1}^{n_{i}}\,\frac{\left[\mathrm{\boldsymbol{\mathrm{P}}}\otimes\mathrm{\boldsymbol{\mathrm{P}}}\otimes\boldsymbol{\mathrm{q}}^{\left(\ell\right)}\right]_{i}}{q^{\left(\ell\right)}}\,,\nonumber \\
\nonumber \\
H_{20} & = & H_{10}\,\left[\mathcal{H}_{1}\left(\mu_{{\scriptscriptstyle R}}^{2}\right)+2\bar{\beta}_{0}\ln\frac{\mu_{{\scriptscriptstyle R}}}{Q}\right]+\left[G_{11}+2S_{1}\right]\sum_{\ell=1}^{n_{i}}\,\frac{\left[\boldsymbol{\mathcal{C}}_{1}\otimes\boldsymbol{\mathrm{q}}^{\left(\ell\right)}\right]_{i}}{q^{\left(\ell\right)}}\label{eq:app3-dsigmaS-FO-expansion-Hprime-20}\\
 & - & \sum_{\ell=1}^{n_{i}}\,\frac{\left[\mathrm{\boldsymbol{\mathrm{P}}}_{2}\otimes\boldsymbol{\mathrm{q}}^{\left(\ell\right)}\right]_{i}}{q^{\left(\ell\right)}}-2\bar{\beta}_{0}\ln\frac{Q}{\mu_{{\scriptscriptstyle F}}}\,\sum_{\ell=1}^{n_{i}}\,\frac{\left[\mathrm{\boldsymbol{\mathrm{P}}}\otimes\boldsymbol{\mathrm{q}}^{\left(\ell\right)}\right]_{i}}{q^{\left(\ell\right)}}\nonumber \\
 & + & \bar{\beta}_{0}\sum_{\ell=1}^{n_{i}}\,\frac{\left[\boldsymbol{\mathcal{C}}_{1}\otimes\boldsymbol{\mathrm{q}}^{\left(\ell\right)}\right]_{i}}{q^{\left(\ell\right)}}-\sum_{\ell=1}^{n_{i}}\,\frac{\left[\boldsymbol{\mathcal{C}}_{1}\otimes\mathrm{\boldsymbol{\mathrm{P}}}\otimes\boldsymbol{\mathrm{q}}^{\left(\ell\right)}\right]_{i}}{q^{\left(\ell\right)}}\nonumber \\
 & - & \sum_{\ell_{1}}^{n_{i}}\sum_{\ell_{2}<\ell_{1}}^{n_{i}}\left(\frac{\left[\boldsymbol{\mathrm{P}}\otimes\boldsymbol{\mathrm{q}}^{\left(\ell_{1}\right)}\right]_{i}}{q^{\left(\ell_{1}\right)}}\,\frac{\left[\boldsymbol{\mathcal{C}}_{1}\otimes\boldsymbol{\mathrm{q}}^{\left(\ell_{2}\right)}\right]_{i}}{q^{\left(\ell_{2}\right)}}+\frac{\left[\boldsymbol{\mathrm{P}}\otimes\boldsymbol{\mathrm{q}}^{\left(\ell_{2}\right)}\right]_{j}}{q^{\left(\ell_{2}\right)}}\,\frac{\left[\boldsymbol{\mathcal{C}}_{1}\otimes\boldsymbol{\mathrm{q}}^{\left(\ell_{1}\right)}\right]_{i}}{q^{\left(\ell_{1}\right)}}\right)\,.\nonumber 
\end{eqnarray}
The factorization scale in $\mathcal{C}_{1}$ in eqs.~\ref{eq:app3-dsigmaS-FO-expansion-Hprime-21}-\ref{eq:app3-dsigmaS-FO-expansion-Hprime-20}
(including the $\mathcal{C}_{1}$ implicit in $\bar{\chi}$) is set
to $\mu_{{\scriptscriptstyle F}}$; it is exactly as written in eq.~\ref{eq:app1-curly-C-decomposition}.
We point out that for the regime $y_{01}\ll m_{{\scriptscriptstyle \mathcal{B}}}^{2}$,
in $\mathcal{B}$\noun{jj-Minlo}, the virtual corrections, $\mathcal{H}_{1}$,
will contain large logarithms of ratios of scales deriving from the
related underlying ($\mathcal{B}$\noun{j}) Born kinematics approaching
a singular region.

The $d\sigma_{{\scriptscriptstyle \mathcal{S}}}$ expansion of eq.~\ref{eq:sect24-dsigmaS-FO-expansion-master-formula},
with $H_{nm}$ as given in eqs.~\ref{eq:app3-dsigmaS-FO-expansion-Hprime-11-10}-\ref{eq:app3-dsigmaS-FO-expansion-Hprime-20},
is invariant under $\mu_{{\scriptscriptstyle R}}$ and $\mu_{{\scriptscriptstyle F}}$
variations up to higher order terms ($\propto d\sigma_{0}\bar{\alpha}_{{\scriptscriptstyle \mathrm{S}}}^{3}$)
beyond NLO accuracy. Also $d\sigma_{{\scriptscriptstyle \mathcal{S}}}$
with these $H_{nm}$ is invariant under variations of the resummation
scale, $Q$, up to and including $\mathrm{NNLL}_{\sigma}$ terms. To make
$d\sigma_{{\scriptscriptstyle \mathcal{S}}}$ invariant under resummation
scale variations also at the $\mathrm{N}^{3}\mathrm{LL}_{\sigma}$
level, requires modifying $H_{20}\rightarrow H_{20}+\left[K+4\mathcal{F}_{2}G_{12}\right]\, H_{11}\,\ln(Q_{qq}^{2}/Q^{2})$
only. Such a modification could be easily generated by simple adjustment
of our initial resummation formula, however, since the latter is only
guaranteed to reproduce terms to $\mathrm{NNLL}_{\sigma}$ accuracy
anyway, we do not consider this.

We have compared our expansion formula, eq.~\ref{eq:sect24-dsigmaS-FO-expansion-master-formula},
to known results for the W/Z and Higgs boson transverse momentum spectra
\cite{Arnold:1990yk,Glosser:2002gm}, as well as to those of the associated
leading jet (derived by expanding the NNLL resummation of Banfi et\emph{
}al \cite{Banfi:2012jm}). To ease comparisons, we note the following
relations between our notation and refs.~\cite{Arnold:1990yk,Glosser:2002gm,Banfi:2012jm}:
\begin{equation}
2G_{12}=-A^{\left(1\right)}\,,\quad G_{11}=-B^{\left(1\right)}\,,\quad2G_{12}K=-A^{\left(2\right)}\,,\label{eq:app3-notational-correspondence-wrt-arnold-glosser-banfi}
\end{equation}
where $A^{\left(1\right)}$, $B^{\left(1\right)}$ and $A^{\left(2\right)}$
are used in the latter articles. We also point out that for $\mathcal{B}$
production $\mathcal{F}_{2}=0$ in both the $\mathcal{B}$ transverse
momentum spectrum and that of the leading jet (eq.~\ref{eq:sect22-F2-general-kt-algo-formula}).
Lastly, in the results of refs.~\cite{Arnold:1990yk,Glosser:2002gm}
the resummation scale is set to the invariant mass of $\mathcal{B}$,
i.e.~$Q=Q_{qq^{\prime}}$ in our notation, leading to $S_{1}\rightarrow0$
here, as well as simplifications in the $\bar{\chi}$ and $\mathcal{H}_{1}$
functions.

With the correspondence in notation understood, we find complete agreement
between our singular NLO expansion formula, eqs.~\ref{eq:sect24-NLO-decomposed-as-S-plus-F}-\ref{eq:sect24-dsigmaS-FO-expansion-master-formula},
and those of refs.~\cite{Arnold:1990yk,Glosser:2002gm,Banfi:2012jm},
up to and including $\mathrm{NNLL}_{\sigma}$ terms. To also have
agreement with refs.~\cite{Arnold:1990yk,Glosser:2002gm} for the
$\mathrm{N^{3}LL}_{\sigma}$ terms in the Higgs/vector boson transverse
momentum spectrum, we need only add to $H_{20}$, in eq.~\ref{eq:app3-dsigmaS-FO-expansion-Hprime-20},
\begin{equation}
\widetilde{R}_{20}=-\left[B^{\left(2\right)}+2\zeta_{3}\left[A^{\left(1\right)}\right]^{2}\right]\,,\label{eq:app3-missing-R20-piece-for-B-pt}
\end{equation}
with $B^{\left(2\right)}$ as given in ref.~\cite{Banfi:2012jm}.
For full agreement with ref.~\cite{Banfi:2012jm}, including $\mathrm{N^{3}LL}_{\sigma}$
terms, we only have to add to $H_{20}$, in eq.~\ref{eq:app3-dsigmaS-FO-expansion-Hprime-20},
\begin{equation}
\widetilde{R}_{20}=-\left[B^{\left(2\right)}-8C\left(f^{\mathrm{clust}}+f^{\mathrm{correl}}\right)\right]\,,\label{eq:app3-missing-R20-piece-for-J1-pt}
\end{equation}
where $f^{\mathrm{clust}}$ and $f^{\mathrm{correl}}$ are corrections
due to jet radius dependent clustering/correlated emission effects,
as given in ref.~\cite{Banfi:2012jm}. We point out that the needed/missing
$\widetilde{R}_{20}$ term here (eqs.~\ref{eq:app3-missing-R20-piece-for-B-pt}-\ref{eq:app3-missing-R20-piece-for-J1-pt}),
for $\mathcal{B}$-production, is just a number with no dependence
on kinematics.

\subsection{Basic \noun{Minlo} prescription for merging two units of multiplicity\label{sub:minlo-steps-for-two-units-of-multiplicity}}

In discussing the merging of two units of multiplicity
(sect.~\ref{sec:Merging-without-a-merging-scale-two-units}), the basic
\noun{Minlo} prescription amounts, in practice, to the following sequence
of operations applied to the input NLO calculation 
\begin{enumerate}
\item [0.]\setcounter{enumi}{0}Define $\mu_{{\scriptscriptstyle R}}=K_{{\scriptscriptstyle R}}\max(Q_{{\scriptscriptstyle \mathcal{B}}},Q_{{\scriptscriptstyle \mathcal{B}\mathrm{J}}})$
and $\mu_{{\scriptscriptstyle F}}=K_{{\scriptscriptstyle F}}Q$, where
$K_{{\scriptscriptstyle R/F}}\in\left[\frac{1}{2},2\right]$.%
\footnote{This is the definition of $Q_{{\scriptscriptstyle \mathcal{B}}}$
given in sect.~\ref{sub:Definitions-jet-resolutions}.%
} 
\item Set $\mu_{{\scriptscriptstyle R}}$ everywhere it occurs and, likewise,
for all $\mu_{{\scriptscriptstyle F}}$ set $\mu_{{\scriptscriptstyle F}}\rightarrow\mu_{{\scriptscriptstyle F}}\,\sqrt{v}$:
\begin{equation}
d\sigma\rightarrow d\sigma^{\prime}=d\sigma\,\left(\mu_{{\scriptscriptstyle R}}=K_{{\scriptscriptstyle R}}\max(Q_{{\scriptscriptstyle \mathcal{B}}},Q_{{\scriptscriptstyle \mathcal{B}\mathrm{J}}}),\,\mu_{{\scriptscriptstyle F}}\rightarrow K_{{\scriptscriptstyle F}}\,\sqrt{y}\right)\,.\label{eq:sect24-minlo-step-1}
\end{equation}

\item Replace the additional power of $\bar{\alpha}_{{\scriptscriptstyle \mathrm{S}}}$
that accompanies the NLO corrections according to 
\begin{equation}
d\sigma^{\prime}\rightarrow d\sigma^{\prime\prime}=d\sigma^{\prime}\,\left(\bar{\alpha}_{{\scriptscriptstyle \mathrm{S}}}^{{\scriptscriptstyle \mathrm{NLO}}}\left(\mu_{{\scriptscriptstyle R}}^{2}\right)\rightarrow\bar{\alpha}_{{\scriptscriptstyle \mathrm{S}}}\left(K_{{\scriptscriptstyle R}}^{2}\, y\right)\right)\,.\label{eq:sect24-minlo-step-2}
\end{equation}

\item Multiply the LO component by the $\mathcal{O}\left(\bar{\alpha}_{{\scriptscriptstyle \mathrm{S}}}\right)$
expansion of the inverse of the Sudakov form factor times $\bar{\alpha}_{s}\left(K_{{\scriptscriptstyle R}}^{2}\, y\right)/\bar{\alpha}_{s}\left(\mu_{{\scriptscriptstyle R}}^{2}\right)$:
\begin{eqnarray}
d\sigma^{\prime\prime}\rightarrow d\sigma^{\prime\prime\prime} & = & d\sigma^{\prime\prime}-\left.d\sigma^{\prime\prime}\right|_{{\scriptscriptstyle \mathrm{LO}}}\bar{\alpha}_{{\scriptscriptstyle \mathrm{S}}}\left(K_{{\scriptscriptstyle R}}^{2}\, y\right)\left(G_{12}L^{2}+\left(G_{11}+2S_{1}+\bar{\beta}_{0}\right)L+2\bar{\beta}_{0}\ln\frac{\mu_{{\scriptscriptstyle R}}}{K_{{\scriptscriptstyle R}}\, Q}\right)\,.\hspace{1em}\hspace{1em}\label{eq:sect24-minlo-step-3-subtracted-xsecn-defn}
\end{eqnarray}

\item Multiply by the\noun{ }Sudakov form factor times $\bar{\alpha}_{s}\left(K_{{\scriptscriptstyle R}}^{2}\, y\right)/\bar{\alpha}_{s}\left(\mu_{{\scriptscriptstyle R}}^{2}\right)$:\label{enu:Minlo-procedure-last-step}
\begin{eqnarray}
d\sigma^{\prime\prime\prime}\rightarrow d\sigma_{{\scriptscriptstyle \mathcal{M}}} & = & \exp\left[-R\left(v\right)\right]\,\frac{\bar{\alpha}_{s}\left(K_{{\scriptscriptstyle R}}^{2}\, y\right)}{\bar{\alpha}_{s}\left(\mu_{{\scriptscriptstyle R}}^{2}\right)}\, d\sigma^{\prime\prime\prime}\,.\label{eq:sect24-minlo-step-4-sudakov-aS-ratio-times-subtracted-xsecn}
\end{eqnarray}

\end{enumerate}


Precisely, the steps outlined above are those used in the construction
of the $\mathcal{B}$\noun{j-Minlo}$^{\prime}$ simulation of
ref.~\cite{Hamilton:2012rf}, adopting the general notation of
sect.~\ref{sec:Merging-without-a-merging-scale-two-units}, so that they
apply equally to $\mathcal{B}$\noun{jj-Minlo}$^{\prime}$ --- at least for
what concerns the discussion on merging two units of multiplicity. 

\subsection{Integral of \noun{Minlo} \texorpdfstring{$v$}{v} spectrum\label{sub:Integral-of-Minlo-v-spectrum}}

The \noun{Minlo} cumulant cross section below $v$ is defined 
\begin{equation}
\frac{d\Sigma_{{\scriptscriptstyle \mathcal{M}}}\left(L\right)}{d\Phi}=\int_{\infty}^{L}dL^{\prime}\,\frac{d\sigma_{{\scriptscriptstyle \mathcal{M}}}}{d\Phi dL^{\prime}}\,.\label{eq:app5-dSigmaM-defn}
\end{equation}
We are interested in the expansion of the latter up to and including
NLO terms. Noting that $d\sigma_{{\scriptscriptstyle \mathcal{R}}}$
is a total derivative with respect to $L$,
and using the definitions of $\bar{\chi}$ in terms of $\mathcal{H}_{1}$
and $\mathcal{C}_{1}$, we have
\begin{eqnarray}
\frac{d\Sigma_{{\scriptscriptstyle \mathcal{M},1}}\left(L\right)}{d\Phi} & = & \left.\frac{d\Sigma_{{\scriptscriptstyle \mathcal{R},1}}\left(L\right)}{d\Phi}\right|_{{\scriptscriptstyle \bar{\chi}_{1}\rightarrow0}}+\frac{d\sigma_{0}}{d\Phi}\,\bar{\alpha}_{{\scriptscriptstyle \mathrm{S}}}\bar{\chi}_{1}+\int_{\infty}^{L}dL^{\prime}\,\frac{d\sigma_{{\scriptscriptstyle \mathcal{F},1}}}{d\Phi dL^{\prime}}+\int_{\infty}^{L}dL^{\prime}\,\frac{d\sigma_{{\scriptscriptstyle \mathcal{MR}}}}{d\Phi dL^{\prime}}+\mathcal{O}\left(\bar{\alpha}_{{\scriptscriptstyle \mathrm{S}}}^{2}\right)\,,\label{eq:app5-dSigmaM-expanded-to-NLO-i}\\
 & = & \frac{d\Sigma_{{\scriptscriptstyle \mathrm{NLO}}}\left(L\right)}{d\Phi}+\int_{\infty}^{L}dL^{\prime}\,\frac{d\sigma_{{\scriptscriptstyle \mathcal{MR}}}}{d\Phi dL^{\prime}}+\mathcal{O}\left(\bar{\alpha}_{{\scriptscriptstyle \mathrm{S}}}^{2}\right)\,.\label{eq:app5-dSigmaM-expanded-to-NLO-ii}
\end{eqnarray}
In determining the equality in eq.~\ref{eq:app5-dSigmaM-expanded-to-NLO-i}
we have made use of the relation in eq.~\ref{eq:sect22-dSigmaR-with-chibar-set-to-zero-expanded-to-NLO}.
In going from eq.~\ref{eq:app5-dSigmaM-expanded-to-NLO-i} to eq.~\ref{eq:app5-dSigmaM-expanded-to-NLO-ii}
we have made use of the $\bar{\chi}_{1}$ definition in eq.~\ref{eq:sect22-dSigmaNLO}.

\subsection{\texorpdfstring{$\delta\left(\Phi\right)$}{d(Phi)} denominator\label{sub:Expression-for-denominator-of-delta-Phi}}

Neglecting $\mathrm{N}^{3}\mathrm{LL}_{\sigma}$ terms (as in sect.~\ref{sub:Removal-of-spurious-terms})
we obtain for the denominator of $\delta\left(\Phi\right)$ in eq.~\ref{eq:sect26-first-delta-defn}
\begin{equation}
\int dL\, h\left(L\right)\,\frac{d\sigma_{{\scriptscriptstyle \mathcal{M}}}}{d\Phi dL}=\frac{d\sigma_{0}}{d\Phi}\,\frac{1-\exp\left[-\left|G_{12}\right|\rho\right]}{\left|G_{12}\right|}\,\bar{\alpha}_{{\scriptscriptstyle \mathrm{S}}}\left(1+\mathcal{O}\left(\sqrt{\bar{\alpha}_{{\scriptscriptstyle \mathrm{S}}}}\right)\right),\label{eq:app6-delta22-denominator-to-NNLLsigma-1}
\end{equation}
where the $\mathcal{O}\left(\sqrt{\bar{\alpha}_{{\scriptscriptstyle \mathrm{S}}}}\right)$
comes from $\mathrm{N}^{3}\mathrm{LL}_{\sigma}$ terms in the integrand.

\subsection{\texorpdfstring{$n=3$}{n=3} soft wide angle radiation coefficient \texorpdfstring{$S_{1}$}{S1}\label{sub:Expressions-for-n-eq-3-soft-wide-angle-coeffs-S1}}

\subsubsection*{Hadronic initial-state}

For hadronic initial-states, plugging in $Q_{{\scriptscriptstyle \mathcal{B}}}^{2}=m_{{\scriptscriptstyle \mathcal{B}}}^{2}$
and $Q_{{\scriptscriptstyle \mathcal{B}\mathrm{J}}}^{2}=y_{01}$,
eq.~\ref{eq:sect22-n-eq-2-S1-defn} for $n=2$ hard legs in the Born
process gives simply $S_{1}=0$, while eq.~\ref{eq:sect22-n-eq-3-S1-defn}
for $n=3$ hard legs in the Born process gives

\begin{eqnarray}
qq^{\prime}\rightarrow Vg: &  & \phantom{\Delta}S_{1}=-\frac{1}{2}\left(C_{q}+C_{q^{\prime}}\right)L_{01}+\Delta S_{1}\,,\qquad\qquad\qquad\qquad\qquad\quad\label{eq:app7-n-eq-3-qqpr-Bg-S1-i}\\
 &  & \Delta S_{1}=+\frac{1}{2}\left(C_{q}+C_{q^{\prime}}\right)\ln\frac{m_{{\scriptscriptstyle \mathcal{B}}}^{2}}{Q_{qq^{\prime}}^{2}}\,,\label{eq:app7-n-eq-3-qqpr-Bg-DeltaS1-i}\\
 &  & \phantom{\Delta S_{1}}=+\frac{1}{2}\left(C_{q}+C_{q^{\prime}}\right)\ln z\,,\label{eq:app7-n-eq-3-qqpr-Bg-DeltaS1-ii-exact}
\end{eqnarray}
~ 
\begin{eqnarray}
qg\rightarrow Vq^{\prime}: &  & \phantom{\Delta}S_{1}=-\frac{1}{2}\left(C_{q}+C_{q^{\prime}}\right)L_{01}+\Delta S_{1}\,,\label{eq:app7-n-eq-3-qg-Bqpr-S1-i}\\
 &  & \Delta S_{1}=+\frac{1}{2}\left(C_{q}+C_{q^{\prime}}\right)\ln\frac{m_{{\scriptscriptstyle \mathcal{B}}}^{2}}{Q_{qg}^{2}}+\frac{1}{2}\left(C_{q}+C_{q^{\prime}}-2C_{g}\right)\ln\frac{Q_{qg}^{2}}{Q_{qq^{\prime}}^{2}}\,,\label{eq:app7-n-eq-3-qg-Bqpr-DeltaS1-i}\\
\lim_{Q_{q^{\prime}g}\rightarrow0} &  & \Delta S_{1}=+\frac{1}{2}\left(C_{q}+C_{q^{\prime}}\right)\ln z-\frac{1}{2}\left(C_{q}+C_{q^{\prime}}-2C_{g}\right)\ln\left(1-z\right)\,,\label{eq:app7-n-eq-3-qg-Bqpr-DeltaS1-ii-small-Qqprg-approx}
\end{eqnarray}
~ 
\begin{eqnarray}
qg\rightarrow Hq^{\prime}: &  & \phantom{\Delta}S_{1}=-\frac{1}{2}\left(C_{g}+C_{g}\right)L_{01}+\Delta S_{1}\,,\label{eq:app7-n-eq-3-qg-Bqpr-S1-alt-i}\\
 &  & \Delta S_{1}=+\frac{1}{2}\left(C_{g}+C_{g}\right)\ln\frac{m_{{\scriptscriptstyle \mathcal{B}}}^{2}}{Q_{qg}^{2}}-\frac{1}{2}\left(C_{q}+C_{q^{\prime}}-2C_{g}\right)\ln\frac{Q_{qg}^{2}}{Q_{q^{\prime}g}^{2}}\,,\label{eq:app7-n-eq-3-qg-Bqpr-DeltaS1-alt-i}\\
\lim_{Q_{qq^{\prime}}\rightarrow0} &  & \Delta S_{1}=+\frac{1}{2}\left(C_{g}+C_{g}\right)\ln z+\frac{1}{2}\left(C_{q}+C_{q^{\prime}}-2C_{g}\right)\ln\left(1-z\right)\,,\label{eq:app7-n-eq-3-qg-Bqpr-DeltaS1-alt-ii-small-Qqqpr-approx}\\
\nonumber 
\end{eqnarray}
where $V$ refers to a $W/Z$ vector boson and $z\equiv m_{{\scriptscriptstyle \mathcal{B}}}^{2}/\hat{s}$.
For convenience we note that in the $qq^{\prime}\rightarrow Vg$ channel,
without approximations, $y_{01}=Q_{qg}^{2}Q_{q^{\prime}g}^{2}/Q_{qq^{\prime}}^{2}$,
while in $qg\rightarrow Vq^{\prime}$ and $qg\rightarrow Hq^{\prime}$,
also without any approximation, $y_{01}=Q_{qq^{\prime}}^{2}Q_{q^{\prime}g}^{2}/Q_{qg}^{2}$.
For the $gg\rightarrow Hg$ process $S_{1}$ is exactly as in eq.~\ref{eq:app7-n-eq-3-qqpr-Bg-S1-i}
with the replacements $\left\{ q,\, q^{\prime},\, g\right\} \rightarrow\left\{ g_{1},\, g_{2},\, g_{3}\right\} $,
where $g_{1}$ and $g_{2}$ refer to the two incoming gluons; it follows
that in $gg\rightarrow Hg$ we have exactly $y_{01}=Q_{g_{1}g_{3}}Q_{g_{2}g_{3}}/Q_{g_{1}g_{2}}$.

Lastly we note the following approximations used in arriving at the
limit $Q_{q^{\prime}g}\rightarrow0$ in eq.~\ref{eq:app7-n-eq-3-qg-Bqpr-DeltaS1-ii-small-Qqprg-approx}
\begin{equation}
\hat{s}=Q_{qg}^{2}\,,\qquad\hat{t}=-Q_{q^{\prime}g}^{2}\rightarrow-p_{{\scriptscriptstyle \mathrm{T}}}^{2}\,\frac{1}{1-z}\,,\qquad\hat{u}=-Q_{qq^{\prime}}^{2}\rightarrow-m_{{\scriptscriptstyle \mathcal{B}}}^{2}\,\frac{1-z}{z}\,,\label{eq:app7-Qqprg-to-zero-approximations}
\end{equation}
and for the limit $Q_{qq^{\prime}}\rightarrow0$ used in eq.~\ref{eq:app7-n-eq-3-qg-Bqpr-DeltaS1-alt-ii-small-Qqqpr-approx}
\begin{equation}
\hat{s}=Q_{qg}^{2}\,,\qquad\hat{t}=-Q_{qq^{\prime}}^{2}\rightarrow-p_{{\scriptscriptstyle \mathrm{T}}}^{2}\,\frac{1}{1-z}\,,\qquad\hat{u}=-Q_{q^{\prime}g}^{2}\rightarrow-m_{{\scriptscriptstyle \mathcal{B}}}^{2}\,\frac{1-z}{z}\,.\label{eq:app7-Qqqpr-to-zero-approximations}
\end{equation}

\subsection{\texorpdfstring{$\mathrm{NLL}_{\sigma}$}{NLLo} resummation formula in \noun{Ckkw} notation\label{sub:NLLsigma-resummation-in-CKKW-notation}}

In the notation commonly used for the coherent parton branching formalism
and \noun{Ckkw} method, Sudakov form factors for quark and gluon evolution
are typically written as 
\begin{equation}
\Delta_{q}\left(q,Q\right)=\exp\left[-\int_{q}^{Q}dq\,\Gamma_{q}^{\prime}\left(q,Q\right)\right]\,,\qquad\Delta_{g}\left(q,Q\right)=\exp\left[-\int_{q}^{Q}dq\,\Gamma_{g}^{\prime}\left(q,Q\right)\right]\,,\,\,\label{eq:app8-CKKW-Sudakovs}
\end{equation}
with $\Gamma_{q/g}^{\prime}$ given by 
\begin{equation}
\Gamma_{g}^{\prime}\left(q,Q\right)=\Gamma_{g}\left(q,Q\right)+\Gamma_{f}\left(q\right)\,,\qquad\Gamma_{q}^{\prime}\left(q,Q\right)=\Gamma_{q}\left(q,Q\right)\,,\label{eq:app8-CKKW-Gammag-and-Gammaq-prime}
\end{equation}
and $\Gamma_{q/g/f}$ functions therein defined as 
\begin{eqnarray}
\Gamma_{g}\left(q,Q\right) & = & \frac{2}{q}\,\bar{\alpha}_{{\scriptscriptstyle \mathrm{S}}}\left(q\right)\,\left(C_{\ell}\ln\frac{Q^{2}}{q^{2}}+2B_{\ell}C_{\ell}\right)-\Gamma_{f}\left(q\right)\,,\qquad\Gamma_{f}\left(q\right)=\frac{2}{q}\,\bar{\alpha}_{{\scriptscriptstyle \mathrm{S}}}\left(q\right)\,\frac{N_{f}}{3}\,,\qquad\label{eq:app8-CKKW-Gammag-Gammaf}\\
\Gamma_{q}\left(q,Q\right) & = & \frac{2}{q}\,\bar{\alpha}_{{\scriptscriptstyle \mathrm{S}}}\left(q\right)\,\left(C_{\ell}\ln\frac{Q^{2}}{q^{2}}+2B_{\ell}C_{\ell}\right)\,.\label{eq:app8-CKKW-Gammaq}
\end{eqnarray}

\subsection{Expansion of \texorpdfstring{$\mathrm{NLL}_{\sigma}$}{NLLo} formula\label{sub:Fixed-order-expansion-of-conjectured-resummation-formula}}

The coefficients for the fixed order expansion of our conjectured
resummation formula in section \ref{sub:Extended-Bjj-Minlo} (eq.~\ref{eq:sect-32-expanded-resummation-formula})
are given by 
\begin{eqnarray}
R_{12}^{{\scriptscriptstyle \left[01\right]}} & = & G_{12}^{{\scriptscriptstyle \left[01\right]}}\,,\quad\quad R_{11}^{{\scriptscriptstyle \left[01\right]}}=2\bar{\beta}_{0}+G_{11}^{{\scriptscriptstyle \left[01\right]}}+2S_{1}^{{\scriptscriptstyle \left[01\right]}}\,,\label{eq:app9-conjectured-formula-expansion-coeff-a}\\
\nonumber \\
H_{11}^{{\scriptscriptstyle \left[12\right]}} & = & 2G_{12}^{{\scriptscriptstyle \left[12\right]}}\,,\quad\quad H_{10}^{{\scriptscriptstyle \left[12\right]}}=G_{11}^{{\scriptscriptstyle \left[12\right]}}+2S_{1}^{{\scriptscriptstyle \left[12\right]}}-\sum_{\ell=1}^{n_{i}}\,\frac{\left[\mathrm{\boldsymbol{\mathrm{P}}}\otimes\boldsymbol{\mathrm{q}}^{\left(\ell\right)}\right]_{i}\left(x_{\ell}^{{\scriptscriptstyle \left[12\right]}},y_{01}\right)}{q^{\left(\ell\right)}\left(x_{\ell}^{{\scriptscriptstyle \left[12\right]}},y_{01}\right)}\,,\label{eq:app9-conjectured-formula-expansion-coeff-b}\\
\nonumber \\
H_{23}^{{\scriptscriptstyle \left[12\right]}} & = & 2G_{12}^{{\scriptscriptstyle \left[12\right]}2}\,,\quad\quad H_{22}^{{\scriptscriptstyle \left[12\right]}}=\bar{\beta}_{0}\, H_{11}^{{\scriptscriptstyle \left[12\right]}}+3G_{12}^{{\scriptscriptstyle \left[12\right]}}\, H_{10}^{{\scriptscriptstyle \left[12\right]}}\,.\label{eq:app9-conjectured-formula-expansion-coeff-c}
\end{eqnarray}
 \bibliographystyle{jhep}
\bibliography{paper}

\end{document}